%% file: Sumrule2.tex
\documentclass[a4paper, twoside, 12pt]{book}
\usepackage{epsfig}
\usepackage{mathrsfs}
\usepackage{psfrag}
\usepackage{bbm}
\usepackage{a4}
\usepackage{wasysym}
\usepackage[footnotesize,bf]{caption}
\begin{document}
\renewcommand{\thepage}{}

\begin{titlepage}

 \centering
  \vspace*{\fill}
  {\LARGE\bf Faculty of Physics and Astronomy}
  \vfill
  {\Large\bf University of Heidelberg}
  \vfill
  \vfill
  \vfill
  \vfill
  \vfill
  \vfill
   Diploma thesis\\%[0.5cm]
   in Physics\\[0.5cm]
   submitted by\\%[0.5cm]
  {\bf Felix Schwab}\\%[0.5cm]
   born in Lich\\%[0.5cm]
   2002
\end{titlepage}
\thispagestyle{empty}

\begin{titlepage}
  \centering
  \vspace*{\fill}
  {\huge\sc Strange Quark Mass Determination From\\ Sum Rules For Hadronic $\mathbf{\tau}$-Decays\\ \ }
  \vfill
  \vspace{1cm}
  \vfill
  \vfill
  {\bf This diploma thesis has been carried out by Felix Schwab at the\\[0.5cm]
   Institute for Theoretical Physics\\[0.5cm]
   under the supervision of\\[0.5cm]
   Priv. Doz. Matthias Jamin}
\end{titlepage}
\thispagestyle{empty}
\cleardoublepage
\thispagestyle{empty}
\cleardoublepage

\centerline{\large\bf Strange Quark Mass Determination from}
\centerline{\large\bf Sum Rules for Hadronic $\mathbf{\tau}$-Decays}\vspace{0.5cm}
\centerline{ Abstract}\vspace{0.5cm}
\noindent
We discuss the ratio of hadronic to leptonic $\tau$-decays, that can be expanded in an operator product expansion. 
The sensitivity to the strange mass is increased, if only the flavor-breaking difference of strange to 
non-strange currents is considered, which leaves us with terms that are at leading order proportional to $m_s^2$. In this 
sum rule the scalar channel exhibits a very bad convergence behavior in the perturbation series, that we control by replacing the 
scalar OPE with a phenomenological ansatz for the scalar and pseudoscalar spectral functions. The spectral ansatz is 
 compared to the theoretical (OPE) expressions and we show that the results agree within the 
uncertainties and that the uncertainties of our ansatz are smaller than those of the theoretical expressions. This 
allows us to considerably reduce the uncertainties in comparison to previous $m_s$ determinations from $\tau$-decays. 
Our final results are $m_s$(2~GeV) = 117$\pm$17~MeV for the unitarity fit of $\left|V_{us}\right|$ and $m_s$(2~GeV) = 103$\pm$17~MeV for 
the Leutwyler-Roos value of $\left|V_{us}\right|$.
\vfill
\centerline{\large\bf Strange-Quark Massenbestimmung aus}
\centerline{\large\bf Summenregeln f\"ur Hadronische $\mathbf{\tau}$-Zerf\"alle}\vspace{0.5cm}
\centerline{Zusammenfassung}\vspace{0.5cm}
\noindent
Wir untersuchen das Verh\"altnis von hadronischen zu leptonischen $\tau$-Zerf\"allen, das in einer Operator Produkt 
Entwicklung entwickelt werden kann. Um die Sensititivit\"at auf die Masse des Strange-Quark zu erh\"ohen, wird nur die flavor 
brechende Differenz aus $ud$ and $us$ Str\"omen betrachtet, die in f\"uhrender Ordnung proportional zu $m_s^2$ ist. Der skalare 
Kanal der so erhaltenen Summenregel zeigt ein extrem schlechtes Konvergenzverhalten in der St\"orungsreihe, das wir umgehen, indem wir die 
skalare OPE durch einen phenomenologischen Ansatz f\"ur die skalaren und pseudoskalaren Spektralfunktion ersetzen. 
Unser Ansatz wird daraufhin mit den theoretischen (OPE) Ausdr\"ucken verglichen,  
und wir zeigen, da{\ss} die Ergebnisse im Rahmen der Unsicherheiten \"ubereinstimmen und 
die Unsicherheiten unseres Ansatzes kleiner sind als die der theoretischen Seite, was uns erlaubt, die Fehler im 
Vergleich zu fr\"uheren $m_s$ Bestimmungen aus $\tau$-Zerf\"allen signifikant zu reduzieren. Unsere Ergebnisse sind $m_s$(2~GeV) = 117$\pm$17~MeV 
f\"ur den Unitarit\"atsfit von $\left|V_{us}\right|$ und $m_s$(2~GeV) = 103$\pm$17~MeV f\"ur den Leutwyler-Roos Wert von $\left|V_{us}\right|$ .
\vfill
\clearpage
\thispagestyle{empty}
\cleardoublepage

\thispagestyle{empty}
\tableofcontents
\thispagestyle{empty}

\chapter{Introduction}
\pagenumbering{arabic}
\input{intro2.tex}

\newcommand{\Mtau}{M_{\tau}}
\input{Theory.tex}

\input{FESR2.tex}

\input{spectr2.tex}

\input{ms2.tex}

\begin{appendix}

\input{AppA2.tex}

\input{AppB.tex}

\input{taudecay.tex}

\end{appendix}

\bibliography{Citations}

\addcontentsline{toc}{chapter}{Bibliography}
\bibliographystyle{../physics.bst}
\thispagestyle{empty}

\chapter*{Acknowledgements}
\thispagestyle{empty}
First and foremost I would like to thank Matthias Jamin for the oppertunity to do this work as well 
as for guidance and illuminating discussions.    \\\\
\noindent
I thank Prof. Hans-G\"unter Dosch for taking over the co-correction.\\\\
\noindent
Markus M\"uller, Christian M\"uller, Thomas Auer and Martin Pospischil for fighting over Peskin and 
Weinberg, collecting quotes and being crazy enough to actually count the days up to TTT, in short, 
simply for being ``\underline{Das} Westzimmer''; there can only be one conclusion: you guys are the 
greatest.\\\\
\noindent
Frank Steffen: \twonotes \, Thank You for the Music \twonotes\\\\
\noindent
Thanx again to Markus and Christian for proof-reading.\\\\
\noindent
Finally, I owe thanks to my parents for support during all of my studies.

\clearpage

\thispagestyle{empty}
\cleardoublepage
\thispagestyle{empty}
%\noindent\underline{\large\bf Erkl\"arung:} \\[1cm]

%\noindent Ich versichere, da{\ss} ich diese Arbeit selbst\"andig verfa{\ss}t und 
%keine anderen als die angegebenen Hilfsmittel benutzt habe. \\[0.5cm]
%Heidelberg, den \dotfill \hspace{4.5cm} \dotfill\\
%\hspace*{\fill} Felix Schwab

\end{document}

%% file: intro2.tex
\hspace*{\fill}\begin{minipage}[t]{8.2cm}
\begin{footnotesize}
{\em We have a simple, definite theory that is supposed to explain all the properties of protons and neutrons,
yet we can't calculate anything with it because the mathematics is too hard for us.}\\
(Richard P. Feynman, QED)
\end{footnotesize}
\end{minipage}
\vspace{1cm}

\noindent
If one were to ask a modern day elementary particle physicist about his assessment on the status of the field, the answer would 
almost certainly be ambivalent: on the one hand we have an extremely well tested theory, the standard model (SM), that seems to 
accurately describe all observed phenomena. On the other hand this resistance against falsification has, to an extent, turned into a
problem: there are quite a few ``unsatisfactory'' elements in the theory, that demand further investigation or should not be 
present in a fundamental theory: the most glaringly obvious is of course the fact that gravitation is not included
at all; furthermore we do not understand the Higgs sector, and one would hope for a smaller number of fundamental parameters that 
have to be fixed by experiment. In the simplest version of the standard model there are 18 free parameters, among these parameters 
the quark masses are the ones less precisely known, along with the Higgs mass and the CKM matrix 
parameters. These uncertainties in the quark masses are mainly due to the fact that quarks (except for the top, that shall be 
excluded from the following discussion) are not observed in nature as free particles, and their determination is then linked to another, more 
technical, problem of the SM: the theory that is supposed to describe all effects of quarks and their ``binding forces'', the 
gluons, Quantum Chromodynamics (QCD) {\cite{Fritzsch73}}, while perfectly tested and calculable at high energies due to asymptotic freedom 
(the smallness of the coupling constant)
fails at low energies when the QCD coupling becomes large and perturbation theory breaks down. This is known as confinement and leads to the 
fact that the observed degrees of freedom are the hadrons not the fundamental quarks and gluons. It is then clear that, in order 
to determine quark masses, nonperturbative methods will have to be used.

Today, the most important of these are:
\begin{itemize}
\item{\bf Chiral Perturbation Theory:} Introduced by Weinberg {\cite{Weinberg79}}, Gasser and Leutwyler {\cite{Gasser84,Gasser85}},
               Chiral Perturbation Theory is an effective field theory that begins from the notion
               that the QCD Lagrangian is symmetric under chiral transformation in the limit of massless quarks. The lowest lying meson octet
               is seen as the Goldstone bosons of the spontaneous breaking of this chiral symmetry. Additionally, the symmetry is
               also broken explicitely by the quark masses and the idea of the theory is to calculate for example mesonic scattering
               by constructing a perturbation series in small momenta and quark masses. Furthermore, (and for our purposes more 
               interestingly) one can also calculate mass ratios {\cite{Leutwyler96}}. An absolute scale, however, can only be 
               determined by another method.
\item{\bf QCD Sum Rules:} Introduced by Shifman, Vainshtein and Zakharov {\cite{SVZ,SVZ2}}(SVZ), sum rules are up to 
               now the most successful method for the determination of quark masses. The basic idea is to relate hadronic 
               quantities to fundamental QCD parameters by using dispersion relations
               and the operator product expansion and to include nonperturbative effects by defining operator vacuum expectation 
               values as condensates. It is interesting to note that originally sum rules were used to determine
               hadronic properties by giving the fundamental parameters as an input; this has been turned around when experimental 
               precision on the hadronic side improved over the years.
\item{\bf Lattice QCD:} Based on the idea of the Wilson Loop {\cite{Wilson74}} it is possible to discuss QCD on a lattice instead of 
               in the continuum
               and this will almost certainly lead to the most precise determinations in the future. Up to now the calculations
               have been performed mainly in the so called ``quenched'' approximation that neglects internal fermion loops.
               Mathematically this corresponds to setting the fermion determinant to 1, which leads to a significant improvement
               in computational speed. It is, however, not entirely clear how good this approximation is.
\end{itemize}           
Further nonperturbative methods that have less impact on quark mass determination are phenomenological quark models, in which the
quarks are assumed to move in an effective potential (examples for this would be the MIT Bag or the Isgur Karl model) and expansions
in 1/$N_c$, where the number of colors $N_c$ is assumed to be large.
\\\\
In this work we will concentrate on the strange quark mass and determine it from QCD sum rules for $\tau$-decays. 
There has recently been some activity in both the sum rule {\cite{jm2,PP2,mutau,JOP2,MK}} and the lattice 
{\cite{Lubicz2000,Eicker01,Pleiter2000}} community for a determination of this parameter, but the relative uncertainties 
remain higher than for the heavier quarks, while the best determinations for $m_u$ and $m_d$ use chiral perturbation theory and 
$m_s$, which causes their uncertainties to depend on those of the strange mass. 
It is noteworthy that $m_s$ is not only interesting as a fundamental parameter but also shows up in the standard model calculations of the 
directly CP violating parameter $\varepsilon^{\prime}/\varepsilon$. 
\\\\
The first chapter of this work deals with the basic theoretical background: we begin by very briefly sketching renormalization from the QCD
point of view in order to motivate the renormalization group equations and the running of couplings and masses. Additionally we 
introduce the main ingredients for QCD sum rules, the dispersion relations and operator product expansion mentioned above as well as 
the types of sum rules most commonly used today; among these are the finite energy sum rules that 
the rest of this work will be dealing with. 

In the next chapter we show how finite energy sum rules can be applied to $\tau$-decays in 
order to determine $m_s$. The advantage of these sum rules is the availability of rather precise data from $\tau$-decay studies 
for example at ALEPH {\cite{Buskulic:1993sv,Barate98,Ackerstaff:1998yj,Barate:1999hj}}. The OPE for both scalar and vector channel of the sum rule are then given, before we go 
on to examine the convergence of the expressions in both channels. We observe an extremely bad convergence in the perturbative 
contribution to the scalar channel, that causes large uncertainties in a strange mass determination from the $\tau$-sum rule.

In the third chapter we present the phenomenological ansatz that we make to avoid this problem of convergence: we use 
spectral functions for scalar and pseudoscalar $us$ and $ud$ currents. These spectral functions can be considered as resonance
dominated in both pseudoscalar channels, while the scalar $us$ channel has to be treated dynamically. Previously calculated 
scalar form factors enable us to provide such an ansatz. The sum of all phenomenological expressions is dominated by the kaon
pole and our phenomenological ansatz is thus afflicted with smaller uncertainties than the corresponding theoretical 
expressions. In this chapter we also investigate briefly the possibility of using our spectral function for an $m_s$ determination 
from pure scalar sum rules. 

Finally, in the last chapters we give the experimental data and the numerical results that we find from them. Here we discuss
also the use of the input parameter $\left|V_{us}\right|$ and the significant impact on our determination as well the results 
of other determinations.

%% file: Theory.tex
\chapter{Theoretical Background}
\hspace*{\fill}\begin{minipage}[t]{8.2cm}
\begin{footnotesize}
{\em There is nothing so practical as a good theory.}\\
(Richard P. Feynman, QED)
\end{footnotesize}
\end{minipage}
\section[The three R's: Renormalization and All That]{The three R's: Regularization, Renormalization and the Renormalization Group}
\subsection{Regularization}
\begin{figure}[bt]
\psfrag{p}{$p$}
\psfrag{k}{$k$}
\psfrag{   p-k}{$p-k$}
\begin{center}
\epsfig{file=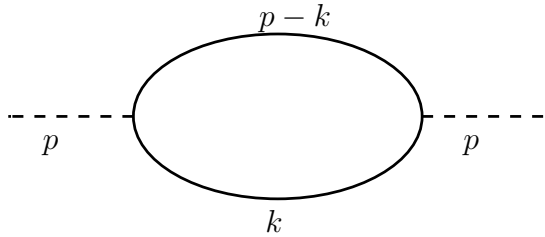,height=3cm,clip=}
\begin{minipage}[t]{12cm} 
\caption{\label{Feynmangraph}A next-to-tree-level Feynman graph of an arbitrary theory; it contains a loop and leads to divergent integrals.}
\end{minipage}
\end{center}
\end{figure}

A thorough discussion of the complete renormalization program can be found in any good textbook on quantum 
field theory of which there are many, so only the most important issues shall be reviewed briefly, emphasizing those methods that are 
relevant to QCD. The general approach for any field 
theory is to start from a Lagrangian density that consists of kinetic and interaction terms. From this Lagrangian one can 
derive Feynman rules that allow transition amplitudes to be written as a sum of Feynman graphs. In an ideal theory (which QCD 
is not, we will see why later in this chapter) the coupling constant $\alpha$ is small enough to admit a perturbative expansion in 
$\alpha$. All is well as long as one is satisfied with tree-level diagrams that do not involve loops. It is, 
however, in the human nature not to be content with this approximation and to strive for further precision.
Unfortunately, a naive computation of a diagram like in \mbox{Fig.~\ref{Feynmangraph}} that contains loops turns 
out to be divergent, because it involves integrals of the form
\begin{equation}
\label{divint}
\alpha \int \, \frac{d{}^4 k}{( 2 \pi )^4} \frac{1}{(k^2-m^2)((k-p)^2-m^2)}\,.
\end{equation}
After Feynman parametrization, change of variables \mbox{$b^2 \rightarrow m^2-p^2x(1-x)$}, and \mbox{$k \rightarrow k-xp$}
and Wick rotation we arrive at
\begin{equation}
\mathrm{const.} \int\limits_{0}^{\infty} \, dk \frac{k^3}{(k^2+b^2)^2} \, , 
\end{equation}
which is clearly divergent for large $k$ (more specifically, this type of integral is called 
logarithmically divergent since a straightforward integration leads to logarithmic expressions).

\noindent

At this point one can either condemn field theory altogether or think of a possibility to at least keep 
divergences out of observable quantities\footnote{It is worth noting that singularities can even be encountered in classical 
electrodynamics of point particles.}. The first step of the latter must be to give a meaning to the divergent integrals. We 
will only be concerned with high momentum (ultraviolet) divergences since low energy divergences that appear if 
the loop contains a massless particle can be absorbed into soft-bremsstrahlung processes.
Defining the integrals can be done in several ways, two of which are most commonly used today: Pauli Villars and 
dimensional regularization. For Pauli Villars regularization a cutoff $\Lambda$ is introduced up to which the 
integral should be taken. This type of regularization is most often used in QED and we will therefore not go 
into any more details here. It shall suffice to add that the limit \begin{math} \Lambda \rightarrow \infty \end{math} 
has to yield a finite result for any observable. In QCD nowadays mostly dimensional regularization {\cite{thooft}} is used. 
This works as follows: the integration of equation (\ref{divint}) is taken not in 4 dimensions but in $ D$, where 
\begin{math} D = 4-2\varepsilon \end{math}. The value of the integral is then
\begin{equation} 
 \alpha \frac{i}{(4\pi)^{D/2}} \int\limits_{0}^{1} dx \, \left(\frac{\mu}{b}\right)^{2\varepsilon}  \frac{\Gamma(2-\frac{D}{2})}{\Gamma(2)}  \, ,
\end{equation}
so that the divergence of the diagram is still present (as it has to be) in the pole of the $\Gamma$-function.
Similar results hold for any other diagram, the pole in particular will always appear in the same way. 
In addition a mass scale $\mu$ is introduced, which is necessary to give the whole expression an integer dimension 
(alternatively, this also makes the coupling dimensionless). $\mu$ can be interpreted as the scale at which the theory is 
defined.

\subsection{Renormalization}
The basic idea of renormalization is astonishingly simple: just redefine parameters like masses or coupling 
as renormalized in such a way that the calculations yield finite results when carried out using the physical quantities.
This is most easy to see on the level of the Lagrangian. Let the ``original`` Lagrangian of the theory be 
\begin{equation}
\mathscr{L}_{tot} = \mathscr{L}_{kin} +\mathscr{L}_{int} \, .
\end{equation}
In terms of the ``physical'' parameters this Lagrangian will take the form:
\begin{equation}
\mathscr{L}_{tot} =\mathscr{L}_{tot,ren} + \mathscr{L}_{co} \, , 
\end{equation}
where {\em ``co''} stands for the counterterms in the Lagrangian. It is tempting but untrue to say that one has 
``added'' these counterterms. In fact, the Lagrangian has merely been separated into a divergent part and a renormalized finite one. 
The counterterms also correspond to Feynman diagrams, and their divergences will precisely cancel those of the original terms (the original 
terms now involve renormalized quantities, but the divergences remain, since nothing about the calculations changes). A theory is 
called renormalizable if a finite number of counterterms only is needed to render the results finite in all 
orders of the perturbation series.

Let us look once again at our test integral (\ref{divint}). We have found that the result still includes the 
divergences as seen in the poles of $\Gamma$. We can expand the gamma function for small $\varepsilon$ and 
will, up to constant factors, arrive at
\begin{equation}
(4 \pi)^{\varepsilon} \int\limits_{0}^{1} dx \left(\frac{\mu}{b}\right)^{2\varepsilon} \Gamma (\varepsilon) =
\frac{1}{\varepsilon} + \ln 4\pi - \gamma_E - \ln \frac{-p^2}{\mu^2} +2 +\mathcal{O}(\varepsilon) \, ,
\end{equation} 
in the massless limit, where $\gamma_E$ is the Euler constant. This integral can for example be part of a two point correlation 
function, denoted by $\Pi (p^2)$. The correlation function can be split into a meaningless divergent part and a finite part that 
governs the $p^2$ dependence. At this point it is necessary to impose a renormalization condition. While in QED on-shell renormalization 
is used in order to guarantee that the measured mass is the one that appears in the Lagrangian, this
is less sensible in QCD, since quarks are not free particles and their masses cannot be measured. As for regularization there are 
multiple conventions, but the ones most often used are minimal subtraction $(MS)$ and a modified minimal subtraction scheme known as 
$\overline{MS}$. In these schemes the correlation function is split in the following way, where we have again dropped irrelevant constants:
\[
  \Pi_{ren} (p^2) = \left\{ \begin{array}{r@{\quad:\quad}l}
                             -\ln \frac{-p^2}{\mu^2}+2 & \overline{MS}\\   \ln 4\pi - \gamma_E - \ln \frac{-p^2}{\mu^2}+2 & MS
                            \end{array} \right. \]
\[
  \Pi_{div}  = \left\{ \begin{array}{l@{\quad:\quad}l}
                            \frac{1}{\varepsilon} + \ln 4\pi - \gamma_E \qquad  & \overline{MS}\\ \frac{1}{\varepsilon} \qquad & MS
                            \end{array} \right. \]
A third convention, that we will only mention, is the $\mu$ or Weinberg scheme where, by definition, the renormalized term contains only 
the logarithm. Note that the logarithmic dependence of the renormalized term is the same in the three schemes. All of the following work has 
been done in the $\overline{MS}$ scheme.
    
\subsection{The Renormalization Group}
\label{RGE}
The renormalization group equations can be seen as a connection between statistical mechanics and quantum field 
theory. In short they govern the scaling behavior of the theory. This is dubbed ``group'' because of a groupoid
structure {\cite{PS,PT}}, even though it is not actually a group in the mathematical sense, since renormalization 
is not invertible.

Now the basic statement is that all of the hard work we have put into the renormalization of the obtained expressions is
highly artificial and should not be seen in measurable quantities. An observable should then neither depend on 
renormalization scheme nor scale. This statement of renormalization invariance leads to expressions that govern the 
scaling behaviour of physical quantities such as the correlation functions to be discussed in the next section. 
This, in turn, can sometimes also be true for non-physical quantities such as the Wilson coefficients of the operator 
product expansion, that will be the main topic of the next section. Consider, for example, a Greens' function denoted by $\Gamma$  that shall depend on external 
momenta, the coupling constant and masses. The relation between the the bare Greens function and the renormalized one is then
\begin{equation}
\label{Ren1}  
\Gamma (\{p_i\}, a, m_a,\mu) = \lim_{\epsilon \to 0}
 Z_{\mu}(\mu,\epsilon) \Gamma_{0}(\{p_i\}, a_0, m_{a,0},\epsilon) \, ,
\end{equation}
where $Z_{\mu}(\mu,\epsilon)$ is the appropriate combination of renormalization constants.
It is customary to define $a \equiv \alpha_s/\pi$ and we will do so throughout this
work\footnote{In general a Greens function will also depend on the gauge parameter, unfortunately often also denoted by 
$a$; we will use Landau gauge and thereby omit this factor.}. 
Taking the total derivative of Eq.~(\ref{Ren1}) with respect to $\mu$ and keeping in mind that the bare Greens function must definitely not depend on $\mu$ we obtain
\begin{equation}
\left[\mu \frac{\partial}{\partial \mu} - \beta \, \frac{\partial}{\partial \alpha} - \sum\limits_{a} \gamma_a m_a \frac{\partial}
{\partial m_a} + \gamma_Z \right]\, \Gamma (\{p_i\}, a, m_a,\mu) = 0\,.
\end{equation}
This is known as the Callan-Symanzyk or renormalization group equation and we have introduced the beta function and 
anomalous dimensions as
\begin{equation}
\label{betaf}
\beta  \equiv - \mu \, \frac{da}{d\mu} \, , \qquad \gamma_a \equiv -\frac{\mu}{m_a} \, \frac{dm_a}{d\mu}\, , \qquad \gamma_Z \equiv \frac{\mu}{Z}
\,\frac{dZ}{d \mu}  \,.
\end{equation}
Let us first discuss the running of the coupling constant. In the $\overline{MS}$ scheme the $\beta$ function depends on $a$ only (it 
is therefore called a mass independent scheme)
and can be expanded
\begin{equation} 
\beta(a) = \beta_1 a^2 +\beta_2 a^3 + \beta_3 a^4 + ...\,,
\end{equation}
where nowadays all the coefficients up to $\beta_4$ are known. $\beta_1$ and $\beta_2$ are, in fact, scheme independent. All known coefficients 
are given in the appendix. An explicit expression for $a$ can be obtained by integrating the defining equation. Up to leading 
order the resulting expression reads
\begin{equation}
a(Q) = \frac{ a(\mu)}{1+ \beta_1 a(\mu) \ln \frac{Q}{\mu}} \, .
\end{equation}
In order to remove the renormalization scale one conventionally introduces a  another mass scale $\Lambda$ in such a way that the 
expression simplifies to
\begin{equation}
\label{runc}
a(Q^2) = \frac{2}{\beta_1 \ln \frac{Q^2}{\Lambda^2}} \, .
\end{equation}
For higher orders a further expansion in powers of logarithms holds. The scale $\Lambda$ is at several hundred MeV and
will depend on the number of quarks included in the theory which causes it to change at quark thresholds. This is achieved by imposing 
matching conditions for the effective theories with the appropriate number of quarks. Conventionally $a$ is given at the scale of 
the Z boson mass for which the Particle Data Group value is {\cite{PDG2000}}
\begin{equation}
a(M_Z) = 0.119 \pm 0.002 \, .
\end{equation}
From Eq.~(\ref{runc}) we can see that for positive $\beta_1$ (as is 
the case for ``real life QCD'', see appendix A) the coupling exhibits both confinement and asymptotic freedom and the scale 
$\Lambda$ becomes meaningful as the scale at which confinement is effective.

In this work we will define the coupling implicitely, by directly integrating Eq.~(\ref{betaf}). We arrive at
\begin{equation}
\ln \left (\frac{Q}{\mu}  \right) =  \int\limits_{a(\mu)}^{a(Q)} \frac{1}{\beta(a)} \,, 
\end{equation}
which can be numerically solved for given values of the $\beta$-function coefficients. We will be using the known 
4-loop expression that should be a very good approximation, unless the $\beta$ function has some unexpected higher-order 
pathologies.

The treatment of the running mass is in perfect analogy to that of the running coupling: again the anomalous dimension $\gamma$ 
can be expanded into a power series in $a$
\begin{equation} 
\label{gammaf}
\gamma(a) = -\frac{\mu}{m} \, \frac{dm_a}{d\mu} = \gamma_1 a +\gamma_2 a^2 +... \,,
\end{equation}
where the coefficients are known up to fourth order and given in the appendix. A simple integration of (\ref{gammaf}) gives a
series for $m$ in powers of $a$. Up to next-to-leading order this is:
\begin{equation}
m(Q) = m(\mu) \left[\frac{a(Q)}{a(\mu)} \right]^{\gamma_1/\beta_1} \left( 1+ \left[a(Q)-a(\mu) \right] \left( \frac{\gamma_2}
{\beta_1}-\frac{\beta_2\gamma_1}{\beta^2_1} \right) \right) \, .
\end{equation}
The coefficient $\beta_1$ is positive as noted above and the same holds for $\gamma_1$, so we conclude that masses will 
decrease along with the coupling constant for high energies.

As for the coupling we will choose to solve the resulting integral equation numerically. It is
\begin{equation}
\ln \frac{m(Q)}{m(\mu)} = \int\limits_{a(\mu)}^{a(Q)} dx \frac{\gamma(a)}{\beta(a)} \,.
\end{equation}
Taking the exponential of both sides of this equation shows that the ratio of two quark masses at the same scale is scale-independent:
\begin{equation}
m(Q) =m(\mu) \exp \int\limits_{a(\mu)}^{a(Q)} dx \frac{\gamma(a)}{\beta(a)} \,,
\end{equation} 
since the exponential factor on the right hand side will drop out if we take the mass ratio.
 
For the light quarks ($u,d,s$) it has become customary to quote values at either 2 or 1 GeV, since perturbation 
theory breaks down at about this scale. On the other hand, masses for heavy quarks are usually given 
at the scale of their own mass.  
The current values for the quark masses are then {\cite{privdisc}}:
\par
\begin{center}
%\begin{footnotesize}
\[\begin{array}{rclrcl}
 m_u(\mathrm{2\,GeV}) & = & 2.9 \pm 0.6 \,\mathrm{MeV}\,  , & m_d(2\,\mathrm{GeV}) & = &  5.2\pm0.9\, \mathrm{MeV}\,  , \\
 m_s(\mathrm{2\,GeV}) & = & 98\pm18 \, \mathrm{MeV} \, , & m_c(m_c)& = & 1.28\pm0.09\, \mathrm{GeV}\,  , \\ 
 m_b(m_b)& = & 4.21\pm0.06\, \mathrm{GeV} \, , & m_t(m_t)& = &  165 \pm 5 \,\mathrm{GeV}\, . \\ 
\end{array} \]
%\end{footnotesize}
%\clearpage
\end{center}
%There is another definition of the quark masses that is mort often used for heavy quarks and is closer related to the intuitive mass 
%definitions of for example QED, the pole mass. The mass is defined as the pole of the   

\section{Operator Product Expansion (OPE)}
The operator product expansion was introduced in 1969 by Wilson {\cite{W}} as an alternative to the current algebra 
methods used at the time. 
He proposed to expand nonlocal 
operator products into a sum of local operators, thereby relegating the 
nonlocality into the expansion coefficients and thus separating long and short
distance effects:

\begin{equation}
A(x)B(y) \stackrel{x \rightarrow y}{\sim}\sum_{n}C_n(x-y)O_n(y)\,,
\end{equation}  
where $A, B$ and $O_n$ are operators and the $C_n(x-y)$ are known as the Wilson coefficients of the OPE. For simplicity $y$ is 
usually set to zero, a convention that we will adopt from here. It is, however, important to note 
that the above equality is valid only in a weak sense, meaning that the 
operators have to be sandwiched into initial and final states.
In general the sum on the right hand side involves operators of different mass 
dimensions, which causes the coefficient dimensions to vary accordingly. As
 we will see it turns out that, in the specific cases discussed in sum rules, 
this leads to a suppression of higher dimensional operators, a fact without 
which the OPE would not be of much use. 

Since the product of two operators at arbitrarily small distances can in 
general be singular one expects the same to be true for the coefficients.
However, the nature of the these singularities is governed by the underlying exact and broken symmetries, namely 
scale invariance; for a complete discussion the interested reader is referred to Ref. {\cite{PS}} or {\cite{PT}}. 
At this point we will only mention that the scaling dimension (as opposed to mass dimension) of the coefficients is 
\begin{equation}
C_n(x) \stackrel{x^\mu \rightarrow 0}{\sim} x^{-\lambda_n},\qquad           
\lambda_n=d(A)+d(B)-d(O_n)\,,
\end{equation}
just as one would expect from dimensional analysis, only that $d(X)$ is not the 
conventional mass dimension $d_{c}$ but
\begin{equation}
d(X) = d_{c} + \gamma(a^*)\,,
\end{equation}
where $\gamma(a)$ is the anomalous mass dimension of the operator and $a^*$ is the fixed point of the 
renormalization group equation.
Operators that generate symmetries as well as the identity operator $\mathbbm{1}$ 
among others will keep their naive dimensions through this process.  

The operator product investigated in sum rules is the two point correlation function 
of two-quark currents:
\begin{equation}
\label{gencorr}
\Pi^{\mu\nu}(q)=i\int\,d{}^4 x \, e^{iqx} \langle \Omega | T(j^\mu(x) j^{\nu}(0)) | \Omega 
\rangle \, .  
\end{equation} 
Here the current $j^\mu$ is built from any two quark-antiquark flavors.
\begin{equation}
j^{\mu}(x)=\bar{q}_i\Gamma q_j(x) \, ,
\end{equation}
with $\Gamma$ any combination of Dirac matrices introduced in order to obtain the desired quantum numbers. The 
coefficients for various currents have been calculated and can be found in the literature (see e.g.  
Ref. {\cite{PP2}} for a compilation of coefficients and references to the original works). Taking for example 
\begin {equation}
j^{\mu}(x)=\frac{1}{2} \left(:\bar{u}\gamma^{\mu}u(x) - \bar{d}\gamma^{\mu}d(x): \right)\,,
\end{equation}
allows one to model the $\rho$-meson with rather good accuracy. As usual, the notation :  : 
implies a normal ordering. Indeed, this vector current sum rule is the standard example of an OPE 
and has already been discussed in the original paper by SVZ {\cite{SVZ,SVZ2}}, while the explicit calculation to this correlation 
function is presented in great detail in Ref. {\cite{PT}}. The lowest dimensional scalar operators 
(\mbox{Dim $\leq$ 6}) that will appear in the OPE are 
\begin{equation}
\mathbbm{1} ,\hspace{0.5cm} m_i\bar{q}q,\hspace{0.5cm} F^{\mu\nu}_aF_{\mu\nu}^a,\hspace{0.5cm} m_i\bar{q}\sigma_{\mu\nu}\frac{\lambda^a}{2} F^{\mu\nu a}q,\hspace{0.5cm} 
 \bar{q}\Gamma  q\bar{q}\Gamma q\,.
\end{equation}
A normal ordering is understood in all of these operators. In ordinary perturbation theory these vacuum expectation values 
vanish by definition except for the unit operator. For the others they are introduced as ``condensates'' in 
nonperturbative QCD in order to describe the nontrivial vacuum structure. A natural origin of the quark condensate 
can also be found in chiral perturbation theory, where it arises as the order parameter of spontaneous chiral symmetry 
breaking and is then responsible for the pion mass (the famous Gell-Mann Oakes Renner (GMOR) relation) {\cite{GOR,WB}}.
At this point an additional problem turns up: the standard proof of OPE relies on Feynman
diagrams and perturbation theory, so its use in a nonperturbative regime is far from obvious. This 
causes the expansion series to break down at a certain dimension $d_{crit}$ that has already been estimated by SVZ. In practise the calculations 
are usually performed up to dimensions 6-8, depending on the strength of the suppression. 

\begin{figure}[bt]
\psfrag{p}{$p$}
\psfrag{q}{$q$}
\psfrag{p-q}{$q-p$}
\psfrag{a}{$a$)}
\psfrag{b}{$b$)}
\begin{center}
\epsfig{file=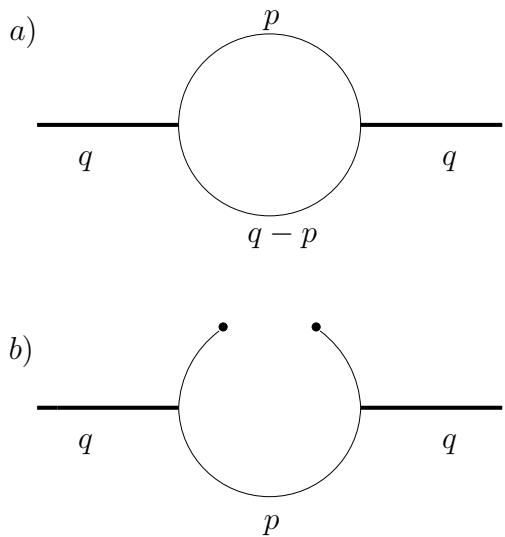,height=7cm,clip=}
\begin{minipage}[t]{12cm} 
\caption{\label{condensate}Quark condensates: pictorially the nonlocal condensate $b$) corresponds to cutting one propagator in the Feynman graph for the 
perturbative operators $a)$. }
\end{minipage}
\end{center}
\end{figure}

Now let us see how the higher dimensional operators arise in a semiperturbative expansion. First, however, we need to 
decompose the correlator into parts of different Lorentz structure by way of the Ward identities. This can 
be done as follows:
\begin{eqnarray}
\label{corr}
\Pi^{\mu\nu}(q) & = & (q^{\mu}q^{\nu} - q^2 g^{\mu\nu})\Pi^{V,A}(q^2) + \frac{g^{\mu\nu}}{q^2} (m_i \mp m_j)^2\Pi^{S,P}(q^2) \\
                &   & {} + g^{\mu\nu}\frac{(m_i \mp m_j)}{q^2}[\langle \bar{q}_iq_i \rangle \mp \langle \bar{q}_jq_j \rangle] \nonumber\, ,
\end{eqnarray}
$\Pi^{S,P}(q^2)$ being the correlator of scalar (pseudoscalar) currents.
The proof for this relation can be found in Ref. {\cite{B}} but the structure can also be understood intuitively:
the term in the first brackets is the ``ordinary'' vector current correlator that implies transversality, the second can be 
understood by looking at the correlator of the divergence of a vector current. Using the Dirac equation this can be 
transformed into a scalar correlator multiplied precisely by the given expression in the masses. The last two terms are subtraction 
constants for the scalar current. All of this will be discussed in more detail when we study the scalar two point function in
chapter 3.
In our example of the vector current from above only the first term contributes, and we will
 begin by sketching the calculation for the unit operator (Fig {\ref{condensate}} a)). Contracting both sides of Eq.~(\ref{corr})
with $g_{\mu\nu}$ we get:
\begin{eqnarray}
\nopagebreak
\label{Feynm}
\Pi(q^2) & = & -\frac{i}{4q^2(d-1)}\int\,d{}^D x e^{iqx} \langle \Omega | T(:\bar{u}(x)\gamma^{\mu}u(x) - \bar{d}(x)\gamma^{\mu}
d(x):\\\nopagebreak
         &   & \times :\bar{u}(0)\gamma_{\mu}u(0) - \bar{d}(0)\gamma_{\mu}d(0): ) | \Omega \rangle \nonumber \,,
\end{eqnarray} 
to which we apply Wick's theorem:
\begin{equation}
\Pi(q^2)=-\frac{N_ci}{4q^2(D-1)}\int\frac{\,d{}^D x}{(2\pi)^D}[Tr[\gamma^\mu S^u(p) \gamma_\mu 
S^u(q-p)] + u \rightarrow d ]\,.
\end{equation}
Completing the calculation gives a coefficient of \begin{math} \frac{3}{8\pi^2}(\frac{5}{3}-\ln(-\frac{q^2}{\mu^2} \end{math})) 
where $\mu$ is the renormalization scale that arises in dimensional regularization.

\noindent

 But let us go back one step to Eq.~(\ref{Feynm}) and 
leave one quark-antiquark pair uncontracted. Pictorially this corresponds to cutting one propagator in the 
Feynman graph (see Fig.~\ref{condensate}b)). One is left with expressions including terms like  \mbox{\begin{math} \langle: \!\bar{q}(x)q(0)\!: \rangle \end{math}}
that can be Taylor expanded in order to arrive at well defined local condensates 
\mbox{\begin{math} \langle: \!\bar{q}(0)q(0)\!: \rangle \end{math}}, often written as \begin{math} \langle \bar{q}q 
\rangle \end{math}. Note that this simple derivation naturally leads to normal ordered condensates, a 
subtility that becomes important in calculations: writing the OPE in terms of operators that are 
renormalized in a non minimal scheme (i.e. normal ordering) leads to coefficients including powers of \begin{math} m
\ln(\frac{m^2}{\mu^2}) \end{math}. The usual procedure is to set $-q^2=\mu^2$ because all of the logarithms 
of the type \begin{math} \ln(\frac{-q^2}{\mu^2}) \end{math} are then set to zero (this procedure is known 
as ``summing up'' the logarithms and can be done because, as we will see, the correlator is related to an 
observable quantity and has to obey a renormalization group equation). Unfortunately, this summing up 
process does not get rid of mass logarithms and in fact causes them to be large, which makes calculations 
of nonleading contributions impossible. Mass logarithms are then interpreted as remnants of long distance effects so that the separation 
of long and short distance effects in non-minimal schemes is incomplete and working in minimal schemes appears advantageous. A 
more elaborate discussion of this subject can be found in Ref. {\cite{jm}}, where the relation between normal ordered and minimally 
subtracted condensates are also given.

%\clearpage
\section{Dispersion Relations and Sum Rules}

This section will be dealing with the analytic properties of the two point correlators introduced above 
and their consequences. It is shown for example in \mbox{Ref. {\cite{EdR}}} that a general two point function can be written as an 
integral over a spectral function $\rho$(s)
\begin{equation}
\label{corr2}
\Pi(s) = \int\limits_{0}^{\infty} \,dt \rho(t) \frac{1}{t-s-i\varepsilon} \, ,
\end{equation}
where the spectral function is the sum over all possible final states $\lambda$:
\begin{equation}
\rho(q) = (2\pi)^3 \, \int\limits_{\lambda} \hspace{-17pt} \sum \hspace{5pt}  \left| \langle \Omega | 
j_{\lambda}(0) | \lambda \rangle \right|^2 \delta^4 (q - p_{\lambda})\,.
\end{equation}
We have also introduced the notation $q^2 = s$ that we will be using frequently 
from now on. The correlator is then an analytic function except for possible poles on the positive real axis.
Now remember that according to Cauchy's theorem  every analytic function $f(z)$, that we will also assume to fall off 
fast enough for $s \rightarrow \infty$, obeys the following relation for any closed curve $\mathcal{C}$ with $t$ as an 
interior point: 
\begin{equation}
f(t) = \frac{1}{2\pi i} \int\limits_{\mathcal{C}} \,dz \frac{f(z)}{z-t} \, .
\end{equation}

\begin{figure}[bt]
\psfrag{Im s}{Im(s)}
\psfrag{Re s}{Re(s)}
\psfrag{C1}{$\mathcal{C}_1$}
\psfrag{C2}{$\mathcal{C}_2$}
\psfrag{C3}{$\mathcal{C}_3$}
\psfrag{C4}{$\mathcal{C}_4$}
\psfrag{R}{R}
\begin{center}
\epsfig{file=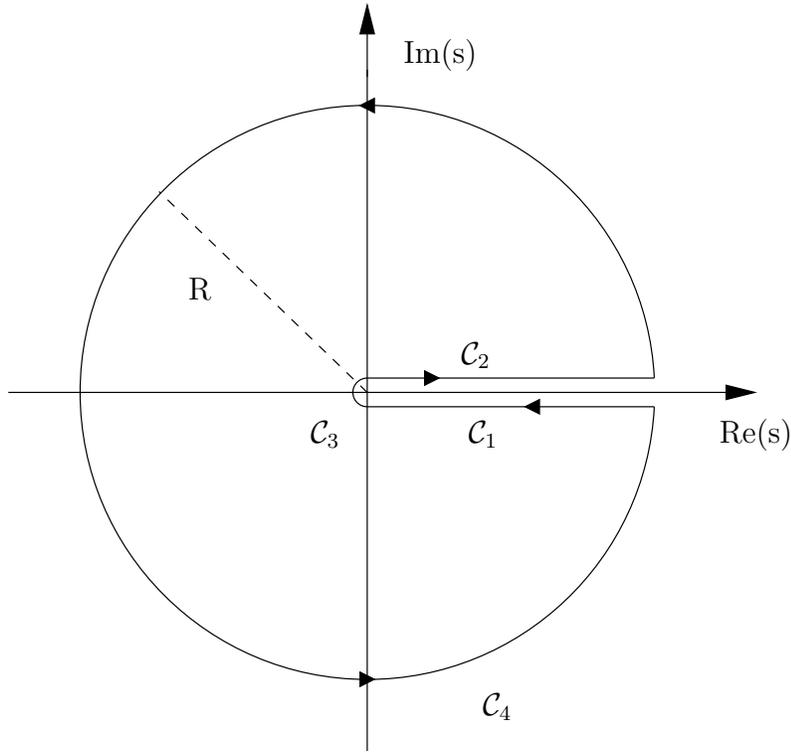,height=10cm,clip=}
\begin{minipage}[t]{12cm} 
\caption{Integration contour for dispersion relations.}
\end{minipage}
\end{center}
\label{contour}
\end{figure}
\noindent
Let $\mathcal{C}$ be the curve from Fig.~\ref{contour}: it consists of a circle of radius R around the origin, with an interval 
of 2$\varepsilon$ left out around the real axis. The curve runs above and below the axis and is finally closed by a circle of 
radius $\varepsilon$ around the origin. Integrating the integrand of Eq.~(\ref{corr2}) over this curve and applying Cauchy's theorem, 
we find:
\begin{eqnarray}
        \int\limits_{\mathcal{C}} \,dt \frac{\rho(t)}{t-s-i\varepsilon} 
  & = & 2\pi i \rho(s+i\varepsilon) \\
  & = & \int\limits^{}_{\mathcal{C}_1} \,dt \frac{\rho(t)}{t-s-i\varepsilon} + \int\limits_{\mathcal{C}_2} \,dt 
\frac{\rho(t)}{t-s-i\varepsilon} + \int\limits_{\mathcal{C}_3} \,dt \frac{\rho(t)}{t-s-i\varepsilon} \nonumber \\
  &   & {} + \int\limits_{\mathcal{C}_4} \,dt \frac{\rho(t)}{t-s-i\varepsilon} \, . \nonumber
\end{eqnarray}
The integral over the large circle vanishes for R$ \rightarrow \infty$, as does the integral over 
the small one, if the integrand is analytic into the origin, which we will assume. The remaining 
terms are
\begin{eqnarray}
  2 \pi i \, \rho(s) & = & 2 \int\limits_{0}^{\infty} \,dt \, \rho(t) \pi i \, \delta(t-s) \\\nonumber   
                     & = & \lim_{\varepsilon\to0} \int\limits_{\mathcal{C}_1 +\mathcal{C}_2} dt \,  \frac{\rho(t)}{t-s-i\varepsilon} , 
\end{eqnarray}
or, written in a more symbolic form:
\begin{equation}
\label{Princ}
\frac{1}{t-s-i\varepsilon} = \mathcal{P} \, \frac{1}{t-s} + i \pi \delta (t-s).
\end{equation}
where $\mathcal{P}$ indicates that the principle value is to be taken. We can now apply Eq.~(\ref{Princ}) to the correlator
and compare real and imaginary parts of both sides. Beginning with the imaginary part, we find
\begin{equation}
\rho(s) = \frac{1}{\pi} \mathrm{Im} \, \Pi (s) \, .
\end{equation}
Analogously, the real part is given by
\begin{equation}
\mathrm{Re} \, \Pi(s) = \int\limits_{0}^{\infty} \, dt \frac{\rho(t)}{t-s} \, ,
\end{equation}
where convergence of the integral will have to be checked. The same holds for the integral in equation Eq.~(\ref{corr2}), 
which may not converge either if $\rho(s)$ does not fall of fast enough. In order to make the integrals convergent  one can
expand both sides into a Taylor series and subtract the first term before integrating (this is 
generally referred to as ``subtractions''):
\begin{eqnarray}
\Pi(s) - \Pi(s_0) & = & \int\limits_{0}^{\infty} \, dt \rho(t) \left(\frac{1}{t-s} - \frac{1}{t-s_0} \right) \\ \nonumber 
 & = & (s-s_0) \int\limits_{0}^{\infty} \, dt \frac{\rho(t)}{(t-s)(t-s_0)}\, .
\end{eqnarray}
This integral will obviously converge better due to the higher power of $t$ in the numerator. In certain situations more 
than one subtraction is needed (the number of necessary subtractions can easily seen by power counting since the integrand needs 
a negative mass unit to converge). The generalization of the above result can then be shown by induction to be
\begin{equation}
\Pi(s) = \sum_{k=0}^{n-1} a_k (s-s_0)^k + (s-s_0)^n \int\limits_{0}^{\infty} \, dt \frac{\rho(t)}{(t-s)(t-s_0)^n} \, ,
\end{equation}
where
\begin{equation}
a_k = \frac{1}{k!} \frac{d^k \Pi(s)}{ds^k} \Big|_{s=s_0}\,.
\end{equation}
A very simple sum rule can already be obtained by taking the OPE of the correlator on the left side of the equation and 
using the measured spectral function on the right side. Often the spectral function can be directly related to a cross 
section or decay rate by the optical theorem: in the case of the ratio of quark anti-quark pair production to muon pair production 
in $e^{+}e^{-}$ collision for example this relation is given by
\begin{equation}
R(s) := \frac{\sigma_{e^{+}e^{-} \rightarrow q \bar{q}}}{\sigma_{e^{+}e^{-} \rightarrow \mu^{+}\mu^{-}}} = 12 \pi \mathrm{Im} 
\Pi(s) \, .
\end{equation}
At this point it is worthwhile to pause and recapitulate what we have essentially done: we have, so to speak, built a bridge from 
phenomenological parameters (the spectral function) to the fundamental parameters of QCD (coupling constants and quark masses).
Vacuum effects have been included in a rather small number of parameters that represent static vacuum properties. It is crucial
to assume universality for these parameters in order to make sensible predictions.

Applying this sum rule to our ``toy model'' of the $\rho$ meson turns out to be a bit disappointing because of conflicting 
choices for $s$ on each side. We expect the spectral function to be dominated by the $\rho$ meson at low energies and would thus 
like to choose a very small $s$, while on the other hand the OPE involves terms that are suppressed by factors of quark masses 
over momenta which would favor the choice of large $s$. The first idea would be to use a compromise value of, say 
1 \,$\mathrm{GeV^2}$, which is okay for the OPE but still not good enough as $\rho$ meson saturation approximation. Clearly the sum rule will have 
to be improved. There are different ways of doing this and we will discuss them in the next section.

%\clearpage
\section{Different Types of Sum Rules}
We have seen above that the simplest sum rules can often be only mediocre approximations and we will now try to improve 
on them. There are basically three types of sum rules that are widely used {\cite{EdR}}. They all differ in the way the analytic 
properties of the spectral function are exploited or certain limits are taken.

\subsection{Borel Sum Rules}
This is the kind of sum rule that was used for light quarks in the original paper by SVZ and enhances the low 
energy region of the spectrum. We will, however, choose another motivation: going back to the vector meson 
sum rule of the last section for a minute, dimensional counting shows us that we need one subtraction constant to arrive 
at a well defined expression. In a more general case the correlation function can have dimension $d$ and the asymptotic 
behavior is
\begin{equation}
\lim_{s \to \infty} \frac{1}{\pi} \mathrm{Im}\,\Pi(s) \Rightarrow As^d \left(1+c\,\alpha_s(s)...\right)\,.
\end{equation}
Then $d+1$ subtraction constants are needed. The subtraction constants generally depend on the renormalization scale and 
drop out of any physical observable, but for calculations it is necessary to get rid of them. The easiest way to achieve 
this is to take $d+1$ derivatives with respect to $Q^2 \equiv -s$. We are, of course, not limited to $d+1$ derivatives, since any 
number $N \ge d+1$ of derivatives removes the subtraction constants.
The resulting expression is sometimes called the moment sum rule:
\begin{equation}
\frac{(-1)^{N}}{(N-d-1)!}(Q^2)^{N-d}\frac{\partial^{N}}{(\partial Q^2)^{N}} \Pi(s) \hspace{7cm}\nonumber
\end{equation}
\begin{displaymath}
\hspace{3cm}= \int\limits_{0}^{\infty}dt 
\frac{N(N-1) \cdots (N-d)}{(t+Q^2)^{d+1}} \left(\frac{Q^2}{t+Q^2}\right)^{(N-d)} \frac{1}{\pi} \mathrm{Im}\Pi(t)\,.
\end{displaymath}
To get rid of any arbitrary number of subtraction constants we can take the limit $N\rightarrow \infty$; additionally 
we will let the energy $Q^2$ go to infinity so that the ratio $\frac{Q^2}{N}=M$ remains fixed.
This corresponds to taking the Borel transformation, defined as 
\begin{equation}
\hat{B} = \lim_{Q^2\to\infty}\lim_{N\to\infty} |_ {\frac{Q^2}{N} = M^2} \, \frac{1}{\Gamma(N)} (-Q^2)^N \left(\frac{d}{dQ^2}\right)^N \, ,
\end{equation}
of the first well defined moment.
It can be shown (see Ref. {\cite{Bertl85}} and references therein) that the Borel transformation is the inversion of 
the Laplace transform\footnote{This fact leads to the name Laplace transform sum rules that can also be found in 
the literature instead of Borel sum rules.} so that the Borel transformed dispersion relation reads
\begin{equation}
M^2 \hat{B} \,  \Pi(Q^2) = \int\limits_{0}^{\infty} \, dt \, \rho(t) \, e^{-t/M^2} \,.
\end{equation}
It is easy to see that the low energy region of the spectral density is enhanced, as mentioned above, so that this type 
of sum rule is particularly suitable when studying low energy resonances. In practise the phenomenological spectral 
function is known only up to some threshold energy $s_0$. Above this threshold the imaginary part of the OPE for the 
correlator is taken as a theoretical spectral function. If the threshold energy is large enough this should be a rather 
good approximation. There have already been several successful attempts to extract light quark masses from Borel sum 
rules; the results for the strange mass shall be reviewed and discussed at the end of this work.  

\subsection{Finite Energy Sum Rules}
For finite energy sum rules (FESR) consider again the contour from Fig.~\ref{contour}, only this time we will not let the 
radius of the large circle go to infinity. Instead, we will integrate over a circle of finite energy $s_0$ (hence the 
name). The function to be integrated is a weighted correlation function:
\begin{equation}
\frac{1}{2 \pi \, i} \oint\limits_{\mathcal{C}} \, ds \, w(s) \, \Pi(s) \, ,
\end{equation}
where $w(s)$ is an arbitrary analytic function. According to Cauchy's theorem the integral over the whole contour has 
to vanish. The parts of the contour along the axis will pick up the discontinuity in the spectral function ($\rho$ 
has to obey Schwarz' reflection principle) and we find, if we split the contour into two parts:
\begin{equation}
\label{fesr}
-\frac{1}{2 \pi \, i} \oint\limits_{s=s_0} \, ds \, w(s) \, \Pi(s)  = \int\limits_{0}^{s_0} \, ds \, w(s) \, \rho(s)\,.
\end{equation}
The obvious advantage of this approach is that the spectral functions needs to be known only up to the energy $\sqrt{s_0}$, while for Borel sum rules it 
has to be known to arbitrary high energies. Unfortunately, the convergence behaviour of FESR may not be as good as that of BSR, depending on the
weight function. 
The question of subtractions in finite energy sum rules is dealt with by partial integration:
\begin{equation}
\label{sub}
\oint\limits_{s=s_0} \, ds \, w(s) \, \Pi(s)  = - \oint\limits_{s=s_0} \, ds[W(s)-W(s_0)] \, 
\frac{d\Pi(s)}{ds} \, ,
\end{equation}
where $W(s)$ is the integral over the weight function. For higher order subtraction polynomials this procedure can be repeated and all subtraction constants will then be 
successively removed. Further subtleties of FESR will be discussed in the next chapter that shall deal exclusively with
the strange mass FESR.  

\subsection{Gaussian Transform Sum Rules}
Here we will consider {\cite{Bertl85}} the Gaussian transform of the spectral function 
\begin{equation}
G(s,\tau) = \frac{1}{\sqrt{4 \pi \tau}} \int\limits_{0}^{\infty} \, dt \, \exp \left(-\frac{(s-t)^2}{4 \tau} \right) 
\rho (t) \, .
\end{equation}
Depending on the choice of $s$ and $\tau$ any energy region can be probed, so one is free to investigate the region that is 
most easy to parametrize. In order to calculate $G$  we will define the following combination, where we assume that the dispersion 
relation has at most one subtraction constant (the calculation can easily be generalized to arbitrary subtraction polynomials):
\begin{equation}
\frac{\Pi (t)}{(t-s)^2+\Delta^2} = \frac{\Pi (t)}{(t-s+i\Delta)(t-s-i\Delta)}\,.
\end{equation}
Once again this function is integrated over our contour from Fig~{\ref{contour}} with $R \rightarrow \infty$, and we find 
\begin{equation}
\frac{\Pi (s+i\Delta)}{i\Delta} + \frac{\Pi (s-i\Delta)}{-i\Delta} = \int\limits_{0}^{\infty} \, dt \, \frac{\frac{1}{\pi} \mathrm{Im} \Pi (t)}{(t-s)^2+
\Delta^2}\,.
\end{equation}
Now we can apply the Borel transformation with respect to $\Delta$ to obtain the final result:
\begin{eqnarray}
\frac{1}{\sqrt{4 \pi \tau}} 2\tau  \hat{B} \, \frac{\Pi (s+i\Delta)}{i\Delta} + \frac{\Pi (s-i\Delta)}{-i\Delta} & = &  \frac{1}{\sqrt{4 \pi \tau}} \int\limits_{0}^{\infty} \, dt \,
\exp \left(-\frac{(s-t)^2}{4 \tau} \right) \rho (t) \, \nonumber\\
 & = & G(s,\tau)\,.
\end{eqnarray}
Looking once again at the definition of the Gauss transform we observe two things: the first is that for $\tau$ equal to zero the 
Gaussian transform merely reproduces the spectral function, the second is that it obeys a differential equation:
\begin{equation} 
\frac{ \partial^2 G(s,\tau)}{\partial s^2} = \frac{ \partial G(s,\tau)}{\partial \tau} \, .
\end{equation}
This equation is known in another context as heat evolution equation and we can try to apply this analogy  to our spectral 
function. Taking $\tau$ as our analog to a time variable, we can call the spectral function the initial condition. In pQCD the 
Gaussian transform is an expansion in $\alpha (\sqrt{\tau})$ which is not calculable for small $\tau$ because of confinement and 
thus comparisons of hadronic spectral functions with pQCD cannot be made. But one can evolve the spectral function ``in time'' by 
going to larger $\tau$ and then compare the evolved spectral function with predictions from the corresponding pQCD expressions.

%% file: FESR2.tex
\chapter{$\mathbf{\tau}$-Decay Sum Rules}
\label{FESR}
\hspace*{\fill}\begin{minipage}[t]{8.2cm}
\begin{footnotesize}
{\em Hopefully you will understand the simple examples, because if you do you will understand the generalities at once - that's the way {\em I} understand things anyway.}\\
(Richard P. Feynman, Elementary Particles and the Laws of Nature)
\end{footnotesize}
\end{minipage}

\section{$\mathbf{\tau}$-Decays and Finite Energy Sum Rules}
Finite energy sum rules were first applied to \mbox{$\tau$-decays} systematically by Braaten et.~al.~{\cite{BNP}} when studying 
inclusive $\tau$-decays, following the ideas of Schilcher and Tran {\cite{Tran83}}. Since their work is the basis of our studies and we need to fix our notation, we will review it briefly,
sometimes taking a more modern point of view. At the same time the problems of this approach will become evident. It is interesting to note that
at the time these sum rules were used to determine the QCD coupling at the scale of $M_{\tau}^2$, another example of how the application of sum 
rules often goes beyond the original intention. 

Throughout this section we will be considering the ratio of hadronic to leptonic $\tau$-decays defined as
\begin{equation}
R_{\tau} = \frac{\Gamma (\tau \rightarrow \mathrm{hadrons})}{\Gamma (\tau \rightarrow e \bar{\nu}_e \nu_{\tau})} \stackrel {Exp.}{=} 3.647\pm 0.014{\cite{Barate98}}\,\,.
\end{equation}
A first estimate gives, neglecting strong and elektroweak corrections:
\begin{equation}
R_{\tau} = N_c \left[ |V_{ud}|^2 + |V_{us}|^2 \right] \simeq 3\,.
\end{equation}
Because of the inclusive nature of $\tau$-decays this ratio can be decomposed into several contributions associated with 
different quark currents. These are vector, axial vector and strange contributions. The strange decay ratio cannot be 
decomposed further because parity is not a conserved quantum number in weak interactions. We can now write the Lorentz 
decomposition (\ref{corr}) in a slightly different way, where we will also absorb the subtraction constant into the scalar 
term:
\begin{equation}
\label{corrw2}
\Pi^{\mu\nu}(q)  =  (q^{\mu}q^{\nu}- q^2 g^{\mu\nu})\Pi^{(T)}(q^2) + q^{\mu}q^{\nu} \Pi^{(L)}(q^2) \, .
\end{equation}
The relation between the component $\Pi^{(L)}(q^2)$ and the scalar spectral function $\Pi^{(S,P)}(q^2)$ from Eq.~(\ref{corr}) can be obtained by multiplying 
both equations with $q_{\mu}q_{\nu}$. For Eq.~(\ref{corr}) we find:
\begin{equation}
q_{\mu}q_{\nu} \Pi^{\mu\nu}(q) = (m_i \mp m_j)^2\Pi^{S,P}(q^2) + (m_i \mp m_j)[\langle \bar{q}_i q_i \rangle  \mp \langle \bar{q}_jq_j \rangle] \,,
\end{equation}
and for Eq.~(\ref{corrw2}):
\begin{equation}
q_{\mu}q_{\nu} \Pi^{\mu\nu}(q) = \Pi^{(L)}(q^2) s^2 \, ,
\end{equation}
so that finally the relation reads
\begin{equation}
\label{scallong}
s^2 \Pi^{(L)}(q^2) = (m_i \mp m_j)^2\Pi^{S,P}(q^2) + (m_i \mp m_j)[\langle \bar{q}_iq_i \rangle \mp \langle \bar{q}_jq_j \rangle]\, .
\end{equation}
The second term on the right hand side is, as mentioned before, merely a subtraction constant, but the first term will be 
needed later for the relation between the spectral functions. Additionally, we need the relation between the transversal and the vector component. This can be seen by adding and subtracting 
$q^2 g^{\mu\nu}$ in equation (\ref{corrw2}); the result of this exercise is
\begin{equation}
\Pi^{V,A}(s) = \Pi^{(L)}(s) + \Pi^{(T)}(s) \equiv\Pi^{(L+T)}(s) \,.
\end{equation}  
Using the optical theorem\footnote{The reader is invited to convince himself of the correctness of this formula by 
following the proof given in appendix \ref{proof}.} the decay rate can be written as an integral of the imaginary part of this 
correlator
\begin{equation}
\label{mainf}
R_{\tau}  =  12 \pi \int\limits_{0}^{M_{\tau}^2} \, \frac{ds}{M_{\tau}^2} \left(1-\frac{s}{M_{\tau}^2} \right)^2 \left[ 
\left(1+2 \frac{s}{M_{\tau}^2} \right)  \mathrm{Im} \Pi^{(T)}(s) +  \mathrm{Im} \Pi^{(L)} (s)
\right] \, .
\end{equation}
This expression looks very much like the right hand side of Eq.~(\ref{fesr}), but before we go on we have to 
discuss weight functions in FESR. The problem is that the integral runs along the real axis where perturbation theory completely breaks down
and the OPE is no longer valid. Since we will want to turn the above expression into one that involves an integral of $\Pi(s)$ 
over a circle around the origin we had better make sure that the contributions along the axis are suppressed. In case of the 
$\tau $-decay we naturally arrive at a weight function that meets this requirement, so we could have proceeded without thinking 
about this topic at all. It is, however, necessary to know what one is doing when working with general FESR. 
That said, we can now go on along the reasoning that led to Eq.~(\ref{fesr}) and obtain: 
\begin{equation}
R_{\tau} = 6 \pi \, i \oint\limits_{|s| = M_{\tau}^2} \frac{ds}{M_{\tau}^2} \left(1-\frac{s}{M_{\tau}^2} \right)^2 \left[ 
\left(1+2 \frac{s}{M_{\tau}^2} \right) \Pi^{(T)}(s) + \Pi^{(L)} (s) 
\right] \,.
\end{equation}
We see that the integral goes over the full circle but suppresses contributions along the axis by a double zero. Experience 
shows that this is enough to make the sum rule well satisfied. Sum rules of this type are called pinch weighted finite 
energy sum rules or pFESR.
 
The vector snd scalar contributions can be cleanly separated by rearranging the terms:
\begin{equation}
\label{rearrange}
R_{\tau} = 6 \pi \, i \oint\limits_{|s| = M_{\tau}^2} \frac{ds}{M_{\tau}^2} \left(1-\frac{s}{M_{\tau}^2} \right)^2 \left[ 
\left(1+2 \frac{s}{M_{\tau}^2} \right) \Pi^{(L+T)}(s) -\left(\frac{2s}{\Mtau^2}\right) \Pi^{(L)} (s) 
\right] \,.
\end{equation} 
To get rid of renormalization scheme dependent subtraction constants\footnote{Dimensional 
analysis shows that one subtraction constant is indeed enough.} we define
\begin{eqnarray}
\label{logdiv}
D^{(L+T)}(s) & = & -s\frac{d}{ds} \left( \Pi^{(L+T)}(s) \, \right) \,, \\\nonumber
D^{(L)}(s) & = & \frac{s}{M^2_{\tau}} \frac{d}{ds} \left( s\Pi^{(L)}(s) \right)\,,
\end{eqnarray}
and use partial integration to arrive at
\begin{equation}
{\label{masterf}}
R_{\tau} = - \pi i \oint\limits_{|s|=M_{\tau}^2} \frac{ds}{s} \left( 1-\frac{s}{M_{\tau}^2}\right)^3 \left[3 \left( 1+\frac
{s}{M_{\tau}^2}\right) D^{(L+T)}(s) + 4 D^{(L)}(s) \right]\,,
\end{equation}
which can be calculated in QCD.

Further information on different parts of the spectrum can be obtained by defining moments {\cite{DP2}}, that include additional weighting factors:
\begin{equation}
R_{kl}  =  \int\limits_{0}^{M_{\tau}^2} \, \left(1-\frac{s}{M_{\tau}^2} \right)^k  \left (\frac{s}{M_{\tau}^2} \right)^l \frac{dR_{\tau}}{ds}ds\,.
\end{equation}
Just as above these moments can be converted into weighted contour integrals over the derivatives defined in (\ref{logdiv}):
\begin{equation}
R_{kl} = - \pi i \oint\limits_{|x|=1} \frac{dx}{x} \left( 3 \, \mathcal{F}^{kl}_{L+T}(x) D^{(L+T)}(M_{\tau}^2 x) +4\, \mathcal{F}^{kl}_{L}(x)
 D^{(L)}(M_{\tau}^2 x) \right) \,,
\end{equation}
where the functions $\mathcal{F}^{kl}(x) $ contain the weight functions as well as all the factors that arise because of partial integration. It 
turns out that for nonzero values of $l$ the sum rule is not very well satisfied, so we will study only $l=0$ moments for different $k$. 
The relevant functions $\mathcal{F}^{kl}(x)$ are given in the appendix as well as a general representation. 

At this point we need to remember that we have decomposed our decay rate into several contributions, as well as to 
keep in mind that the strange channel is Cabibbo-suppressed, while the axial and vector channels are allowed. We 
will also include the conventional electroweak radiative correction terms for which the numerical values is  
\begin{math} S_{EW} = 1.0201 \pm 0.0003\end{math} {\cite{Marciano:1988vm,Braaten90,Erler:2002mv}}. The correction term 
$\delta_{EW}^{\prime}$ {\cite{Braaten90}} is absorbed into $S_{EW}$. 

Inserting the operator product expansion we then obtain
\begin{equation}
R_{\tau} = 3 \left[ |V_{ud}|^2 + |V_{us}|^2 \right] S_{EW} \left( 1 + \delta^{(0)} + \sum\limits_{D \ge 2}
 \left( \cos^2 \theta_c \delta^{(D)}_{ud} + \sin^2 \theta_c \delta^{(D)}_{us} \right)  \right)\,,
\end{equation} 
where $\theta_c$ is the Cabibbo angle. The coefficients of the operator product expansion are ``hidden'' in the summed 
terms $\delta_{ij}^{(D)}$, which also include the integrals over weight functions and so on. We will be concerned with the precise 
structure of these terms in the next section of this chapter. Even now we can notice several details: the dimension-zero 
term does not involve quark masses and is then independent of the flavors. It only ``sees'' them through the elements of 
the CKM matrix (or in fact is does not because the sum of the two squared elements is very close to one). 
The perturbative contribution as well as all other SU(3) invariant effects such as instantons will drop out, if we only consider the 
SU(3) breaking part of $R_{\tau}$:
\begin{equation}
\delta R_{\tau} =  \frac{R_{\tau,V+A}}{|V_{ud}|^2} - \frac{R_{\tau,S}}{|V_{us}|^2} = 3 S_{EW} \sum\limits_{D \ge 2} [\delta^{(D)}_{ud} - 
\delta^{(D)}_{us}] \, ,
\end{equation}
which then allows us to considerably reduce the errors.

\section{The OPE for SU(3)-breaking Contributions}

In this section we will collect all the terms needed for the OPE side of the sum rule. We will only give general expressions and refer 
to the appendix for the explicit coefficients. Additionally, the dimension-zero perturbative contributions are omitted, since they cancel if we 
study SU(3) breaking effects. The interested reader is referred to Ref. {\cite{DP}} for a detailed discussion of these perturbative contributions. 
The only dimension-two operators that can be constructed are masses so that the dimension-two corrections to the power series will give the needed 
sensitivity to the strange mass. We will separately present the longitudinal and the transversal components of the spectral function, or 
equivalently, as we have seen, the scalar and the vector components. Especially, we will need to discuss the scalar contributions very carefully,
for reasons that will become apparent later. 

\subsection{Dimension-two Corrections for the Scalar Contribution}
The dimension-two terms could in principle consist of any combination of squared mass terms so it is instructive to first 
consider the form that the resulting terms are expected to have. To do that we need to discuss the correlator of 
scalar currents; it is more convenient, however, to start from the correlator of divergences of vector and axial vector 
currents respectively, since we are then working with renormalization invariant objects. To see how they are related to scalar correlators we 
will show more explicitely what we have already mentioned below Eq.~(\ref{corr}): applying the Dirac equation to  
\begin{equation}
j^s(x) = \partial^{\mu}(\bar{s} \gamma_{\mu} u(x)) \, , \quad  j^p(x) = \partial^{\mu}(\bar{s} \gamma_{\mu} \gamma^5 u(x) \, 
\end{equation}
immediately gives
\begin{equation}
\label{currdiv}
j^s(x) = i(m_s-m_u)(\bar{s} u(x)) \, , \quad j^p(x) = i(m_s+m_u)(\bar{s} \gamma^5 u(x))\,,
\end{equation}
so that the correlator should appear multiplied by the sum or the difference of the masses squared (we could have gained this insight by simply looking 
at Eq.~(\ref{scallong}) but now we understand better how the scalar correlator appears as part of a vector one). The dimension-two contribution is then
\begin{equation}
\label{dim2}
D_{ij}^{L,2}(s) = -\frac{3}{8 \pi^2 M_{\tau}^2} \left(m_i \mp m_j\right)^2 \sum d_n (q^2/\mu^2) \, a^n\,,
\end{equation}
where $ a$ and $ m$ are taken at the renormalization scale $\mu$.
The coefficients $d_n (q^2/\mu^2)$ are in general a sum over powers of logarithms:
\begin{equation}
\label{coeff}
d_n (q^2/\mu^2) = \sum\limits_{j=0}^{n} d_{nj} L^j \, ,
\end{equation}
where $L \equiv \ln(-q^2/\mu^2)$. The quantity $D^{(L)}(s)$ is related to an observable (the spectral function), so it has to obey a homogeneous renormalization 
group equation, which leads to relations between the $d_{nj}$.
The RGE is:
\begin{equation}
\left( -2 \frac{\partial}{\partial L} - \beta(a)  \frac{\partial}{\partial a}
-\gamma(a) \left(m_i \frac{\partial}{\partial m_i} + m_j  
\frac{\partial}{\partial m_j} \right) \right) D_{ij}^{L,2}(s) = 0  \, ,
\end{equation} where the factor of two comes from the fact that $L$ involves
$\mu^2$. Inserting Eq.~(\ref{dim2}) with the representation (\ref{coeff}) into
the RGE and comparing the terms of same order in $L$ and $a$ gives the
following scaling relations: \begin{eqnarray} \label{scaling}
d_{11} & = & - \gamma_1 d_{00}  \, ,  \\
d_{21} & = & -\frac{1}{2} \Big[ \left( \beta_1 + 2\gamma_1 \right) d_{00} + 2 \gamma_2 d_{00} \Big]\,,  \nonumber\\
d_{22} & = & \frac{1}{4} \gamma_1 \left( \beta_1 + 2\gamma_1 \right) d_{00}  \, ,  \nonumber\\
d_{31} & = & -\frac{1}{2} \Big[2 \left( \beta_1 + \gamma_1 \right) d_{20} + \left( \beta_2 + 2\gamma_2 \right) d_{10} +2\gamma_3 d_{00} \Big] \, ,  \nonumber\\ 
d_{32} & = & \frac{1}{4} \Big[ \left( \gamma_1\beta_2 + 2 \gamma_2 (2\gamma_1+\beta_1) \right) d_{00} + \left( \beta_1 + \gamma_1 \right)(2\gamma_1+\beta_1) d_{10} \Big]  \, ,  \nonumber\\
d_{33} & = & -\frac{1}{12} \gamma_1 \left( \beta_1 + \gamma_1 \right)(2\gamma_1+\beta_1) d_{00}  \, .  \nonumber
\end{eqnarray}
This means that all that needs to be calculated are the $d_{i0}$ coefficients; all others are completely determined by the scaling laws of the 
renormalization group. We could now go on and just integrate the above expression for $ D_{ij,V/A}^{L,2}(s)$ over our circular contour. Then 
we have to set the renormalization scale at $\mu^2 = M_{\tau}^2$. It turns out, however, that this leads to a rather bad convergence behavior 
due to large logarithms when integrating over the circle. This can be avoided by using the so called ``contour improved perturbation theory''(CIPT), 
which means the following: both the renormalization group and the operation of integrating are linear, and we can set our 
renormalization scale as $\mu^2 =-q^2$ before doing the integration. We will thus sum up the contributions point by point along the contour. 
As the masses and couplings are given at the renormalization scale the resulting expression will now involve integrals over both the running 
coupling as well as the running mass. This not only leads to simpler expressions for the integrand but also to a better convergence, as has 
been shown in Refs. {\cite{DP,PP1}}. For illustration purposes we will quote here the perturbation series as given in Ref. {\cite{PP1}} for ordinary 
perturbation theory and CIPT. The quoted values are the sum of the longitudinal and the vector part that we have not yet discussed. The 
numerical values are:
\begin{equation}
1 + 0.5866 + 0.5566 + 0.5150 +...
\end{equation}
for ordinary perturbation theory and
\begin{equation}
0.9840+ 0.4613 + 0.3421 + 0.298 +...
\end{equation}
for CIPT. The better convergence of the latter should be obvious.

Finally we can write $\delta R_{\tau}$ in a form that is more transparent for the numerical evaluation:
\begin{equation}
\delta R_{kl}^{L,2} = 6 S_{EW} \frac{m_s^2(\Mtau^2)}{\Mtau^2} (1-\varepsilon_d^2) \sum\limits_{n=0} d_n(\xi)  B_{kl}^{L(n)}\,,
\end{equation}
where
\begin{equation}
 B_{kl}^{L(n)}  =  \frac{1}{2 \pi i} \oint\limits_{|x|=1} \frac{dx}{x}  \mathcal{F}^{kl}_{L}(x) \left[\frac{m(-\xi^2 M_{\tau}^2 x)}{m(M_{\tau}^2)}
\right]^2 a^n(-\xi^2 M_{\tau}^2 x)\, ,
\end{equation}
and $\varepsilon_d$ is the ratio of of strange and down quark masses calculated rather accurately in Chiral Perturbation Theory~{\cite{Leutwyler96}}: 
\mbox{$\varepsilon_d \equiv m_d/m_s = 0.053 \pm 0.002$}. Later we will also encounter $\varepsilon_u$, for which the value is
\mbox{$\varepsilon_u \equiv m_u/m_s = 0.029 \pm 0.003$}. We have also introduced a parameter $\xi$ that is of order one and will be varied later as an estimate for errors.  
This is necessary because the truncation of the perturbative OPE contributions leaves a residual 
scale dependence in the obtained expressions (the reason being that the RGE for the correlation function ``mixes'' different orders of $a$ and $L$,
as can be seen from the scaling relation for the Wilson coefficients). In the language of $s$ and $\mu$ we find that $\xi^2=-\mu^2/s$ so that all logarithms
are zero for $\xi=1$.

\subsection{Dimension-four Corrections and Higher}
While there was only one operator of dimension two, namely the mass, we will have to consider several contributions with dimension four.
First of all there is obviously a term with the fourth power of the masses, next we have to take into account quark condensates 
multiplied by a mass and finally there could also be a gluon condensate operator. Again we can begin by trying to get an intuitive feeling about 
the structure that we ought to expect. Remember that the longitudinal contribution is just the scalar correlator times a squared term 
in the masses so that the dimension-4 terms of the longitudinal contribution will be related to the dimension-2 
contributions of the scalar series. This means that there are no condensate terms, except for those contributions from the subtraction 
constants in Eq.~(\ref{scallong}).  
A general expression for the dimension-2 series of the scalar correlator is given in Ref. \cite{jm2}, from which we can easily calculate the following 
expressions relevant to us:
\begin{eqnarray}
D_{ij}^{L,4}& =& \frac{1}{\Mtau^2 s} \bigg[ -\langle \left(m_i \mp  m_j\right)\left(\bar{q}_i q_i \mp \bar{q}_j q_j \right) \rangle 
 \\\nonumber&& + \frac{3}{8\pi^2} \left(m_i^2(- \xi^2 s) \mp m_j^2(- \xi^2 s)\right) 
 (m_i^2(- \xi^2 s)+m_j^2(- \xi^2 s))
\\\nonumber&&\times \Big(4+ 4\ln \xi +a (- \xi^2 s)\Big(-8\zeta(3)+ \frac{82}{3}+\frac{112}{3}\ln \xi 
+16\ln^2 \xi\Big)\Big) \\ \nonumber
 && \pm m_i(- \xi^2 s) m_j(- \xi^2 s) \Big(6 + 4 \ln \xi + a (- \xi^2 s) (48-8\zeta(3)\\\nonumber &&
+\frac{160}{3}\ln \xi + 16\ln^2 \xi)\Big)
\bigg]\,,
\end{eqnarray}
that has to be integrated over the contour according to Eq.~(\ref{masterf}). While the masses will be scaled according to the renormalization group 
equations for this integration, we will take the combination of condensate times mass to be scale independent. This is because they are given 
to a very good approximation in chiral perturbation theory by the GMOR relation and should then indeed be scale invariant.

Next we have to include terms of dimension 6. The same rationale that fixed the structure of the dimension 4 can again be applied and shows that
we have to take into account every possible operator of dimension 4 and multiply them by the standard mass prefactor. We will then consider $m^6$ terms, 
condensates multiplied by cubic mass terms and finally a gluon condensate term. The relevant expressions, calculated again from those given in 
Ref.~{\cite{jm2}} are:
\begin{eqnarray}
D_{ij}^{L,6} & = &-\frac{3}{4 \pi^2 \Mtau^2 s^2} \left(m_i^2(- \xi^2 s) \mp m_j^2(- \xi^2 s)\right) \Big( \pm m_i^3(- \xi^2 s)m_j(- \xi^2 s) 
\\\nonumber&&- 2m_i^2(- \xi^2 s) m_j^2(- \xi^2 s) \pm m_i(- \xi^2 s) m_j^3(- \xi^2 s)  \\
&& \nonumber+ 2\ln \xi\Big(m_i^4(- \xi^2 s) \pm 2m_i^3(- \xi^2 s)m_j(- \xi^2 s) \\\nonumber&& 
 \pm2m_i^3(- \xi^2 s)m_j(- \xi^2 s)+m_j^4(- \xi^2 s) \Big) \Big),\\
D_{ij}^{L,6}& = &-\frac{1}{\Mtau s^2} \left(m_i^2(- \xi^2 s) \mp m_j^2(- \xi^2 s)\right)   \\\nonumber
&&\times \left[ \left(-1-\frac{4}{3} \left(\frac{14}{4} +3\ln \xi \right) 
a (- \xi^2 s) \right) \Big( m_i \langle {\bar q_i}q_i \rangle + m_j \langle {\bar q_j}q_j \rangle \Big)  \right.  \\\nonumber
&& \pm \left. 2\left(-1-\frac{4}{3} \left(\frac{17}{4}+ 3\ln \xi \right) a (- \xi^2 s) \right) 
\Big( m_i \langle {\bar q_j}q_j \rangle + m_j \langle {\bar q_i}q_i \rangle \Big) \right] ,\\
D_{ij}^{L,6}& = &\frac{1}{4 \Mtau^2 s^2} \left(m_i^2(- \xi^2 s) \mp m_j^2(- \xi^2 s)\right) \langle aFF \rangle\,.
\end{eqnarray}
Finally we will include operators of dimension 8. The complete expressions will obviously become quite lengthy, but we will only consider the terms
that are numerically most important. The largest would be a leading order 4 quark condensate but this turns out to be zero {\cite{jm2}}, so we shall
include the mixed condensate, that is suppressed by only one mass power (apart from, obviously, the squared terms in the prefactors) and the first order four
quark operators. Since these four quark operators cannot be reliably estimated, they are conventionally reduced to the chiral
condensates using the vacuum saturation approximation {\cite{SVZ}}.

For that we shall take a general example of a four quark condensate with arbitrary Dirac structure $\Gamma$ and fixed color structure
$t^a$, where $t^a$ are the Gell-Mann matrices that generate the SU(3) multiplied by an overall factor of one half, in other words we 
discuss
\begin{equation}
\langle {\bar q^A} \Gamma^1 t^a q^B {\bar q^C} \Gamma^2 t^a q^D\rangle\,,
\end{equation}
or
\begin{equation}
\Gamma^1_{\mu\nu} \Gamma^2_{\rho\lambda}t^a_{kl} t^a_{mn} \langle {\bar q^A_{\mu k}} q^B_{\nu l} {\bar q^C_{\rho m}} 
                                         q^D_{\lambda n}\rangle\,.
\end{equation}
Dirac indices are denoted by Greek letters and run from one to four, while the Roman indices, corresponding to the color, run only from 
one to three. A sum over flavors, denoted by capital Roman indices, is not included.

We will now insert a complete set of eigenstates
into the bracket and assume that the contribution from the vacuum state is the largest. Since the vacuum is a flavor- and 
colorless scalar state, we arrive at
\begin{equation}
\label{vacsat}
\Gamma^1_{\mu\nu} \Gamma^2_{\rho\lambda}t^a_{kl} t^a_{mn}  \delta_{AB} \frac{\delta_{\mu\nu}}{4} \frac{\delta_{kl}}{3} 
                                                           \delta_{CD} \frac{\delta_{\rho\lambda}}{4} \frac{\delta_{mn}}{3}
                                                         \langle {\bar q^A_{\mu k}} q^B_{\nu l}\rangle 
                                                      \langle {\bar q^C_{\rho m}}q^D_{\lambda n}\rangle\,,
\end{equation}
where the factors in the denominator have to be included to ensure proper vacuum normalization.
This expression can be more compactly written as:
\begin{equation}
\frac{1}{144} \mathrm{Tr}\Gamma^1 \mathrm{Tr}\Gamma^2   \mathrm{Tr} t^a \mathrm{Tr} t^a \delta_{AB}\delta_{CD}  
                                                        \langle {\bar q^A} q^B\rangle 
                                                      \langle {\bar q^C}q^D\rangle\,.
\end{equation}
The trace of the color matrices is zero, but we will keep track of the explicit expressions, since the final result we will obtain can 
then more easily be generalized.
Additionally, we have to consider the expression where the quarks in (\ref{vacsat}) have been exchanged, i.e.
\begin{equation}
-\Gamma^1_{\mu\nu} \Gamma^2_{\rho\lambda}t^a_{kl} t^a_{mn} \langle {\bar q^A_{\mu k}} q^D_{\lambda n} {\bar q^C_{\rho m}} 
                                         q^B_{\nu l}\rangle\,,
\end{equation}
where the additional minus sign stems from the interchange of the fermion fields. Again we insert the vacuum and find
\begin{equation}
\label{vacsat2}
-\Gamma^1_{\mu\nu} \Gamma^2_{\rho\lambda}t^a_{kl} t^a_{mn}  \delta_{AD} \frac{\delta_{\mu\lambda}}{4} \frac{\delta_{kn}}{3} 
                                                           \delta_{CB} \frac{\delta_{\rho\nu}}{4} \frac{\delta_{ml}}{3}
                                                         \langle {\bar q^A_{\mu k}} q^D_{\lambda n}\rangle 
                                                      \langle {\bar q^C_{\rho m}}q^B_{\nu l}\rangle\,,
\end{equation}
or
\begin{equation}
-\frac{1}{144} \mathrm{Tr}\Gamma_1\Gamma_2   \mathrm{Tr} t^a t^a \delta_{AD}\delta_{CB}  
                                                        \langle {\bar q^A} q^D\rangle 
                                                      \langle {\bar q^C} q^D\rangle\,,
\end{equation}
Combining the two we eventually arrive at
\begin{eqnarray}
\lefteqn{
\frac{1}{144} \Big(\mathrm{Tr}\Gamma^1 \mathrm{Tr}\Gamma^2   \mathrm{Tr} t^a \mathrm{Tr} t^a \delta_{AB}\delta_{CD}  
                                                        \langle {\bar q^A} q^B\rangle 
                                                      \langle {\bar q^C}q^D\rangle } \\ \nonumber
             && - \mathrm{Tr}\Gamma^1\Gamma^2   \mathrm{Tr} t^a t^a \delta_{AD}\delta_{CB}  
                                                        \langle {\bar q^A} q^D\rangle 
                                                      \langle {\bar q^C} q^D\rangle \Big)\,.
\end{eqnarray}
Finally, for calculations we need the relation
\begin{equation}
t^a t^a = C_f \mathbbm{1},
\end{equation} 
where $C_f$ is the Casimir invariant of the SU(3) so that Tr $t^a t^a$ = 4 in the fundamental representation.

The four quark operators we are dealing with are {\cite{jm2}}
\begin{equation}
\langle {\bar q_i} \sigma_{\mu\nu}t^a q_j {\bar q_j\sigma^{\mu\nu}t^a}q_i\rangle\,  \mbox{ \qquad and \qquad}  
\langle {\bar q_i}\gamma_{\mu}t^a q_i \sum\limits_{k=u,d,s} {\bar q_k}\gamma^{\mu}t^a q_k \rangle   \,,
\end{equation}
for which the vacuum saturation gives 
\begin{equation}
-\frac{4}{3}\langle {\bar q_i}q_i \rangle \langle {\bar q_j}q_j \rangle \,  \mbox{ \qquad and \qquad} 
-\frac{4}{9}\langle {\bar q_i}q_i \rangle^2   \,,
\end{equation}
so that finally the complete expressions read:
\begin{eqnarray}
D_{ij}^{L,8}& = & (m_i \mp m_j)^2\left(\mp \frac{1}{2} m_i \langle g {\bar q_i} \sigma Fq_i \rangle + 
 m_j \langle g {\bar q_j} \sigma F q_j \rangle \right)  \\\nonumber 
&& \pm 4\pi^2 {\bar a} \frac{4}{3} \langle {\bar q_i}q_i \rangle \langle {\bar q_j}q_j \rangle  
+\frac{4}{3}\pi^2{\bar a} \frac{4}{9} \left( \langle {\bar q_i}q_i \rangle^2 + \langle {\bar q_j}q_j \rangle^2 \right)\,, 
\end{eqnarray}
where ${\bar a}$ is the coupling at the scale $\Mtau^2$ and $\langle g {\bar q_j} \sigma F q_j \rangle$ is the mixed
condensate, for which we take $\langle g {\bar q_i} \sigma F q_i \rangle = 0.8 \langle {\bar q_i}q_i \rangle$.

There is one last effect that we have to include in the OPE, namely instantons. Maltman and Kambor have argued in 
Refs. {\cite{MK,MK1}} that these 
effects are important in (pseudo)scalar currents and should preferably be accounted for. This was done by taking a fit to 
several FESR and comparing the obtained result to the one from a BSR; it was found that consistency of the results strongly 
favors inclusion of instantons. Unfortunately, it is not entirely clear how to parametrize them and the model conventionally 
used in sum rules is the instanton liquid model (ILM), that goes back to Shuryak et.~al.~{\cite{Shuryak84}}. The OPE expressions 
for the FESR can be evaluated from
\begin{equation}
\label{Inst}
\frac{-1}{2\pi i} \oint\limits_{|s| = s_0} ds \, s^k \, \Pi_{ij}^{ILM}(s) = \frac{\mp3(m_i\pm m_j)^2 \eta_{ij}}{4 \pi} \int\limits_{0}^{s_0} ds \, s^{k+1} 
J_1(\rho_I \sqrt{s})\, Y_{1}(\rho_I \sqrt{s})\,,
\end{equation}
i.e.
\begin{equation}
R_{kl}^{ILM} = \frac{\mp 18 \pi(m_i \pm m_j)^2 \eta_{ij}}{\Mtau^4} \int\limits_{0}^{s_0} 
ds \, \left(1-\frac{s}{\Mtau^2}\right)^{2+k} \left(\frac{s}{\Mtau^2}\right)^l J_1(\rho_I \sqrt{s})\, Y_{1}(\rho_I \sqrt{s})\,,
\end{equation}
where $\eta_{ud} \equiv 1$, while  $\eta_{us} \approx 0.6$; $Y_1(s)$ and  $J_1(s)$ are the respective Bessel functions and $\rho_I$ is 
the average instanton size, which we will take to be $\rho_I \simeq$ 1/0.6 as taken by Maltman and Kambor. In their calculations they add an 
additional error to their results, which they estimate from consistency. To us it is less obvious how to estimate this error, and we will 
not give one, especially since this uncertainty would not appear in our final result anyway.

\section{OPE for the Vector Contribution}
As we will see later, the OPE expressions we actually need for our calculations are the vector contributions. Leaving out dimension
 zero, the expression for dimension two is:  
\begin{eqnarray}
\label{L+T}
D_{ij}^{L+T,2}(s) & = & \frac{3}{4 \pi^2 s} \left[ \left(m_i^2 + m_j^2 \right) \sum\limits_{n=0} c_n(q^2/\mu^2)  a^n  \right. \\
                      &   &  \pm m_i m_j \sum\limits_{n=1} e_n (q^2/\mu^2) a^n  \nonumber \\
                      &   & \left. + \left(\sum\limits_{k} m_k^2 \right) \sum\limits_{n=2} f_n (q^2/\mu^2) a^n \nonumber \right]\,.
\end{eqnarray}
As in the longitudinal contribution the quantity $D^{L+T}(s)$ will have to obey a RGE and the scaling of 
the coefficients will be according to the relations~(\ref{scaling}). The $c_{i0}$, $e_{i0}$ and $f_{i0}$ are given in the appendix along with the $d_{i0}$.
We are now in the position to give precisely the vector structure of the term $\delta_{ij}^{(2)}$ from the previous section.
Defining
\begin{eqnarray}
B_{kl}^{L+T(n)} = -\frac{1}{4 \pi i} \oint\limits_{|x|=1} \frac{dx}{x^2}  \mathcal{F}^{kl}_{L+T}(x) \left[\frac{m(-\xi^2M_{\tau}^2 x)}{m(M_{\tau}^2)}
\right]^2 a^n(-\xi^2 M_{\tau}^2 x)\, ,
\end{eqnarray}
we get an expansion analogous to Eq.(\ref{L+T}):
\begin{equation}
\delta_{ij}^{(2),L+T} = 6 \frac{(m_i^2(M_{\tau}^2) + m_j^2(M_{\tau}^2))}{M_{\tau}^2} \sum\limits_{n=0} c_n(\xi)  
B_{kl}^{L+T(n)}\,.
\end{equation}
In writing down this line we have already added vector and axialvector terms $\delta_{ij}^{(2)}=(\delta_{ij,V}^{(2)}+ \delta_{ij,A}^{(2)})$ which 
gets rid of the $m_im_j$ terms; since we discuss only the flavor symmetry breaking $\delta R_{\tau}^{kl}$ the term proportional to $f_n(\xi)$ 
drops out, as do the contributions of $m_u$, so that we finally arrive at  
\begin{equation}
\delta R_{\tau}^{kl,L+T} = 18 S_{EW} \frac{m_s^2(M_{\tau}^2)}{M_{\tau}^2}(1 - \varepsilon_d^2 ) \sum\limits_{n=0} c_n(\xi)  
B_{kl}^{L+T(n)}\,.
\end{equation}
In accordance with Ref. {\cite{PP2}} we will refer to the sum in this equation as $\Delta^{(L+T)}_{kl}$.

The dimension-4 structure of the vector current is more complicated than that of the scalar and we have therefore 
relegated a thorough discussion into the appendix and quote here the results. Adding vector and axial vector
channels, the SU(3) breaking piece is given by
\begin{equation}
\label{vectordim4}
s^2 \left(D^{L+T,4}_{ud}(s)-D^{L+T,4}_{us}(s) \right) =  -4~\delta O_4 (-s) \sum\limits_{n=0} q_n (q^2/\mu^2)\, a^n \nonumber
\end{equation}
\begin{displaymath}
+\frac{6}{\pi^2} m_s^4 \left(1-\varepsilon_d^2\right) \sum\limits_{n=0} \left[ \left(1+\varepsilon_d^2 \right) h^{L+T}_{n}(s/\mu^2)
-\varepsilon_u^2 g^{L+T}_{n}(s/\mu^2) \right]~a^n \,,
\end{displaymath}
which leads to the following equation for the theoretical side of the $\tau$-decay rate:
\begin{eqnarray}
\delta R^{L+T}_{kl}& = &12 S_{EW} \left\{3\frac{m_s^4(\Mtau^2)}{\Mtau^4}(1-\varepsilon_d^2)\left[\left(1+\varepsilon_d^2\right)H_{kl}^{L+T}-2\varepsilon_u^2
G_{kl}^{L+T}\right]  \right. \nonumber \\
&& - \left.4\pi^2 \frac{ \delta O_4(\Mtau^2)}{\Mtau^4}Q^{L+T}_{kl} \right\} \,,
\end{eqnarray}
where all the coefficients and the definition of the expressions $H_{kl}^{L+T}, G_{kl}^{L+T}$ and $Q^{L+T}_{kl}$ are given in the appendix and we have defined the
 quark condensate operator as $\delta O_4 = m_s \langle\bar{s}s-\varepsilon_d \bar{d} d \rangle$. We emphasize 
that we have only made use of SU(3) symmetry and cancellation between axial and vector contributions and have not neglected any small terms, even though
terms quartic in masses are expected to be negligible. We have, however, neglected the scaling of the quark condensate operator as discussed before. 
For this operator we have taken the numerical values for the quark condensates and $\varepsilon_d$ as given
in the appendix and estimated it as in Ref. {\cite{PP2}.

Our values for $H^{L+T}_{kl}$ and $G^{L+T}_{kl}$ as well as the $Q^{L+T}_{kl}$ are given in 
Table {\ref{QGH}} for all relevant moments. In principle one could use a combined fit for different moments to determine not only the strange mass, 
but also a more precise value of the quark condensates (for a thorough discussion of subtraction constants and numerical corrections
to the GMOR relation see, for example, Ref. {\cite{J}}). This is limited by experimental accuracy nowadays.

The complete sum rule up to dimension-four is:
\begin{eqnarray}
\lefteqn{\delta R^{kl,L+T} = 12 S_{EW} \left\{\frac{3}{2} \frac{m_s^2(\Mtau^2)}{\Mtau^2}  \left(1-\varepsilon_d^2\right) \Delta^{L+T}_{kl}+
3\frac{m_s^4(\Mtau^2)}{\Mtau^4}(1-\varepsilon_d^2) \right.} \quad\\ \nonumber
&& \times \left.\left[\left(1+\varepsilon_d^2\right)H_{kl}^{L+T}-2\varepsilon_u^2G_{kl}^{L+T}\right]- 
4\pi^2 \frac{ \delta O_4(\Mtau^2)}{\Mtau^4} Q_{kl}^{L+T} \right\}\,.
\end{eqnarray}
Before we go on to combine the scalar and vector contributions let us discuss briefly the implications of dimension 6 contributions. We will have to include them in our analysis
to be consistent since the leading dimension 6 terms turn out to be larger than the quartic mass terms of dimension 4. We expect the leading
terms to be four quark operators that are not suppressed by quark masses and will estimate these along the lines of 
Ref. {\cite{PP2}: if we take the OPE to be
\begin{equation}
\label{dim61}
s^3 \left(D^{L+T,6}_{ud}(s)-D^{L+T,6}_{us}(s) \right) = -3 \delta O_6^{L+T} (\mu^2) \sum\limits_{n=0} q_n (q^2/\mu^2)\, a^n\,
\end{equation}
and neglect the scaling of 4 quark condensate operators as for the regular quark condensates\footnote{The reason is somewhat different however: when
discussing the renormalization group equations we have seen that the scaling is given by higher orders in the coupling as well as higher order 
logarithmic terms, both of which we do not include, so we must neglect the scaling altogether. The same argument holds for the scaling of the gluon condensate.}
we find for $\delta R^{kl}$
\begin{equation}
\delta R^{k0,6} = c_6^k \delta R^{00,6}\,,
\end{equation}
where the $c_6^k$ are given by
\begin{equation}
c_6^k=\frac{1}{4 \pi i} \oint\limits_{|x|=1} \frac{dx}{x} \frac{1}{x^3} \mathcal{F}^{kl}_{L+T}\,.
\end{equation}
\pagebreak

\noindent
Then the final estimate for these terms is:
\begin{eqnarray}
\label{Dunno}
\delta R^{00,6}& \approx & -\frac{12 \pi^2}{\Mtau^6} \, 3\,\delta O_6^{L+T} \,, \\
\delta R^{10,6}& \approx & -\frac{12 \pi^2}{\Mtau^6} \,3\,\delta O_6^{L+T}\,, \nonumber\\
\delta R^{20,6}& \approx & -\frac{12 \pi^2}{\Mtau^6} \,2\,\delta O_6^{L+T} \,, \nonumber\\
\delta R^{30,6}& \approx & -\frac{12 \pi^2}{\Mtau^6} \,0\,\delta O_6^{L+T}\,, \nonumber\\
\delta R^{40,6}& \approx & +\frac{12 \pi^2}{\Mtau^6} \,3\,\delta O_6^{L+T} \,. \nonumber
\end{eqnarray}
Now we need to estimate the dimension 6 operators: again we will reduce them to the chiral condensates using the vacuum saturation hypothesis
\begin{eqnarray}
\delta R^{00,6} & \approx & -S_{EW} \frac{36 \pi^2}{\Mtau^6} \delta O_6^{L+T} \approx S_{EW} \frac{256\pi^4}{9} \bar{a} \frac{\langle \bar{s} s \rangle^2-
\langle \bar{d} d \rangle^2}{\Mtau^6} \nonumber \\
& \approx & 3 S_{EW} \delta^{00,6}_{ud} \frac{1-v_s^2}{2} \approx (0.6\pm2.3) \times 10^{-3}\,,
\end{eqnarray}
where the numerical coefficients for the OPE are given in Ref. {\cite{BNP}}
and we have taken the measured value of $\delta^{00,6}_{ud} = 0.001\pm 0.004$ as in Ref. {\cite{PP2}} and $v_s$ is the ratio of strange to non strange condensates, 
for which the current estimate is
\begin{equation}
v_s = \frac{\langle \bar{s} s \rangle}{\langle \bar{d} d \rangle} \approx 0.8 \pm 0.2\,.
\end{equation}
Our numerical result confirms the necessity to include these terms since the quartic mass terms are of order 
$10^{-5}$ and thus much smaller, as noted above. 

With all the tools given in this section it is generally possible to calculate the strange mass from the existing $\tau$-decay data. These 
determinations have been performed rather successfully {\cite{PP2,mutau,Chetyrkin:1998ej,Korner:2000wd,Kam00}}, but they all suffer from the same problem; to see this we shall very briefly go back to the
perturbative series of dimension 2 and examine its convergence: 
we define
\begin{equation}
\Delta^{(L)}_{kl}=\sum\limits_{n=0} d_n(\xi)  B_{kl}^{L(n)}
\end{equation}
and
\begin{equation}
\Delta^{(2)}_{kl} = \frac{1}{4} \left( 3 \Delta^{(L+T)}_{kl} + \Delta^{(L)}_{kl} \right)\,.
\end{equation}
Above we have stated that a 
contour improved calculation leads to better convergence and given the series for $\Delta^{(2)}$ as an example. Now 
let us discuss the separate contributions. For the vector contribution we find
\begin{equation}
\Delta^{(L+T)}_{00} = 0.7532 + 0.2137 + 0.0650 - 0.0002 c_3+...
\end{equation}
Taking $c_3 \approx c_2 \left( c_2 / c_1 \right) = 323$ as a naive estimate gives a last term of -0.0501, so that this series 
is rather well behaved. Unfortunately the same is not true for the longitudinal contribution. Even the improved 
series is unsatisfactory in its convergence:
\begin{equation}
\Delta^{(L)}_{00} = 1.6334 + 1.2813 + 1.2892 + 1.5026 +...
\end{equation}
This bad convergence does not manifest itself too greatly in the central value of $\delta R_{\tau}^{kl}$ because it is 
suppressed by a factor three when both contributions are added and the bad behavior is dominant only in higher orders. It does, 
however, have a significant impact on the errors obtained. Table \ref{Deltanum} lists our values 
for the dimension 2 contributions for the moments we will study.
Throughout this work we will be using for $\alpha_s(\Mtau)$ the value obtained by the ALEPH collaboration in Ref {\cite{Barate98}}, 
meaning
\begin{equation} 
\alpha_s \left( \Mtau \right) = 0.334\pm0.022\,.
\end{equation}
The origin of the errors is as follows: The theoretical error is taken by varying the parameter $\xi$ in an interval from \mbox{0.75 - 2.0}.
Additionally we have also included an error due to the uncertainty in the coupling constant.
\begin{table}
\begin{center}
\[
\begin{array}{|c|ccc|}\hline
(k,l) & \Delta^{(L+T)}_{kl} &  \Delta^{(L)}_{kl} & \Delta^{(2)}_{kl}  \\  \hline
(0,0) & 0.98 \pm 0.13 \pm 0.10 &  5.71 \pm 1.52 \pm 0.65 & 2.16 \pm 1.22 \pm 0.09\\ 
(1,0) & 1.49 \pm 0.12 \pm 0.07 &  5.73 \pm 1.62 \pm 0.83 & 2.55 \pm 1.55 \pm 0.16\\
(2,0) & 2.02 \pm 0.25 \pm 0.03 &  3.98 \pm 1.76 \pm 0.54 & 2.37 \pm 1.89 \pm 0.14\\
(3,0) & 2.60 \pm 0.49 \pm 0.11 &  2.48 \pm 2.05 \pm 0.65 & 1.94 \pm 2.23 \pm 0.20\\ 
(4,0) & 3.23 \pm 0.78 \pm 0.21 &  2.47 \pm 1.59 \pm 0.35 & 2.13 \pm 1.88 \pm 0.13\\\hline
\end{array}
\]
\begin{minipage}[t]{12cm} 
\caption{\label{Deltanum}Numerical values for the dimension 2 contributions. The first error shows the uncertainties 
for fixed coupling, while the second shows the uncertainty induced by the coupling constant.}
\end{minipage}
\end{center}
\end{table}

As mentioned before, the combined series is dominated in its convergence by the vector part of the sum for low orders. For higher 
orders the bad convergence of the longitudinal series causes the coefficients to increase again so that the series is merely asymptotic in 
some sense and we have to find a convention to truncate it. We have have followed the standard procedure to truncate the series after the minimum value coefficient and 
taken the full term of the next order as an additional error. 
To keep this discussion more transparent we will show exactly how the values for $\Delta^{(2)}_{kl}$ arise and give 
the series explicitly
\begin{eqnarray} 
\Delta^{(2)}_{00} & = & 0.9732 + 0.4806 + 0.3710 + 0.3377 +...\\
\Delta^{(2)}_{10} & = & 1.0396 + 0.5578 + 0.4829 + 0.4732 +...\\
\Delta^{(2)}_{20} & = & 1.1167 + 0.6443 + 0.6111 + 0.6382 +...\\
\Delta^{(2)}_{30} & = & 1.2008 + 0.7393 + 0.7569 + 0.8361 +...\\
\Delta^{(2)}_{40} & = & 1.2907 + 0.8430 + 0.9223 + 1.0713 +...\,
\end{eqnarray}
so that we will take the (0,0) and (1,0) series up to third order, the (2,0) up to second and truncate the 
(3,0) and (4,0) series after first order.  
For the longitudinal series itself one could also adopt this prescription but the 
behavior becomes extremely bad for higher moments, so that this does not seem meaningful. We have instead used an ad hoc definition and 
truncated the series at the same order as the total series. On the other hand we have seen that the vector series converges rather 
nicely and we will use the complete series with the estimated third order coefficient whenever we discuss $\Delta^{(L+T)}_{kl}$.
\begin{table}
\begin{center}
\[
\begin{array}{|c|ccc|}\hline
(k,l) & \ H^{L+T}_{kl} &  G^{L+T}_{kl} & Q^{L+T}_{kl}  \\  \hline
(0,0) & -1.83 \pm 0.40 \pm 0.08 & -1.30 \pm 1.32 \pm 0.14 & 0.076 \pm0.047 \pm 0.162\\ 
(1,0) & -1.34 \pm 0.75 \pm 0.20 & -1.03 \pm 1.76 \pm 0.28 & 0.514 \pm0.038 \pm 0.014\\
(2,0) & -0.38 \pm 1.01 \pm 0.30 & -0.43 \pm 2.02 \pm 0.42 & 0.932 \pm0.030 \pm 0.008\\
(3,0) & 1.09  \pm 1.10 \pm 0.39 & 0.54 \pm 1.96 \pm 0.53 & 1.332 \pm0.026 \pm 0.007\\ 
(4,0) & 3.11  \pm 1.36 \pm 0.44 & 1.91 \pm 1.43 \pm 0.57 & 1.714 \pm0.077 \pm 0.024\\\hline
\end{array}
\]
\begin{minipage}[t]{12cm} 
\caption{\label{QGH}Numerical values for the dimension 4 contributions; again we have split the uncertainty into the theoretical uncertainty and the uncertainty in the coupling.}
\end{minipage}
\end{center}
\end{table}

%% file: spectr2.tex
\chapter{Scalar and Pseudoscalar Spectral Functions}
\label{spectr}
\hspace*{\fill}\begin{minipage}[t]{8.2cm}
\begin{footnotesize}
{\em You can know the name of the bird in all the languages of the world, but when you're finished, you'll know 
absolutely nothing whatsoever about the bird.}\\
(Richard P. Feynman, What Do \underline{You} Care What Other People Think?)\\
\end{footnotesize}
\end{minipage}

\noindent
We have seen in the last chapter that and how the strange mass can be determined from $\tau$-decays. We have also 
seen that large uncertainties in this determination come from the bad convergence of the dimension two longitudinal 
perturbative series. This could be avoided by measuring longitudinal and transversal decays separately 
and then considering only the $L+T$ terms of the sum rule. Unfortunately, a complete spin disentanglement has not been 
performed over the entire allowed kinematic range. If the scalar and pseudoscalar spectral functions are known well 
enough up to an energy of $\Mtau^2$, however, they can simply be subtracted from the decay rate using 
\mbox{Eqs.~(\ref{mainf}) and (\ref{rearrange}):}
\begin{equation}
\label{speclong}
\delta R^{L}_{kl} = -24 \pi^2 \int\limits_{0}^{\Mtau^2} \frac{ds}{\Mtau^2} \left( 1-\frac{s}{\Mtau^2} \right)^{2+k}
\left(\frac{s}{\Mtau^2}\right)^{1+l} \left[ \rho_{ud}^{(L)}(s) - \rho_{us}^{(L)}(s) \right] \,,
\end{equation}
where $\rho_{ud}^{(L)}(s)$ and $\rho_{us}^{(L)}(s)$ are the longitudinal spectral functions for $ud$ and $us$ currents respectively.
From Eq.~(\ref{scallong}) we see that
\begin{equation}
\rho_{ij}^{(L)}(s) = \left(m_i + m_j\right)^2 \frac{\rho_{ij,ps}}{s^2} +  \left(m_i - m_j \right)^2 \frac{\rho_{ij,s}}{s^2}\,,
\end{equation}
which allows us to relate the scalar($s$) and pseudoscalar($ps$) spectral functions to the longitudinal ones we need (we 
will in fact relate them to those of the divergence of axial and vector currents, but this is equivalent, as shown in 
\mbox{Eq.~(\ref{currdiv})).} 

\section{The Spectral Function for the ud Current}
As a general rule the $ud$ current gives much smaller contributions to $\delta R$ than the $us$ current and 
we could neglect the scalar channel entirely. This can be seen if we approximate the spectral functions by their 
theoretical expressions. The OPE of the pseudoscalar and scalar currents differ only in a general minus sign in the 
prefactor as well as some higher order terms that do not contribute much. The pseudoscalar 
spectral function is multiplied by $(m_i+m_j)^2$, while the scalar comes with a factor of $(m_i-m_j)^2$. This means the scalar 
spectral function for the $ud$ current is suppressed by a factor of 
$(m_u-m_d)^2$. Additionally, it is not entirely clear how to parametrize the scalar spectral function
cleanly, so we take the theoretical expression as an estimate of how much the contribution might be.

There is one more noticeable feature: dimensional analysis tells us that both spectral functions will rise linearly for 
large $s$.
A resonance approximation can then only be justified for small $s$, where the theoretical expression is not yet a good 
approximation. This is to be compared to the vector sum rule that has a constant spectral density for large $s$, thereby making 
resonance saturation more intuitive.

In any case the pseudoscalar channel gives sizable contribution and we will have to include it in our expressions. It is 
dominated by the pion pole and we will also add the next two higher lying resonances, namely the $\pi(1300)$ and 
the $\pi(1800)$. In the narrow width approximation the spectral function is simply a sum of $\delta$ functions,
\begin{equation}
\rho_{ud}^{(L)}(s) = \sum\limits_{\pi_{n}} 2f_n^2 \frac{M_{n}^4}{s^2} \delta (s-M_n^2)\,,
\end{equation} 
where from the definition of the spectral function we see that
\begin{equation}
\langle \pi_n |\partial^{\mu} j_{\mu} | 0 \rangle = \sqrt{2} f_n M_n \,.
\end{equation}
Of course narrow width approximation is only good for the pion and we will include the excited resonances as Breit-Wigner 
functions:
\begin{equation}
\rho_{ud}^{(L)}(s) = 2 f_{\pi}^2 \frac{M_{\pi}^4}{s^2}\delta(s-M_\pi^2) + 2\frac{f_1^2 M_{1}^4}{s^2} B_1(s) +  2\frac{f_2^2 M_{2}^4}{s^2} B_2(s)\,,
\end{equation}
where $B_i(s)$ is
\begin{equation}
B_i(s) = \frac{1}{\pi} \frac{\Gamma_i M_i}{\left[ \left(s-M_i^2 \right)^2 +\Gamma_i ^2 M_i^2\right]}\,,
\end{equation}
and 1,2 refers to the $\pi(1300)$ and $\pi(1800)$ respectively. The pion decay constant is known rather precisely to
be $f_{\pi} = 92.4\pm0.3$ MeV, while the couplings for the higher resonances have been calculated by Maltman and Kambor in
a combined fit of pseudoscalar sum rules {\cite{MK1}}. They find
\begin{equation}
f_1 = 2.20 \pm 0.47\, \mathrm{MeV} \,, \qquad 0 < f_2 < 0.37 \, \mathrm{MeV},
\end{equation} 
where we have added the errors they give in quadrature. 

\section{The Spectral Function for the us Current} 
Since the contributions in the $us$ channel are generally larger than those from the $ud$ current, we have to include both 
scalar and pseudoscalar channels. The discussion of the pseudoscalar channel can proceed along the lines of the 
pseudoscalar channel of the $ud$ current. The three lowest lying resonances are the $K$ meson itself, the $K(1430)$ and the $K(1830)$. 
As for the $\pi$ a narrow width ansatz is again justified for the $K$, while for the excited resonances we need to use 
Breit-Wigner forms. The decay constants are 
\begin{equation}
f_K = 0.113 \pm 0.001\, \mathrm{MeV}   \,, \quad \,\qquad f_{K(1460)} = 21.4 \pm 2.8 \, \mathrm{MeV} \,,\\\nonumber 
\end{equation}
\begin{displaymath} 
0 < f_{K(1830)} < 8.9\, \,\mathrm{MeV}\,,
\end{displaymath}
that have been taken again from Maltman and Kambor {\cite{MK}} and the PDG. 
Numerical values for all resonance masses and widths are compiled in the appendix. There are rather large 
uncertainties attached to the parameters of the excited $K$ resonances (in fact, both are listed by the PDG as ``needing 
confirmation''), but these large uncertainties do not manifest themselves dramatically in the final values for 
$\delta R^{kl}_{L}$: varying the width of the $K$(1460) where the induced error should be most obvious gives 
an uncertainty of 0.000056, which is about 1\% of the uncertainties in the coupling constants that give the 
dominant uncertainties as we will show in the next section.

The scalar $us$ channel could in principle also be described by a resonance ansatz as above, being dominated by the 
$K_{0}^{*}(1430)$. 
The authors of {\cite{jm2}} make such an approach by adding this as well as the next higher resonance, the $K_{0}^{\ast}(1930)$, and
normalizing to threshold. Additionally the scalar form factor is included in an $\mathrm{Omn \grave{e} s }$ 
representation. The problems of this are twofold: on a simple level, trouble already arises when
parametrizing the resonance (of course the same combined fit that was used for the pseudoscalar could be and has been
used {\cite{MK2}}, but the uncertainties will be larger because there is no well known pole). Additionally there are 
significant dynamical contributions that interfere destructively in this channel and have to be included in a 
description. We will then have to think of something better, especially because the scalar $us$ contributes significantly
when adding up all four channels

\begin{figure}[bt]
\begin{center}
\psfrag{s[GeV]}{$\sqrt{s}$[GeV]}
\psfrag{rho[GeV]}{$\rho(s) \times 10^3$ [$\mathrm{GeV}^4$]} 
\psfrag{0}{0}
\psfrag{1}{1}
\psfrag{2}{2}
\psfrag{3}{3}
\psfrag{4}{4}
\psfrag{5}{5}
\psfrag{6}{6}
\psfrag{7}{7}
\psfrag{1,5}{1.5}
\psfrag{2,5}{2.5}
\epsfig{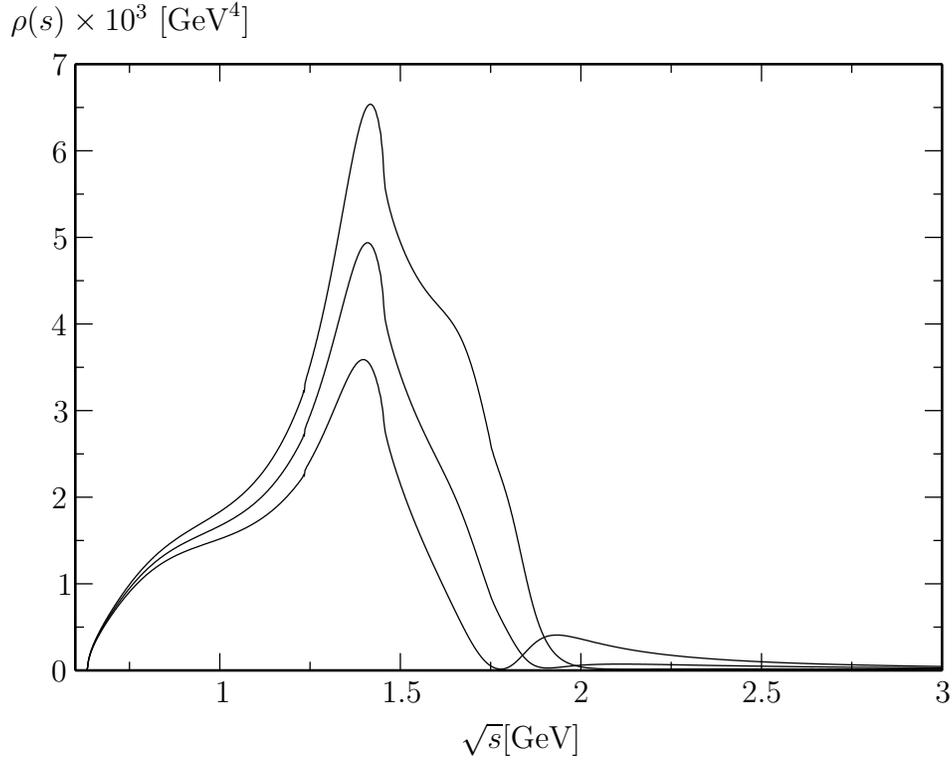}
\begin{minipage}[t]{12cm} 
\caption{\label{Phenspec}The phenomenological spectral functions for the divergence of a vector current. We will use the central curve
for our central values and take the others as error estimates.}
\end{minipage}
\end{center}
\end{figure}

Following the approach for a dynamical ansatz by Refs.~{\cite{JOP0,JOP}}, we take into account the lowest lying scalar states, which are 
excited $K \pi$, $K \eta$ and $K \eta'$ resonances, and the phenomenological spectral function is given by
\begin{equation}
\rho(s) = \frac{3\Delta^2_{K\pi}}{32\pi^2} \left[\sigma_{K\pi}(s) \left|F_{K\pi}(s)\right|^2 + \sigma_{K\eta}(s) \left|F_{K\eta}
(s) \right|^2 + \sigma_{K\eta'}(s) \left|F_{K\eta'}(s)\right|^2 \right] \,,
\end{equation}
where $\Delta_{K\pi}=M_K^2-M_{\pi}^2$, the $\sigma_{KP}(s)$ are the threshold phase space factors for a two body final state
\begin{equation}
\sigma_{KP}(s) = \theta \left(s-\left(M_K+ M_{P}\right)^2\right) \sqrt{ \left( 1- \frac{(M_K+ M_{P})^2}
{s}\right) \left( 1- \frac{\left(M_K- M_{P}\right)^2}{s}\right)} \,, 
\end{equation}
and the form factors are defined by
\begin{equation}
\langle 0| \partial_{\mu} \left(\bar{s}\gamma^{\mu}u\right) | KP\rangle \equiv -i \sqrt{\frac{3}{2}}\Delta_{K\pi}
F_{KP}(s)\,.
\end{equation}
These form factors have been calculated in Ref. {\cite{JOP}} from $KP$ scattering data, which allows us to give a 
phenomenological spectral function that should be valid up to about 2\,GeV. It is important to note that all the spectral 
functions given above have already been written as spectral function for the $\tau$-decay current, while the scalar spectral 
function as plotted in Fig.~{\ref{Phenspec}} is specifically given for the divergence of the vector current. We will divide by the necessary 
factor of $s^2$ in later calculations.
\noindent  
%\vspace{1cm}
%\section{Subtracting the phen. spectral functions}
\noindent
We can now go on to subtract the pure longitudinal contribution by evaluating the integral of (\ref{speclong}) with 
the spectral functions discussed above. We shall begin with the pole contribution that can be directly seen to be
\begin{eqnarray}
\label{pole}
\delta R^{L,pole}_{k,0}&  = & 48 \pi^2 \left[ \left(\frac{f_K^2}{\Mtau^2}\right) \left(\frac{M_K^2}{\Mtau^2}\right)\left
(1-\frac{M_K^2}{\Mtau^2}\right)^{2+k} \right. \\ \nonumber
 & &- \left.\left(\frac{f_{\pi}^2}{\Mtau^2}\right) \left(\frac{M_{\pi}^2}{\Mtau^2}\right)\left(1-\frac{M_{\pi}^2}
{\Mtau^2}\right)^{2+k} \right] \,,
\end{eqnarray}
where we have already taken $l=0$. The integral for the remaining resonance contribution should only be taken from a threshold 
value $s_{thr}$ to $\Mtau^2$:
\begin{equation}
\label{resonance}
\delta R_{kl}^{L} = -24 \pi^2 \int\limits_{s_{thr}}^{\Mtau^2} \frac{ds}{\Mtau^2} 
\left( 1-\frac{s}{\Mtau^2} \right)^{2+k} \left(\frac{s}{\Mtau^2}\right) 
\left[ \rho_{ud}^{(L)}(s) - \rho_{us}^{(L)}(s)  \right] \,.
\end{equation}
The threshold energy for the scalar $us$ channel is $s_{thr} = (M_K + M_{\pi})^2$ while for the pseudoscalar it is $s_{thr} = (M_K + 2 M_{\pi})^2$.
Equivalently, the threshold energy for the pseudoscalar $ud$ current is 3$M_{\pi}$.

\section{Pure (Pseudo)scalar Sum Rules}
In this section we shall study in detail the scalar/pseudoscalar part of the $\tau$-decay sum rule. To do so we will discuss separately the scalar
and the pseudoscalar channel of the $ud$ and $us$ currents and compare the OPE side with the phenomenological one. 
The results are given in Tables \ref{Table0}-\ref{Table4}. They show the contributions $\delta R_{kl}$ (note the minus sign 
added in the strange channel) for the OPE as in
\begin{equation}
R_{kl} =- 4 \pi i \oint \limits_{|s|=\Mtau^2} \frac{ds}{s}  \mathcal{F}^{kl}_{L}(s/\Mtau^2) D^{(L)}(s)  \,,
\end{equation}    
with the OPE discussed in the last chapter\footnote{We find it more appropriate to truncate the perturbation series after the 
second order for all moments, which leads to better agreement between the expressions.}
 and the phenomenological part as in Eqs. (\ref{pole}) and (\ref{resonance}).
We have given all central values plus their respective errors and have additionally shown how the different terms in the 
OPE contribute. Clearly higher dimensional operators are strongly suppressed.
\begin{table}[h]
\begin{center}
\begin{tabular}{|l|c|l|}\hline
Current& Moment & Numerical value\\ \hline \hline
Pseudoscalar $us$ & (0,0) & 0.135 $\pm$ 0.002 $\pm$ 0.002 $\pm$ 0.001\\
Phenomenology     & (1,0) & 0.121 $\pm$ 0.002 $\pm$ 0.001 \\
                  & (2,0) & 0.110 $\pm$ 0.002 $\pm$ 0.001 \\\hline
Scalar $us$       & (0,0) & 0.0281 $\pm$ 0.0040 \\
Phenomenology     & (1,0) & 0.0180 $\pm$ 0.0020  \\
                  & (2,0) & 0.0123 $\pm$ 0.0012 \\\hline
Pseudoscalar $ud$ & (0,0) & -0.00777 $\pm$ 0.00005 $\pm$ 0.00007 \\
Phenomenology     & (1,0) & -0.00776 $\pm$ 0.00005 $\pm$ 0.00005 \\
                  & (2,0) & -0.00762 $\pm$ 0.00005 $\pm$ 0.00003 \\\hline\hline
Pseudoscalar $us$ & (0,0) & 0.144 $\pm \, 0.024_{trunc} \pm 0.009_{s cond} $\\
Theory            & (1,0) & 0.140 $\pm$ 0.026 $\pm$ 0.009 \\
                  & (2,0) & 0.138 $\pm$ 0.029 $\pm$ 0.009 \\\hline
Scalar $us$       & (0,0) & 0.02790 $\pm\, 0.0187_{trunc} \pm 0.0087_{s cond}$\\
Theory            & (1,0) & 0.02758 $\pm$ 0.0200 $\pm$ 0.0088 \\
                  & (2,0) & 0.02897 $\pm$ 0.0222 $\pm$ 0.0088 \\\hline
Pseudoscalar $ud$ & (0,0) & -0.0078 $\pm  0.00015_{trunc} \pm 0.00002 _{m_u} \pm 0.00005_{m_d}$\\
Theory            & (1,0) & -0.0077 $\pm$ 0.00017 $\pm$  0.00001  $\pm$ 0.00004\\
                  & (2,0) & -0.0077 $\pm$ 0.00018 $\pm$   0.00001  $\pm$ 0.00004\\\hline 
Scalar $ud$       & (0,0) & -1.74 $\pm  1.28_{trunc} \pm 0.14_{m_d}\pm 0.32_{m_u} \cdot 10^{-5}$\\
Theory            & (1,0) & -1.73 $\pm$ 1.38 $\pm$ 0.14 $\pm$ 0.32 $\cdot10^{-5}$ \\
                  & (2,0) & -1.81 $\pm$ 1.51 $\pm$ 0.14 $\pm$ 0.33 $\cdot10^{-5}$ \\\hline
\end{tabular}
\begin{minipage}[t]{12cm} 
\caption{\label{Table0}Total values and errors of $\delta R^{kl}$ for all relevant currents and the three lowest moments; The 
errors are ordered as follows: For the phenomenology the dominant errors are from the coupling constants and we have given them in the 
order of the resonance, i.e. the first is for the $\pi$ the next for the $\pi(1300)$ and the third (if given) for the $\pi(1800)$.
The scalar spectral function is varied in the limits given above. The theoretical uncertainty labled by $trunc$ always includes  
scaling and truncation.}
\end{minipage}
\end{center}
\end{table}

\begin{table}[t]
\begin{center}
\begin{tabular}{|l|ccc|}\hline
                      & (0,0)  & (1,0) & (2,0) \\  \hline
Dimension 2                    & 0.0503393      & 0.0480009       &  0.00476702 \\ 
Subtraction constant           & 0.0829935      & 0.0829935       &  0.0829935  \\
Dimension 4 (mass)             & 0.000738       & 0.000983        &  0.001247  \\
Dimension 6 (gluon condensate) & -0.000395      &  -0.000687      &   -0.001024     \\
Dimension 6 (quark condensate) & -0.000262      & -0.000480       &  -0.000744  \\ 
Dimension 6 (mass)             & 1$\cdot10^{-8}$&  3$\cdot10^{-8}$&  6$\cdot10^{-8}$       \\
Dimension 8                    & -3$\cdot10^{-5}$& -0.000121      & -0.000277 \\ 
Instantons                     & 0.00885        &   0.00770       &  0.00603          \\       \hline
Total theory                   & 0.1436         & 0.1395          & 0.1376 \\
Phenomenology                  & 0.1345         & 0.1210          & 0.1102  \\ \hline
\end{tabular}
\begin{minipage}[t]{12cm} 
\caption{\label{Table1}Composition of the  theoretical expressions for the pseudoscalar $us$ current.}
\end{minipage}
\end{center}
\end{table}

\begin{table}[t]
\begin{center}
\label{phenthSus}
\begin{tabular}{|l|ccc|}\hline
                        &(0,0)  & (1,0) & (2,0)  \\  \hline
Dimension 2                      & 0.044778  & 0.0426978 &  0.0424037 \\ 
Subtraction constant            & -0.00869721  &  -0.00869721    & -0.00869721  \\
Dimension 4 (mass)              & 0.0007259  & 0.0009666 & 0.0012272   \\
Dimension 6 (gluon condensate) &  -0.000351   & -0.000611  &  -0.000911  \\
Dimension 6 (quark condensate)  & 0.000522 &  0.000950  &  0.001468 \\ 
Dimension 6 (mass)              &  -1$\cdot10^{-8}$ &  -3$\cdot10^{-8}$ &  -5$\cdot10^{-8}$  \\
Dimension 8                    & 4$\cdot10^{-5}$  & 0.000144  &  0.000330  \\ 
Instantons                     &  -0.009116  & -0.007871 & -0.00685 \\       \hline
Total theory                   &  0.02790 & 0.02758  & 0.02897 \\
Phenomenology                  &  0.02809 & 0.01802  & 0.01231  \\ \hline
\end{tabular}
\begin{minipage}[t]{12cm}
\caption{Composition of the theoretical expressions for the scalar $us$ current.}
\end{minipage}
\end{center}
\end{table}

\begin{table}[t]
\begin{center}
\label{phenthPud}
\begin{tabular}{|l|ccc|}\hline
                        & (0,0)  & (1,0) & (2,0) \\  \hline
Dimension 2                    & -0.000321 & -000306& -0.000306\\ 
Subtraction constant           & -0.007362 & -0.007362   & -0.007362\\
Dimension 4 (mass)             & -5$\cdot10^{-9}$ & -7$\cdot10^{-9}$ & -7$\cdot10^{-9}$\\
Dimension 6 (gluon condensate) & 3$\cdot10^{-6}$ & 4$\cdot10^{-6}$ & 4$\cdot10^{-6}$ \\
Dimension 6 (quark condensate) & 1$\cdot10^{-7}$ & 2$\cdot10^{-7}$ & 2$\cdot10^{-7}$ \\ 
Dimension 6 (mass)             & -3$\cdot10^{-14}$ &-7$\cdot10^{-14}$ & -7$\cdot10^{-14}$\\
Dimension 8                    & 2$\cdot10^{-7}$ &  6$\cdot10^{-7}$ &6$\cdot10^{-7}$ \\ 
Instantons                     & -0.00011 & -9$\cdot10^{-5}$ & -8$\cdot10^{-5}$\\       \hline
Total theory                   & -0.007789 & -0.007757 & -0.007739 \\
Phenomenology                  & -0.007774 & -0.007686 & -0.007615 \\ \hline
\end{tabular}
\begin{minipage}[t]{12cm}
\caption{Composition of the theoretical expressions for the pseudoscalar $ud$ current.}
\end{minipage}
\end{center}
\end{table}

\begin{table}[t]
\begin{center}
\begin{tabular}{|l|ccc|}\hline
                        & (0,0)  & (1,0) & (2,0) \\  \hline
Dimension 2                    & -2.658$\cdot10^{-5}$ & -2.534$\cdot10^{-5}$ & -2.517$\cdot10^{-5}$\\ 
Subtraction constant           & 0&0 &0\\
Dimension 4 (mass)             & -3$\cdot10^{-9}$ & -3$\cdot10^{-9}$ & -4$\cdot10^{-9}$\\
Dimension 6 (gluon condensate) & 2$5\cdot10^{-7}$ & 4$\cdot10^{-7}$ & 5$\cdot10^{-7}$\\
Dimension 6 (quark condensate) & -3$\cdot10^{-8}$& -5$\cdot10^{-8}$ & -7$\cdot10^{-8}$\\ 
Dimension 6 (mass)             & 2$\cdot10^{-16}$& 5$\cdot10^{-16}$ &9$\cdot10^{-16}$\\
Dimension 8                    & -2$\cdot10^{-8}$& -7$\cdot10^{-8}$ & -2$\cdot10^{-7}$\\ 
Instantons                     & 9$\cdot10^{-6}$ & 8$\cdot10^{-6}$  & 7$\cdot10^{-6}$ \\       \hline
Total theory                   & -1.740$\cdot10^{-5}$ & -1.732$\cdot10^{-5}$ & -1.810$\cdot10^{-5}$\\     \hline
\end{tabular}
\begin{minipage}[t]{12cm} 
\caption{\label{Table4}Composition of the theoretical expressions for the scalar $ud$ current.}
\end{minipage}
\end{center}
\end{table}

\begin{figure}[bt]
\psfrag{1}{$\scriptstyle{1}$}
\psfrag{2}{$\scriptstyle{2}$}
\psfrag{3}{$\scriptstyle{3}$}
\psfrag{s[GeV]}{$s[\mathrm{GeV^2}]$}
\psfrag{0}{$\scriptstyle{0}$}
\begin{center}
\epsfig{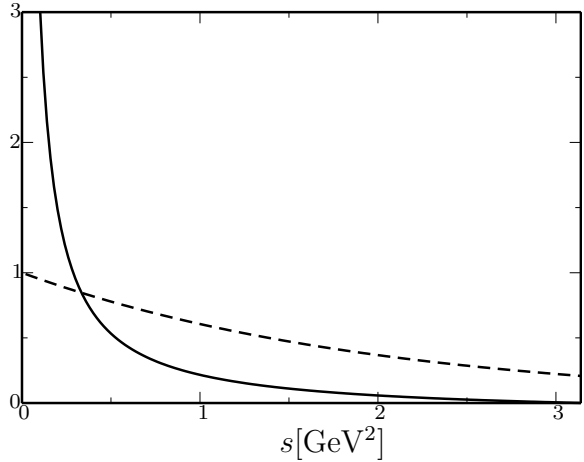}
\begin{minipage}[t]{12cm}
\caption{The $k=0$ weight function for $\tau$-decays (solid); due to the factor $1/s^2$ low energies are weighted very strongly.
As a comparison we have plotted the Borel weight function (dashed) with a Borel parameter of $M^2$=2 GeV.}
\label{weights}
\end{minipage}
\end{center}
\end{figure}

We have used the following numerical values: $m_u = m_s \varepsilon_u$, $m_d = m_s \varepsilon_d$, $\langle \bar{u}u \rangle
= \langle \bar{d}d \rangle = -f_{\pi}M_{\pi}/(m_u+m_d)\cdot(1-0.047)$, $ \langle \bar{s}s \rangle /\langle \bar{d}d \rangle = 0.8 \pm 0.2$ and
 $ \langle aFF \rangle =  (0.5-2) \,2.1\cdot10^{-2} \, \mathrm{GeV}^{4}$. These are also the values that we will use in all subsequent calculations.
Additionally we have taken $m_s(2\,\mathrm{GeV}) = 105 \, \mathrm{MeV}$ in accordance with the current world average to be able to provide a numerical
value. The estimates are performed with the uncertainties given above, adding the uncertainty that arrises by varying the parameter $\xi$
introduced in the last chapter from 0.75--2. We have also followed the standard procedure for asymptotic series and taken the full value of the last coefficient
as an error estimate, where we have truncated the series after the minimal coefficient. We have not taken into account the error induced by 
the strange quark mass for obvious reasons: we want to show that the errors for the theoretical expression are larger than for a 
phenomenological one and then result in larger uncertainties for a strange quark mass determination. It would thus be dishonest to include the
error for the quantity we want to determine and pretend that this error plays a significant role. In any case the dominant uncertainties are those 
coming from truncation, scale dependence and the uncertainties in the quark condensates for the strange currents plus of course
those from the quark masses in the $ud$ currents.
The phenomenological integrals are dominated so strongly by the poles (See Fig.~\ref{weights}) that the main uncertainties come 
from the $\pi$ and $K$ coupling 
constants given by $ f_{\pi} = 92.4 \pm 0.3 \mathrm{MeV} $ and $ f_{K} = 113 \pm 0.001 \mathrm{MeV}$. The uncertainties from the couplings to higher resonances are also
significant, while those from masses and widths are negligible.\\\\
\noindent

\noindent
In the tables given above there are two things that are especially noteworthy: first, that the central values for all currents agree rather well (though
this agreement becomes worse for higher moments) and second, that the uncertainties of the phenomenological spectral functions are smaller than those 
of the theoretical one by an order of magnitude. This gives us much confidence in the general potential of our ansatz. 
There is one more rather important fact that can be seen from those tables: a naive OPE estimate would suggest that the scalar spectral 
function should be very similar to the pseudoscalar because the OPE only differ by the global factor $(m_i \pm m_j)^2$, which is very similar 
for pseudoscalar and scalar if one quark is much heavier than the other. This similarity is obviously not present in the numerical values for
the phenomenological spectral functions. The reason is that subtraction constants differ significantly in both channels (on the phenomenological
side this is of course due to the fact that the weight functions put extreme emphasis on the low energy part of the spectrum and the main contributions come, 
as mentioned above, from the poles). \\\\
\noindent          
Another check of the spectral ansatz can be to attempt an $m_s$ determination from pure scalar sum rules, and we will discuss this for
our spectral ansatz. Let us study the divergence of vector current and denote the correlation function as customary by $\Psi$:
\begin{equation}
\Psi(q^2)=i\int\,d{}^4 x \, e^{iqx} \langle \Omega | T(j^\mu(x) j^{\nu}(0)) | \Omega 
\rangle \, ,
\end{equation}
with 
\begin{equation}
j(x) = \partial_{\mu} \bar{q_i} \gamma^{\mu} q_j(x) = i (m_i-m_j) \bar{q_i} q_j(x) \,.
\end{equation}
Clearly this quantity is well suited for such a determination because of the global mass factor. We can now generally attempt 
both Borel sum rules and finite energy sum rules, but we will concentrate on the scalar FESR since a Borel sum rule with the same 
scalar ansatz was performed by the authors of Ref. {\cite{JOP2}} and pseudoscalar sum rules are discussed in much detail by 
Maltman and Kambor in Refs. {\cite{MK,MK1}} for the $ud$ and $us$ currents.
For finite energy sum rules it will be in interesting to vary the finite energy $s_0$. We can thus check different energy regions 
of the spectral function by using the stability of the obtained $m_s$ values as a criterion for the quality of our ansatz. We expect 
the results to be sensible only for a certain
energy interval, since for low energies the OPE will break down, and for higher energies our spectral ansatz is now longer valid.

\begin{figure}[bt]
\psfrag{ms[GeV]}{\hspace{-0.5cm}$m_s$[GeV]}
\psfrag{0,08}{\hspace{-0.3cm}$\scriptstyle{0.08}$}
\psfrag{0,1}{\hspace{-0.3cm}$\scriptstyle{0.1}$}
\psfrag{0,12}{\hspace{-0.3cm}$\scriptstyle{0.12}$}
\psfrag{0,14}{\hspace{-0.3cm}$\scriptstyle{0.14}$}
\psfrag{0,2}{\hspace{-0.3cm}$\scriptstyle{0.2}$}
\psfrag{0,18}{\hspace{-0.3cm}$\scriptstyle{0.18}$}
\psfrag{0,16}{\hspace{-0.3cm}$\scriptstyle{0.16}$}
\psfrag{0,06}{\hspace{-0.3cm}$\scriptstyle{0.06}$}
\psfrag{2}{$\scriptstyle{2}$}
\psfrag{2,5}{$\scriptstyle{2.5}$}
\psfrag{3,5}{$\scriptstyle{3.5}$}
\psfrag{3}{$\scriptstyle{3}$}
\psfrag{4}{$\scriptstyle{4}$}
\psfrag{4,5}{$\scriptstyle{4.5}$}
\psfrag{5}{$\scriptstyle{5}$}
\psfrag{s0[GeV]}{$s_0 [\mathrm{GeV^2}]$}
\begin{center}
\epsfig{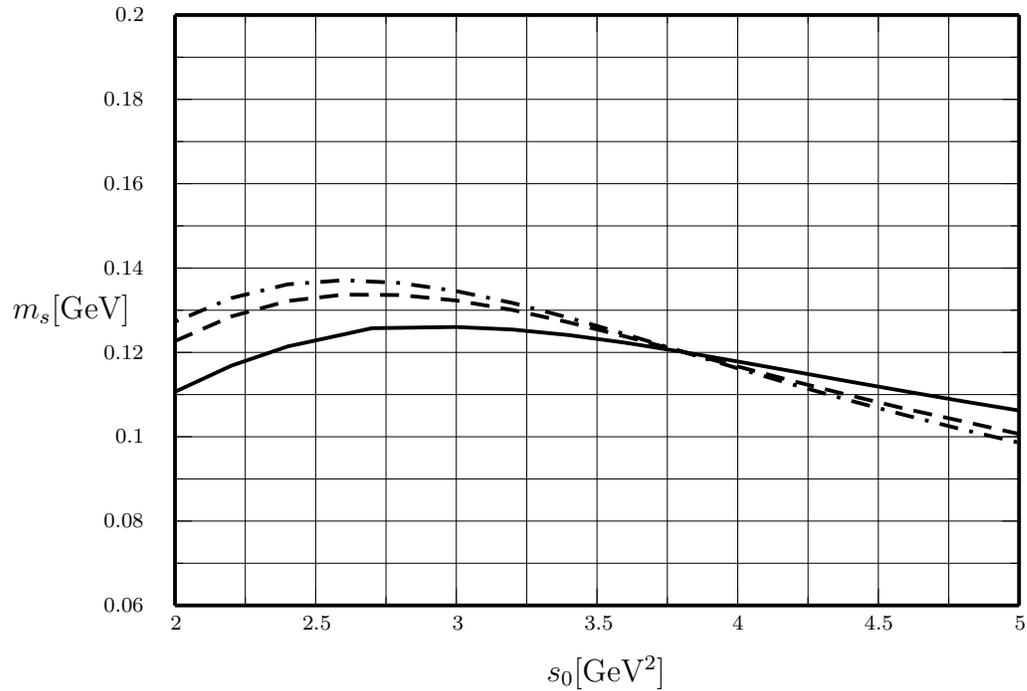}
\begin{minipage}[t]{12cm}
\caption{$m_s$ determination from pure scalar sum rules: we have plotted $m_s$(2\,GeV) for A = 0(solid), 2(dashed), 
4(dot-dashed).}
\label{Scalsum}
\end{minipage}
\end{center}
\end{figure}

\begin{figure}[bt]
\psfrag{ms[GeV]}{\hspace{-0.5cm}$m_s$[GeV]}
\psfrag{0,08}{\hspace{-0.3cm}$\scriptstyle{0.08}$}
\psfrag{0,1}{\hspace{-0.3cm}$\scriptstyle{0.1}$}
\psfrag{0,12}{\hspace{-0.3cm}$\scriptstyle{0.12}$}
\psfrag{0,14}{\hspace{-0.3cm}$\scriptstyle{0.14}$}
\psfrag{0,2}{\hspace{-0.3cm}$\scriptstyle{0.2}$}
\psfrag{0,18}{\hspace{-0.3cm}$\scriptstyle{0.18}$}
\psfrag{0,16}{\hspace{-0.3cm}$\scriptstyle{0.16}$}
\psfrag{0,06}{\hspace{-0.3cm}$\scriptstyle{0.06}$}
\psfrag{2}{$\scriptstyle{2}$}
\psfrag{2,5}{$\scriptstyle{2.5}$}
\psfrag{3,5}{$\scriptstyle{3.5}$}
\psfrag{3}{$\scriptstyle{3}$}
\psfrag{4}{$\scriptstyle{4}$}
\psfrag{4,5}{$\scriptstyle{4.5}$}
\psfrag{5}{$\scriptstyle{5}$}
\psfrag{s0[GeV]}{$s_0 [\mathrm{GeV^2}]$}
\begin{center}
\epsfig{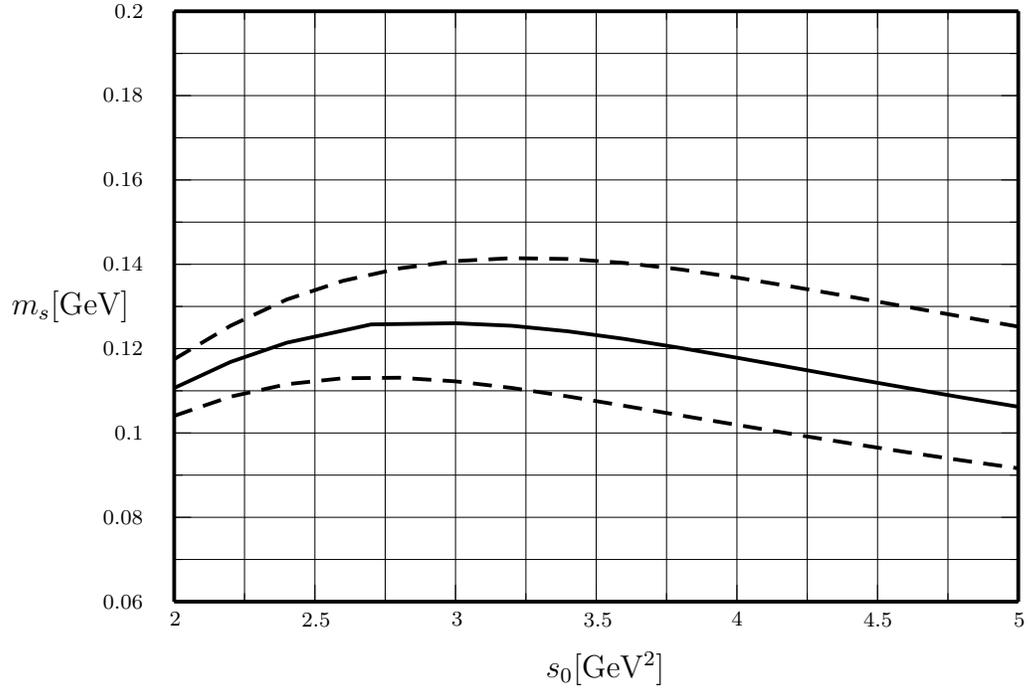}
\begin{minipage}[t]{12cm}
\caption{$m_s$ determination from pure scalar sum rules: we have plotted $m_s$(2\,GeV) for the three scalar spectral function given in 
                  Fig.~{\ref{Phenspec}} for A=0.}
\label{Scalerror}
\end{minipage}
\end{center}
\end{figure}

\begin{figure}[bt]
\psfrag{ms[GeV]}{\hspace{-0.5cm}$m_s$[GeV]}
\psfrag{0,08}{\hspace{-0.3cm}$\scriptstyle{0.08}$}
\psfrag{0,1}{\hspace{-0.3cm}$\scriptstyle{0.1}$}
\psfrag{0,12}{\hspace{-0.3cm}$\scriptstyle{0.12}$}
\psfrag{0,14}{\hspace{-0.3cm}$\scriptstyle{0.14}$}
\psfrag{0,2}{\hspace{-0.3cm}$\scriptstyle{0.2}$}
\psfrag{0,18}{\hspace{-0.3cm}$\scriptstyle{0.18}$}
\psfrag{0,16}{\hspace{-0.3cm}$\scriptstyle{0.16}$}
\psfrag{0,06}{\hspace{-0.3cm}$\scriptstyle{0.06}$}
\psfrag{2}{$\scriptstyle{2}$}
\psfrag{2,5}{$\scriptstyle{2.5}$}
\psfrag{3,5}{$\scriptstyle{3.5}$}
\psfrag{3}{$\scriptstyle{3}$}
\psfrag{4}{$\scriptstyle{4}$}
\psfrag{4,5}{$\scriptstyle{4.5}$}
\psfrag{5}{$\scriptstyle{5}$}
\psfrag{s0[GeV]}{$s_0 [\mathrm{GeV^2}]$}
\begin{center}
\epsfig{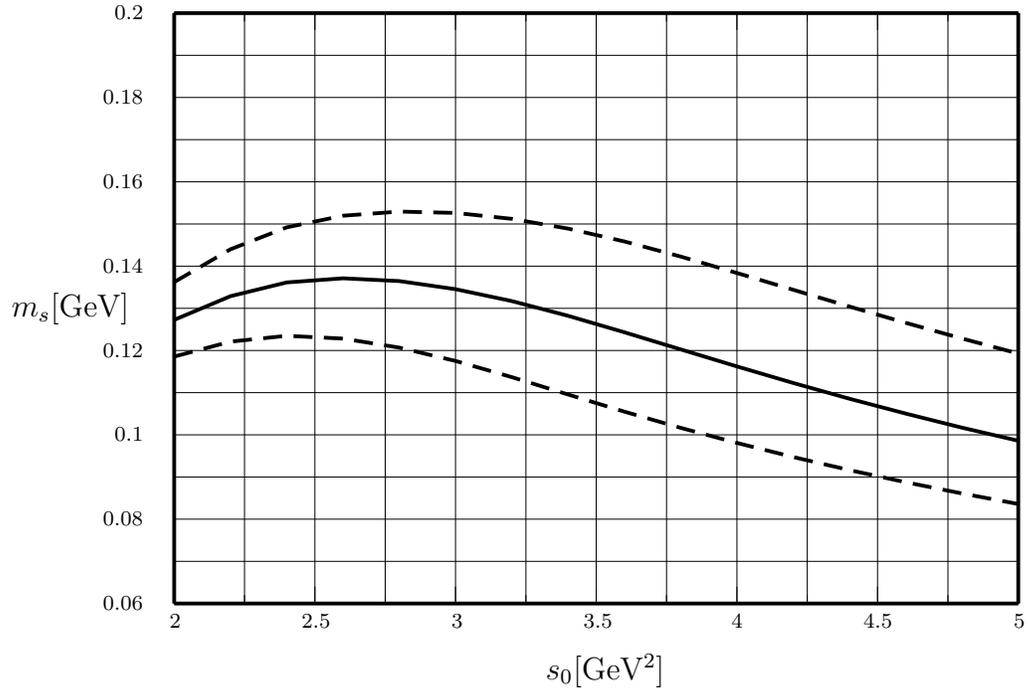}
\begin{minipage}[t]{12cm}
\caption{As in Fig.~{\ref{Scalerror}} but for A=4.}
\label{Scalerror4}
\end{minipage}
\end{center}
\end{figure}

If we now study finite energy sum rules for the correlator of axial vector divergences we see from dimensional analysis that we need two 
subtraction constants, i.e. two partial integrations, which leaves us with the FESR:
\begin{equation}
\label{smassFESR}
-\frac{1}{2 \pi \, i} \oint\limits_{s=s_0} \, ds \, W(s) \, \Psi''(s)  = \int\limits_{0}^{s_0} \, ds \, w(s) \, \rho(s)\,,
\end{equation}
where $W(s)$ is the weight function integrated twice according to Eq.~(\ref{sub}). Since the pseudoscalar sum rules of Maltman and Kambor 
are rather well satisfied, we will choose to work with their weight functions i.e.
\begin{equation}
w(s) = \left(1-\frac{s}{s_0}\right) \left(1+A\frac{s}{s_0}\right)  \Rightarrow  W(s)=-\frac{1}{12} \frac{(s-s_0)^3}{s_0^2} (A(s+s_0)+2s_0)\,,
\end{equation}
and set the parameter $A=0,2,4$. 
Conveniently, Ref.~{\cite{jm2}} has already given some expressions for the second derivative and we need only quote the results.
They are
\begin{eqnarray}
\Psi''_0(s)& = & -\frac{(m_i - m_j)^2}{s}\left( \frac{3}{8 \pi^2} \left[ 1 +\frac{11}{3}a(-s)+ \left(\frac{5071}{144}-
 \frac{35}{2}\zeta(3)\right) a^2(-s)  \right. \right. \\%[\fill]
& &\left. \left.+ \left(\frac{1995097}{5184}-\frac{65869}{216}\zeta(3)-\frac{5}{2}\zeta(4)+\frac{715}{12} \zeta(5) \right) a^3(-s) \right] \right) \,, \nonumber\\%[\fill]
\Psi''_2(s) & = & -\frac{(m_i - m_j)^2}{s^2} \left( \frac{3}{4\pi^2} \left\{ \left[ 1+\frac{28}{3}a(-s) \right]
 \Big(m_i^2+m_j^2 \Big) 
     \right. \right. \\%[\fill]
& & \nonumber \left. \left.  + \left[1+\frac{40}{3}a(-s)\right] m_i m_j \right\} \right)\,,\\%[\fill]
\nonumber
\Psi''_4(s) & = & \frac{(m_i - m_j)^2}{s^3} \left\{ -\frac{1}{4} \langle aFF \rangle + \frac{3}{16 \pi^2 } 
                   \left[ 4(m_i^4 + m_j^4) - 8 m_i^2 m_j^2  \right. \right.\\%[\fill]
\nonumber 
             & & + \left. 12(m_i^3 m_j + m_i m_j^3) \right] + \Big[-1 - \frac{20}{3} a(-s) \Big]
                  \left( m_i \langle \bar{q_i}q_i \rangle+ m_j \langle \bar{q_j}q_j \rangle \right) \\%[\fill]
\nonumber
& &+ \left.2 \Big[-1 - \frac{25}{3} a(-s) \Big] \left( m_i \langle \bar{q_j}q_j \rangle+ m_j \langle \bar{q_i}q_i \rangle \right) \right\}\,,\\%[\fill]
\Psi''_6(s) & = & \frac{6(m_i - m_j)^2}{s^4}\Big[\frac{1}{2}(m_i \langle g {\bar q_i} \sigma F q_i \rangle +
                                                             m_j \langle g {\bar q_j} \sigma F q_j \rangle) \\%[\fill] 
\nonumber
          &&      -       4 \pi^2  a(-s) \frac{4}{3}\langle {\bar q_i}q_i \rangle \langle {\bar q_j}q_j \rangle 
                 -\frac{4}{3} \pi^2  a(-s )\frac{4}{9}(\langle {\bar q_i}q_i \rangle^2+\langle {\bar q_j}q_j \rangle^2) \Big]\,.
\end{eqnarray}
The expressions for dimension 4 and higher are not given explicitly in {\cite{jm2}}, but can be easily calculated, while 
the $\mathcal{O}(a^3)$ coefficient was calculated from the compilation of scalar coefficients given in the appendix. Again, instantons are 
accounted for by using Eq.~(\ref{Inst}).\\\\
The strange quark mass determination will now proceed as follows: insert the spectral ansatz on the right hand side of Eq.~(\ref{smassFESR})
and plug in the OPE on the left. The resulting expressions can then be solved for $m_s$ for different values of $s_0$ and the results 
are shown in Fig.~{\ref{Scalsum}}. We see that the results are reasonable but only moderately stable for large A. The result for 
A=0 appears to be stable enough to possibly allow a determination, but this would obtain large uncertainties; we can show this by plotting the result for constant A and varying the scalar spectral functions in the limits given before. 
This is done in Figs. {\ref{Scalerror}} and {\ref{Scalerror4}} for A=0 and A=4. This results in a variation of $m_s$ of about $\pm$ 20, not 
including any uncertainties from the theoretical expression, a number which is clearly not competitive with other results. 

\begin{figure}[bt]
\psfrag{ms[GeV]}{\hspace{-0.5cm}$m_s$[GeV]}
\psfrag{0,08}{\hspace{-0.3cm}$\scriptstyle{0.08}$}
\psfrag{0,1}{\hspace{-0.3cm}$\scriptstyle{0.1}$}
\psfrag{0,12}{\hspace{-0.3cm}$\scriptstyle{0.12}$}
\psfrag{0,14}{\hspace{-0.3cm}$\scriptstyle{0.14}$}
\psfrag{0,2}{\hspace{-0.3cm}$\scriptstyle{0.2}$}
\psfrag{0,18}{\hspace{-0.3cm}$\scriptstyle{0.18}$}
\psfrag{0,16}{\hspace{-0.3cm}$\scriptstyle{0.16}$}
\psfrag{0,06}{\hspace{-0.3cm}$\scriptstyle{0.06}$}
\psfrag{2}{$\scriptstyle{2}$}
\psfrag{2,5}{$\scriptstyle{2.5}$}
\psfrag{3,5}{$\scriptstyle{3.5}$}
\psfrag{3}{$\scriptstyle{3}$}
\psfrag{4}{$\scriptstyle{4}$}
\psfrag{4,5}{$\scriptstyle{4.5}$}
\psfrag{5}{$\scriptstyle{5}$}
\psfrag{s0[GeV]}{$s_0 [\mathrm{GeV^2}]$}
\begin{center}\
\epsfig{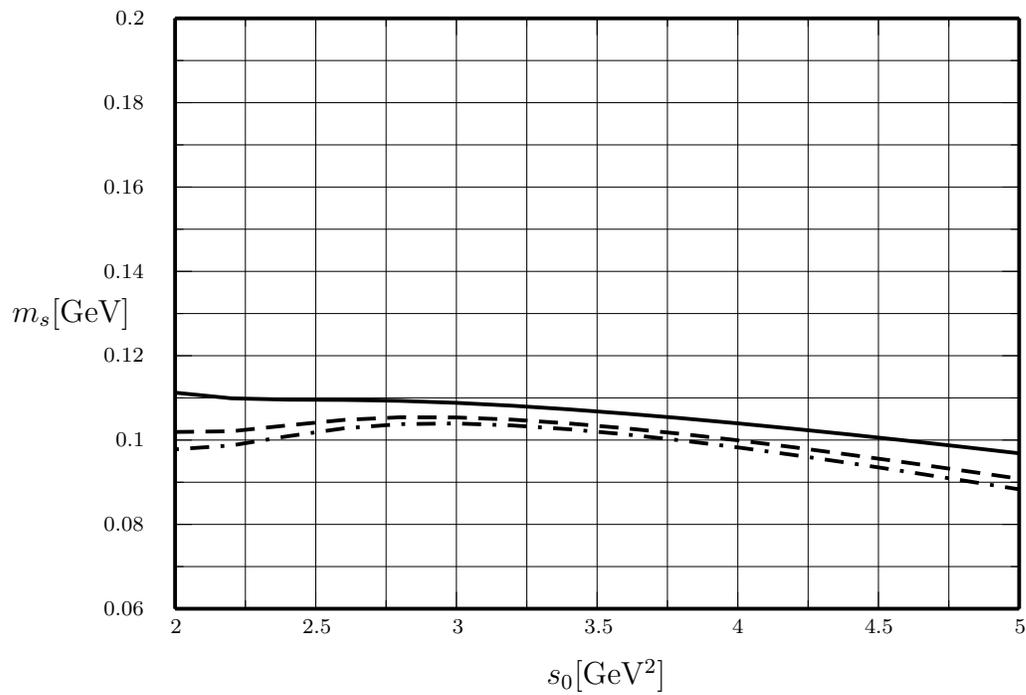}
\begin{minipage}[t]{12cm}
\caption{As in Fig.~{\ref{Scalsum}} but for the pseudoscalar sum rules.}
\label{PScalerror4}
\end{minipage}
\end{center}
\end{figure}

Additionally we have performed the same calculations with the pseudoscalar sum rules discussed by Maltman and Kambor and the 
result is given in Fig.~{\ref{PScalerror4}} for comparison. Note that we have plotted the mass in the $s$ region of 
2-5~$\mathrm{GeV^2}$, while Maltman and Kambor discuss their sum rules only for $s=$ 3-4~$\mathrm{GeV^2}$.

As a conclusion, we hope to have shown that a FESR for pure scalar sum rules would be generally feasible but is only sensible 
with more precise scalar spectral data. Additionally, these plots serve as another confirmation of our spectral functions.
\noindent

\noindent

\noindent

\noindent

%% file: ms2.tex
\chapter[Numerical Values for $\mathbf{m_s}$]{Numerical Values for $\mathbf{m_s}$}
\hspace*{\fill}\begin{minipage}[t]{8.2cm}
\begin{footnotesize}
{\em But if you've ever worked with computers, you understand the disease -- the {\em delight} in being able to see how much you can do.}\\
(Richard P. Feynman,\\ Surely You're Joking, Mr. Feynman)
\end{footnotesize}
\end{minipage}
\vspace{1cm}

\noindent
The starting point of our numerical analysis is the vector side of the sum rule presented in chapter {\ref{FESR}. 
Taking all terms up to dimension six, the complete equation reads:
\begin{eqnarray}
\lefteqn{\delta R^{L+T}_{kl} = 12 S_{EW} \left\{\frac{3}{2} \frac{m_s^2(\Mtau^2)}{\Mtau^2}  \left(1-\varepsilon_d^2\right) \Delta^{(L+T)}_{kl}+
3\frac{m_s^4(\Mtau^2)}{\Mtau^4}(1-\varepsilon_d^2) \right.}\\
&&  \nonumber\times \left.\left[\left(1+\varepsilon_d^2\right)H_{kl}^{L+T}-2\varepsilon_u^2G_{kl}^{L+T}\right]- 
4\pi^2 \frac{ \delta O_4(\Mtau^2)}{\Mtau^4} Q_{kl}^{L+T} \right\}+ c_6^k~\delta R^{00,6}\,,
\end{eqnarray}
where $\delta R^{L+T}_{kl}$ is obtained from the ALEPH data of $\delta R_{kl}$ quoted in Ref.~{\cite{mutau}} as discussed in the last 
chapter by subtracting our ansatz for the scalar channels. From the tables in the last chapter we easily calculate the total values for 
$\delta R^{L}_{kl}$ and we give $\delta R^{L+T}_{kl}$, $\delta R^{L}_{kl}$ and $\delta R_{kl}$ in Table~{\ref{DeltaR}}. Since there is 
ongoing discussion about the size of $|V_{us}|$ due to fact that the value obtained by Leutwyler and Roos {\cite{LeutRoos84}} from 
$K$ decays shows a 2\,$\sigma$ deviation from untarity, we have separated the uncertainties into those from the measured data and 
those induced by the CKM Matrix parameters; this uncertainty is dominated by the error in $|V_{us}|$ since its absolute value is 
smaller. The central values for $m_s$, that we find, are:
\begin{eqnarray}
m_s^{0,0} & = & 0.1973 \, \mathrm{GeV}\nonumber \\
m_s^{1,0} & = & 0.1640  \, \mathrm{GeV}\nonumber \\
m_s^{2,0} & = & 0.1366  \, \mathrm{GeV}\nonumber \\
m_s^{3,0} & = & 0.1152  \, \mathrm{GeV}\nonumber \\
m_s^{4,0} & = & 0.0986  \, \mathrm{GeV}\nonumber 
\end{eqnarray}
\begin{table}[t]
\begin{center}
\[
\begin{array}{|c|ccc|}\hline
(k,l) & \delta R^{L+T}_{kl} &  \delta R^{L}_{kl} & \delta R_{kl}  \\  \hline
(0,0) & 0.217 \pm 0.118 \pm 0.062 & 0.157 \pm 0.005 & 0.374 \pm 0.118 \pm 0.062 \\ 
(1,0) & 0.265 \pm 0.065 \pm 0.042 & 0.133 \pm 0.003 & 0.398 \pm 0.065 \pm 0.042\\
(2,0) & 0.282 \pm 0.044 \pm 0.031 & 0.117 \pm 0.002 & 0.399 \pm 0.044 \pm 0.031\\
(3,0) & 0.292 \pm 0.034 \pm 0.024 & 0.104 \pm 0.002 & 0.396 \pm 0.034 \pm 0.024\\ 
(4,0) & 0.302 \pm 0.028 \pm 0.020 & 0.093 \pm 0.002 & 0.395 \pm 0.028 \pm 0.020\\ \hline
\end{array}
\]
\begin{minipage}[t]{12cm}
\caption{\label{DeltaR}$\tau$-decay widths decomposed into vector and scalar channels 
for the PDG2000 value of $|V_{us}|$.}
\end{minipage}
\end{center}
\end{table}

\noindent
Here we have taken the CKM Matrix elements from the PDG2000 unitarity fit as $|V_{us}| = 0.2225 \pm 0.0021$ and 
$|V_{ud}| = 0.97525 \pm 0.00046$. We can perform the same calculations for 
the Leutwyler-Roos fit of $|V_{us}| = 0.2196 \pm 0.0026$% and $|V_{ud}| = 0.9751 \pm 0.0004$
. Since we do not have access to the ALEPH data we will have to resort to the following procedure: the difference between the two values 
is 1.38 times the PDG error of $|V_{us}|$ and we can subtract from $\delta R_{kl}$ the error induced by $|V_{us}|$ scaled 
accordingly to arrive at the values for $\delta R_{kl}$ with the Leutwyler-Roos fit. The uncertainties induced by $|V_{us}|$ were treated in a similar way, while we 
have not changed the given experimental uncertainties, since we do not know how much of it is actually due to the strange channel.

The results for $\delta R_{kl}$ and $\delta R^{L+T}_{kl}$ are given in Table~{\ref{DeltaRLR}}; the longitudinal contributions are 
obviously equal, so we can omit them. Again we calculate $m_s$ and arrive at significantly lower values:

\begin{eqnarray}
m_s^{0,0} & = & 0.1508  \, \mathrm{GeV}\nonumber \\
m_s^{1,0} & = & 0.1415  \, \mathrm{GeV}\nonumber \\
m_s^{2,0} & = & 0.1225  \, \mathrm{GeV}\nonumber \\
m_s^{3,0} & = & 0.1053  \, \mathrm{GeV}\nonumber \\
m_s^{4,0} & = & 0.0908  \, \mathrm{GeV}\nonumber \,.
\end{eqnarray}\begin{table}[h]
\begin{center}
\[
\begin{array}{|c|cc|}\hline
(k,l) & \delta R^{L+T}_{kl} & \delta R_{kl}  \\  \hline
(0,0) & 0.134 \pm 0.118 \pm 0.077   & 0.288 \pm 0.118 \pm 0.077 \\ 
(1,0) & 0.209 \pm 0.065 \pm 0.052   & 0.340 \pm 0.065 \pm 0.052\\
(2,0) & 0.242 \pm 0.044 \pm 0.038   & 0.356 \pm 0.044 \pm 0.038\\
(3,0) & 0.261 \pm 0.034 \pm 0.030   & 0.363 \pm 0.034 \pm 0.030\\ 
(4,0) & 0.276 \pm 0.028 \pm 0.025   & 0.367 \pm 0.028 \pm 0.025\\ \hline
\end{array}
\]
\begin{minipage}[t]{12cm}
\caption{\label{DeltaRLR}Total and vector contributions to the $\tau$-decay 
widths for the Leutwyler-Roos value of $|V_{us}|$.}
\end{minipage}
\end{center}
\end{table}

\noindent
It is now important to note the effects of replacing the scalar channels by a phenomenological ansatz in comparison to the 
OPE expressions: while the uncertainty in the 
longitudinal contribution could be suppressed to a point of negligibility (see below), we have the drawback of being much more sensitive 
to experimental accuracy. This can be seen easily from the tables above and is particularly drastic for the (0,0) moment: 
after we have subtracted the longitudinal contribution the experimental uncertainty is about one half of the absolute value of 
$\delta R^{L+T}_{00}$. From this observation we conclude that, with the present experimental accuracy, it should be worthwhile to 
include higher moments. These suffer from less experimental uncertainties since they are dominated by the better known low energy 
resonances. In contast, when using the theoretical expressions one usually restricts oneself to lower moments, since for higher 
moments the convergence of the OPE is not as good. 
In any case it is instructive to show this sensitivity by giving the results that arise, if $\delta R_{kl}$ is one experimental 
standard deviation below the central value, explicitly for the unitarity fit:
\begin{eqnarray}
m_s^{0,0} & = & 0.130 \, \mathrm{GeV}\nonumber \, \\
m_s^{1,0} & = & 0.139 \, \mathrm{GeV}\nonumber  \,\\
m_s^{2,0} & = & 0.122 \, \mathrm{GeV}\nonumber \, \\
m_s^{3,0} & = & 0.105 \, \mathrm{GeV}\nonumber \, \\
m_s^{4,0} & = & 0.091 \, \mathrm{GeV}\nonumber  \,.
\end{eqnarray}
These numbers are also in much better agreement with the results of other determinations and show less dependence on $k$, 
thus suggesting systematic experimental bias.
If we now add up all uncertainties, we find for the final results:
\[
\begin{array}{rcrcrcrcrcrl}
     &   &  &    &\scriptstyle  \mathrm{Exp.}&    & \scriptstyle |V_{us}|&   & \scriptstyle \mathrm{ Theor.} & & \scriptstyle \delta R^{L}_{kl}  &  \\
m_s^{0,0} & = & 197 & \pm& 68& \pm  & 33  &\pm & 33  & \pm & 3 & \, \mathrm{MeV}\nonumber\\
m_s^{1,0} & = & 164 & \pm& 25& \pm  & 16  &\pm & 14   &\pm  & 2 & \, \mathrm{MeV}\nonumber \\
m_s^{2,0} & = & 137 & \pm& 15& \pm  & 10   &\pm & 10  & \pm & 1 & \, \mathrm{MeV}\nonumber \\
m_s^{3,0} & = & 115 & \pm &10& \pm  & 7   &\pm & 11  &\pm  & 1 & \, \mathrm{MeV}\nonumber \\
m_s^{4,0} & = &  99 &\pm & 8 & \pm   & 6   &\pm & 15  &\pm  & 1 &  \, \mathrm{MeV}\nonumber  \,,
\end{array}
\]
where the theoretical uncertainties consist mainly of those from truncation and scaling. All others are rather small 
and listed in the table of App. A.4, where we have also compiled all numerical input parameters. Uncertainties arising from 
scaling and truncation of the dimension 4 contributions are not listed, but are all smaller than $1 \,\mathrm{MeV}$.
In our final results will include all uncertainties. 

The weighting of these results has to take into account the correlations between the different moments and we will 
take a conservative route in this work that should not overestimate the uncertainties too much, since these correlations are very large.
We give the central value as the uncorrelated weighted average:
\begin{equation}
\overline{m}_s = \frac{1}{w} \sum  \limits_{i} w_i m_{s,i}\,.
\end{equation}
The weighting factors $w_i$ are given by \begin{math} w_i =  1/\sigma_i^2 \end{math}, with $\sigma_i$ the total uncertainty for 
$m^{i,0}$ and $w$ is just the sum over the $w_i$. In 
case of uncorrelated uncertainties the final uncertainty would simply be given by \begin{math} 1/\sqrt{w} \end{math}, but
in our case this leads to an underestimate. It should be conservative, however, to consider the smallest of the $\sigma_i$
to be our uncertainty. 

This leads to a final value for the strange quark mass of 
\begin{equation}
m_s(\Mtau^2)=122\pm17\, \mathrm{MeV}  \,.
\end{equation}
At the standard scale of 2\,$\mathrm{GeV}$ this gives\\ \begin{center}
\fbox{ \parbox{5cm}{\[ m_s(2 \,\mathrm{GeV}) = 117 \pm 17 \, \mathrm{MeV}\,.\]} } \\
\end{center}
Performing all the error estimates with the Leutwyler-Roos fit, we find the following results:
\[
\begin{array}{rcrcrcrcrcrl}
   &   &    &    &\scriptstyle  \mathrm{Exp.}&    &\scriptstyle  |V_{us}|&   & \scriptstyle \mathrm{Theor.} &  & \scriptstyle \delta R^{L}_{kl}  & \\
m_s^{0,0} & = & 151 & \pm& 103 &\pm&  57  & \pm & 23   &\pm& 3& \, \mathrm{MeV}\nonumber \\
m_s^{1,0} & = & 142 & \pm& 30 &\pm&  24  & \pm &  12  &\pm&1  & \, \mathrm{MeV}\nonumber \\
m_s^{2,0} & = & 123 & \pm& 16 &\pm&  14  & \pm &  10   &\pm&1  & \, \mathrm{MeV}\nonumber \\
m_s^{3,0} & = & 105 & \pm& 11  &\pm&  10  &\pm  &  13  &\pm&1  & \, \mathrm{MeV}\nonumber \\
m_s^{4,0} & = & 91  & \pm& 9  &\pm&   8  &\pm  &  15  & \pm&1  &\, \mathrm{MeV}\nonumber \,,
\end{array}
\]
and the central value turns out to be
\begin{equation}
m_s(\Mtau^2)=107\pm18\, \mathrm{MeV}  \,.
\end{equation}
Again we evolve to a scale of 2\,GeV and find\\ \begin{center}
\fbox{ \parbox{5cm}{\[ m_s(2 \,\mathrm{GeV}) = 103 \pm 17 \, \mathrm{MeV}\,.\]} }\\
\end{center}

\chapter{Conclusion, Discussion and Outlook}

\hspace*{\fill}\begin{minipage}[t]{8.2cm}
\begin{footnotesize}
{\em I have to understand the world, you see.}\\
(Richard P. Feynman,\\ Surely You're Joking, Mr. Feynman)
\end{footnotesize}
\end{minipage}
\vspace{1cm}

\begin{figure}[bt]
\psfrag{100}{$\scriptstyle{100}$}
\psfrag{200}{$\scriptstyle{200}$}
\psfrag{300}{$\scriptstyle{300}$}
\psfrag{ms[MeV]}{$m_s(\mathrm{2\,GeV})[\mathrm{MeV}]$}
\psfrag{50}{$\scriptstyle{50}$}
\psfrag{PFESR[MK]}{    }
\psfrag{PsFESR[MK]}{pseudosc. FESR {\cite{MK}}}
\psfrag{scBR[JOP]}{scalar BSR {\cite{JOP2}}}
\psfrag{eplusSR[Nar]}{Lattice (unquenched)}
\psfrag{realeplus}{$e^{+}e^{-}$ SR{\cite{Narison99}}}
\psfrag{Lattice[Lub]}{Lattice (quenched)}
\psfrag{tau[PP]}{$\tau$-decays {\cite{PP2}}}
\psfrag{tau[multau]]}{$\tau$-decays {\cite{mutau}}}
\psfrag{0}{$\scriptstyle{0}$}
\psfrag{150}{$\scriptstyle{150}$}
\psfrag{250}{$\scriptstyle{250}$}
\psfrag{ourres}{Our values}
\psfrag{UnitarVus}{$\scriptstyle{\mathrm{Unitarity\,} V_{us}}$}
\psfrag{LRVus}{$\scriptstyle{\mathrm{LR\,} V_{us}}$}
\begin{center}
\epsfig{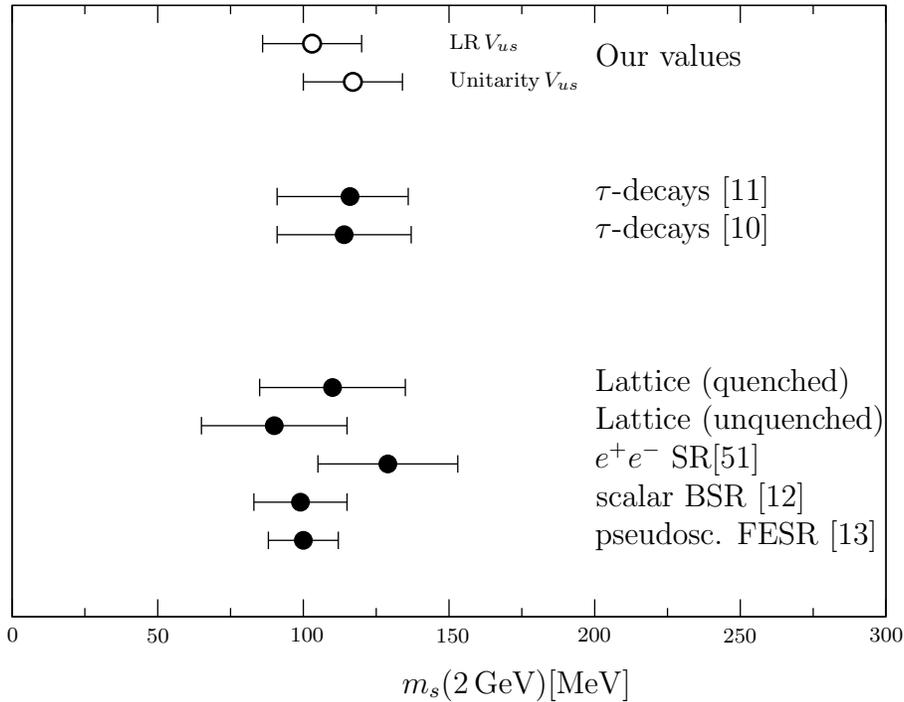}
\begin{minipage}[t]{12cm}
\caption{$m_s$(2\,GeV) for different determination methods. We have always quoted only the latest results, 
considering them an update of the previous ones; further references are given in the main text.}
\label{smass}
\end{minipage}
\end{center}
\end{figure}
\noindent
\noindent
We have calculated the strange quark mass from sum rules for $\tau$-decays and avoided the bad convergence in the 
scalar and pseudoscalar contribution to this sum rule by making a complete phenomenological ansatz for the scalar and 
pseudoscalar spectral functions. This allows us to give not just an upper bound on the value, but to actually 
calculate it. 

As a comparison we show the results of other determinations in Fig.~{\ref{smass}}. Except for earlier results from 
$\tau$-decays we give only the latest results. It seems that the results of different determinations have now finally 
become consistent, in contrast to the situation several years ago, when the results of, for example, sum rule and 
lattice calculations differed significantly. Note that our results too are in good agreement with the others.

The values from $\tau$-decays are from Pich/Prades {\cite{PP2}} and Chen et.~al.~{\cite{mutau}}.
Both use the total inclusive $\tau$-decay sum rule, but Pich/Prades also give an upper bound by making an ansatz for 
the spectral function that consists only of the kaon and pion pole and using the positivity of the spectral function. 
Chen et.~al.~take into account higher moments and use the same $k$ dependent truncation that we discussed in chapter 3, 
while Pich/Prades use the full perturbative series for all moments. Nevertheless both agree very well. 
It is clear that even our conservative estimate has already reduced the uncertainties of the 
$\tau$-decay determination and we will discuss possibilities of further improvements below.

The most promising other sum rule approach uses only the scalar or pseudoscalar channels by considering the correlator of
vector or axial vector divergences very similar to our final discussion in chapter {\ref{spectr}}. As mentioned, one can use Borel sum rules 
{\cite{jm2,JOP2,MK}} or finite energy sum rules {\cite{MK}}. In both cases the main uncertainty  lies in the parametrization of spectral function, since 
higher $K$ resonances are still rather poorly known and the spectral ansatz of {\cite{JOP2}} for the scalar channel also leaves significant uncertainties.
The pseudoscalar channel seems to allow for more precise determinations at the moment since it is strongly dominated by the kaon 
pole. In any case, significant improvements can only be made with experimental progress on the phenomenological spectral functions.
It might, however, be interesting to use $\tau$-like weight functions, replacing $\Mtau^2$ by $s_0$ and varying this parameter 
since the $\tau$-weights are sensitive to low energies and should effectively suppress anything but the poles.

Another sum rule approach is the $\tau$-like $e^{+}e^{-}$ sum rule proposed by Narison {\cite{Narison95}}. Note that this 
 has been criticised {\cite{Maltman98}} due to possibly large isospin breaking corrections. 
In its latest version {\cite{Narison99}} that claims to be less sensitive to these effects one discusses the quantity 
\begin{equation}
R_{\tau,\phi}\equiv \frac{3\left|V_{ud}\right|^2}{2 \pi \alpha^2} S_{EW} \int\limits_{0}^{\Mtau^2} ds \left(1-\frac{s}{\Mtau^2}\right)^2
\left(1+\frac{2s}{\Mtau^2}\right)\frac{s}{\Mtau^2}\sigma_{e^{+}e^{-}\rightarrow \phi^{+}\phi^{-}}\,,
\end{equation}
as well the SU(3)-breaking combination
\begin{equation}
\Delta_{1\Phi}\equiv R_{\tau,1}-R_{\tau,\phi}\,,
\end{equation}
or isopin breaking
\begin{equation}
\Delta_{1,0}\equiv R_{\tau,1}-3R_{\tau,0}\,,
\end{equation}
where $R_{\tau,1}$ is the isovector and $R_{\tau,0}$ the isoscalar decay width. Just like the \mbox{$\tau$-decay} sum rules these sum rules suffer from 
cancellations that cause rather small {\em a priori} experimental uncertainties to become large.

Finally, there are of course lattice results for $m_s$, and we have quoted separately quenched {\cite{Lubicz2000}} and unquenched 
{\cite{privdisc}} values, since unquenched determinations tend to give smaller values and additionally the systematic uncertainty
is unclear. On the other hand the quenched values are given by taking an average over all determinations and adding an uncertainty
of $\pm$ 10 as systematic quenching uncertainty. It shall be very interesting to follow the progress here since the last word is 
most certainly not yet spoken.

\vspace{1cm}
\noindent
Going back to $\tau$-decays, we should ask ourselves about what further possibilities they hold. We have seen that our strange quark 
mass is extremely sensitive to experimental results, and with more precise data, as is expected from the coming charm and $B$ 
factories, our results will improve significantly. This is not only due to purely experimental uncertainties: with better 
experimental precision we can restrict ourselves to lower $k$-moments that also have a smaller theoretical uncertainty.
Additionally, this should eventually get rid of the $k$-dependence of the results, if its origin is indeed experimental,  and then make the weighting a bit more reliable. 
Of course it will then become necessary to investigate in more detail the smaller uncertainties that have become irrelevant in our 
discussion. 

There is one last point that needs to be clarified, and that is the question of the CKM element $\left|V_{us}\right|$. We have shown 
that due to the cancellations that make the experimental uncertainties large, we are also rather sensitive to the value of this 
parameter. This can result in one of the following conclusions: first, that we can improve our strange mass determination if this 
question is answered and second that we can (as has been done before) turn our investigation around and provide input for $m_s$ in 
order to determine $\delta R$ and from that try to obtain values for $\left|V_{us}\right|$. At the moment this is of course 
somewhat speculative because it relies very much on further experimental input, but the possibility should not be entirely 
neglected\footnote{The potential of this approach is discussed in detail in Ref. {\cite{Gamiz:2002nu}}}.

%% file: AppA2.tex
\chapter{Miscellaneous Coefficients and Constants}

\section{Coefficients for Running Masses and Couplings}
There are multiple conventions for defining the $\beta$ function, that mostly differ by the sign and/or a factor 2, that arises
if the derivative is not taken to be with respect to $\mu$ but to $\mu^2$. We have adopted the convention of Pascual and Tarrach
\cite{PT}, except for a minus sign and an extra factor $a$, so that the expansion coefficients for the $\beta$ function in the $\overline{MS}$ scheme are
\begin{eqnarray}
\beta_1 & = & \frac{1}{6} \left(11N - 2f \right) \, , \quad \beta_2 = \frac{1}{12} \left(17 N^2 -5N \, f - 3C_f \, f \right)\, , 
\\ \nonumber
\beta_3 & = & \frac{1}{32} \left( \frac{2857}{54}N^3 - \frac{1415}{54} N^2 f + \frac{79}{54}Nf^2 - \frac{205}{18}NC_ff +\frac{11}{9} 
C_f f^2 +C_f^2f \right) \, ,\\ \nonumber
\beta_4 & = & \frac{140599}{2304} +  \frac{445}{16}\zeta(3) \,  .   
\end{eqnarray}
For $\gamma(a)$ we use
\begin{eqnarray}
\gamma_1 & = & \frac{3}{2} C_f \, , \qquad \gamma_2 = \frac{C_f}{48} \left(97 \, N +9C_f- 10f \right) \,,\\ \nonumber 
\gamma_3 & = & \frac{C_f}{32} \left( \frac{11413}{108} \, N^2 - \frac{129}{4} N C_f -\left( \frac{278}{27} + 24 \zeta(3) \right)
N\,f +\frac{129}{2}C_f^2 \right.\\ 
 & & \left. - \Big(23 - 24\zeta(3)\Big)C_f f -\frac{35}{27}f^2 \right)\, ,\nonumber \\\nonumber
\gamma_4 & = & \frac{2977517}{20736} - \frac{9295}{216} \zeta(3) + \frac{135}{8} \zeta(4) - \frac{125}{6} \zeta(5)\,.
\end{eqnarray}
Here $N$ is the number of colors, $f$ the number of flavors and $C_f$ is the Casimir invariant, $C_f =(N^2-1)/2N$. For simplicity 
we have given both $\beta_4$ and $\gamma_4$ for three flavors and colors only.

\section{Weight Functions for Higher Moments}

A general representation of the Function $\mathcal{F}^{kl}_{j}$, where $j$ is either $L$ or $L+T$, is
\begin{eqnarray}
\mathcal{F}^{kl}_{L+T}(x) & = & 2 \left( 1-x \right)^{3+k} \sum\limits_{n=0}^{l} \frac{l!}{(l-n)!n!} (x-1)^n 
\frac{(6+k+n)+2(3+k+n)x}{(3+k+n)(4+k+n)} \,, \nonumber\\\nonumber
\mathcal{F}^{kl}_{L}(x) & = & 3 \left( 1-x \right)^{3+k} \sum\limits_{n=0}^{l} \frac{l!}{(l-n)!n!} 
\frac{(x-1)^n}{3+k+n} \,.
\end{eqnarray}
Additionally, we give the functions for the values of $k$ and $l$ that we consider:

\begin{center}
\[  \begin{array}{|c|c|c|}\hline
 (k,l) & \qquad \mathcal{F}^{kl}_{L+T}(x) \qquad  & \qquad\mathcal{F}^{kl}_{L}(x) \qquad \\ \hline \hline &&\\
 (0,0) & \qquad (1-x)^3(1+x)\qquad &\qquad (1-x)^3 \qquad \\ &&\\
 (1,0) & \frac{1}{10}(1-x)^4(7+8x)   & \frac{3}{4}(1-x)^4  \\   &&\\
 (2,0) & \frac{2}{15}(1-x)^5(4+5x) & \frac{3}{5}(1-x)^5 \\  &&\\
 (3,0) & \frac{1}{21}(1-x)^6(9 + 12 x)  & \frac{1}{2}(1-x)^6  \\  &&\\ 
 (4,0) & \frac{1}{28} (1 - x)^7(10+14x) & \frac{3}{7}(1-x)^7 \\ &&\\ \hline 
\end{array}   \]
\end{center}

\section{Resonance Parameters}

For the pseudoscalar spectral functions we need the following parameters of low lying resonances, that have been taken from 
{\cite{MK,PDG2000,MK1}}.

\begin{center}
\[  \begin{array}{|cccc|}\hline
  &\mathrm{ Mass (MeV)}  &\mathrm {Width \mathrm(MeV)} & \mathrm {Decay\, constant (MeV)} \\ \hline \hline
\pi  & 138  &  & 92.4 \pm 0.3 \\
\pi(1300)  & 1300 \pm 100 & 400 \pm 200 & 2.20\pm 0.46\\ 
\pi(1800)  & 1800 \pm 13  & 210 \pm 15 & 0.19 \pm 0.19 \\ 
 K  & 495 &  &  113 \pm 1    \\
 K(1460)  & 1460 \pm 60 & 260 \pm 10 & 21.4 \pm 2.8\\
 K(1830)  & 1830 & 250 & 4.5 \pm 4.5 \\ \hline
\end{array}   \]
\end{center}
The uncertainty of the $\pi(1300)$ 
width and coupling are correlated, since some of the uncertainty from the width manifests itself in the determination of the coupling 
constants, but this correlation is not important for our strange mass determination. 
Since the excited $K$ resonances are listed by the PDG as needing confirmation, they do not give uncertainties for widths 
and masses. They do, however, give several central values (in the case of the $K(1460)$), and the numbers we quote are estimated from them.  

\section{Numerical Input Parameters}
For reading convenience we compile here all our numerical input parameters including the corresponding uncertainties. 
Finally, we give the induced uncertainties for $m_s$.
\begin{center}
\[  \begin{array}{|c|c|c|}\hline
\mathrm{Parameter} &\mathrm{ Value} &\mathrm{ \Delta m_s} (\mathrm{MeV})\\\hline \hline
\alpha_s(\Mtau^2)& 0.334 \pm 0.022  & 2 \\&&\\
\langle \bar{s}s \rangle & 0.8 \pm 0.2~ \langle \bar{u}u \rangle  & 7  \\&&\\
m_u & m_s(0.029\pm0.003)& \mathcal{O} (10^{-3})  \\&&\\
m_d  & m_s(0.053\pm0.002) & \mathcal{O} (10^{-1}) \\&&\\
\langle \bar{u}u \rangle & -f_{\pi}M_{\pi}/(m_u+m_d)(1-0.047) & \mathrm{see}\, m_u\,\mathrm{ and}\, m_d \\&&\\
S_{EW}&1.0201 \pm0.0003 &  <\mathcal{O} (10^{-3})\\&&\\
\langle aFF \rangle &(0.5-2) \,2.1 \, 10^{-2} \, \mathrm{GeV}^{4} & <\mathcal{O} (10^{-3})\footnotemark\\&&\\
\eta_{ud}&1& --\\&&\\
\eta_{us}&0.6&--\\&&\\
\rho_I& 1/0.6 &--\\&&\\
\langle g {\bar q_i} \sigma F q_i \rangle & 0.8~ \langle {\bar q_i}q_i \rangle&  --\\\hline
\end{array}   \]
\end{center}
\footnotetext{In our calculations the gluon condensate drops out, but it 
appears in higher dimension contributions suppressed by quark masses.}
According to the way we estimate the uncertainties, we give here those 
uncertainties that contribute to the moment with the smallest total 
uncertainty. 
Additionally we need the CKM matrix elements $|V_{us}|$ and $|V_{ud}|$. For $|V_{ud}|$ we take $0.97525\pm 0.00046$,
while for $|V_{us}|$ we discuss both the PDG unitarity fit \mbox{$|V_{us}| = 0.2225\pm0.0021$} and the Leutwyler/Roos
value of  $|V_{us}|=0.2196\pm0.0026$. Uncertainties for the strange mass determination are discussed in the main part
of the text.

%% file: AppB.tex
\chapter{Higher dimension terms for the OPE}

\section{Dimension-two Coefficients}
The needed coefficients for the longitudinal contribution are:
\begin{equation}
d_{00} = 1 \,, \quad d_{10} = \frac{17}{3} \,, \quad d_{20} = \frac{9631}{144} - \frac{35}{2} \zeta(3) \,,
\end{equation}
\begin{displaymath}
d_{30} = \frac{4748953}{5184} - \frac{91519}{216}\zeta(3) - \frac{5}{2}\zeta(4) + \frac{715}{12}\zeta(5) \, . 
\end{displaymath}
For the transverse:
\begin{equation}
c_{00} = 1 \,, \quad c_{10} = \frac{13}{3} \,, \quad c_{20} = \frac{23077}{432} + \frac{179}{54} \zeta(3) -\frac{520}
{27} \zeta(5) \,,
\end{equation}
\begin{displaymath}
e_{00} = 0 \,, \quad \, e_{10} = \frac{2}{3} \,, \quad \, e_{20} = \frac{769}{54} - \frac{55}{27} \zeta(3) -\frac{5}{27}
 \zeta(5) \,, \qquad
\end{displaymath}
\begin{displaymath}
\; f_{00} = 0 \,, \quad \, f_{10} = 0 \,, \quad \,\, f_{20} = -\frac{32}{9} + \frac{8}{3} \zeta(3) \,.\qquad \qquad \qquad \quad
\end{displaymath}
The notation is mainly according to Ref. {\cite{PP2}}, albeit slightly different, where all of the coefficients have been 
compiled. 

\section{Dimension-four Coefficients}
We will present here the complete $L+T$ dimension 4 contribution. This includes operators $m^4$, quark condensates multiplied
by masses and finally the gluon condensate operator, so that the general expression reads:
\begin{equation}
D^{L+T,4}_{ij}(s)  =   \frac{1}{s^2} \sum\limits_{n} \Omega^{L+T}_{n}(s/\mu^2) a^n\,,
\end{equation}
where 
\begin{eqnarray}
\Omega^{L+T}_{n}(s/\mu^2)&=&\frac{1}{6} \langle aFF \rangle p_n^{L+T}(s/\mu^2) + \sum\limits_{k} \langle m_k \bar{q}_k q_k \rangle r^{L+T}_n (s/\mu^2)  \\
&  & + 2 \langle m_i \bar{q_i} q_i + m_j \bar{q_j} q_j \rangle q^{L+T}_{n}(s/\mu^2)  -\frac{3}{\pi^2} \left(m_i^4+m_j^4\right) h^{L+T}_{n}(s/\mu^2)  \nonumber \\
&  & \pm  \frac{8}{3} \langle m_j \bar{q_i} q_i + m_i \bar{q_j} q_j \rangle t^{L+T}_{n}(s/\mu^2) + \frac{3}{\pi^2} m_i^2 m_j^2 g^{L+T}_{n}(s/\mu^2) \nonumber\\
&  & \mp \frac{5}{\pi^2} m_i m_j \left( m_i^2 +m_j^2 \right) k^{L+T}_{n}(s/\mu^2)-\frac{3}{\pi^2} \sum\limits_{k} m_k^4 j^{L+T}_{n}(s/\mu^2)\nonumber \\
&  & -\frac{6}{\pi^2}\sum\limits_{k\not=l} m_k^2m_l^2  u^{L+T}_{n}(s/\mu^2)\,.\nonumber
\end{eqnarray}
This simplifies considerably when the only SU(3) breaking contributions 
are taken, and we have tried to write the equation in such a way as that these simplifications become most easy to see. We will refer 
to the terms by the letters of the coefficients that they carry. Obviously the 
first line will drop out, as will the $u$ and the $j$ terms in the last two lines. Additionally the $t$ and $k$ terms drop out 
if we add vector and axial vector channels. The remaining terms can be written as
\begin{equation}
\label{DIM4}
s^2 \left(D^{L+T,4}_{ud}(s)-D^{L+T,4}_{us}(s) \right) =  -4 \delta O_4 (-s) \sum\limits_n q_n (q^2/\mu^2)\, a^n+ \nonumber
\end{equation}
\begin{equation}
\frac{6}{\pi^2} m_s^4 \left(1-\varepsilon_d^2\right) \sum\limits_n \left[ \left(1+\varepsilon_d^2 \right) h^{L+T}_{n}(s/\mu^2)
-\varepsilon_u^2 g^{L+T}_{n}(s/\mu^2) \right]a^n \,.
\end{equation}
We can now easily arrive at Eq. (\ref{vectordim4}), if we introduce the following notations:
\begin{eqnarray}
Q^{L+T}_{kl}& = & \frac{\Mtau^4}{4 \pi i} \sum\limits_{n} q^{L+T}_{n}(\xi) \oint\limits_{|s|=\Mtau^2} \frac{ds}{s^3} 
\mathcal{F}^{kl}_{L+T}(s/\Mtau^2) a^n(-\xi^2 s)\,,\\\nonumber
H^{L+T}_{kl} & = & \frac{\Mtau^4}{2 \pi i}\sum \limits_{n} h^{L+T}_{n}(\xi) \oint\limits_{|s|=\Mtau^2} \frac{ds}{s^3} \mathcal{F}^{kl}_{L+T}(s/\Mtau^2)
\left(\frac{m(-\xi^2 s)}{m(\Mtau^2)} \right)^4 a^n(-\xi^2 s)\,,\\\nonumber 
G^{L+T}_{kl} & = & \frac{\Mtau^4}{4 \pi i}\sum \limits_{n} g^{L+T}_{n}(\xi) \oint\limits_{|s|=\Mtau^2} \frac{ds}{s^3} \mathcal{F}^{kl}_{L+T}(s/\Mtau^2)
\left(\frac{m(-\xi^2 s)}{m(\Mtau^2)} \right)^4 a^n(-\xi^2 s)\,.
\end{eqnarray}
We have set $\mu^2/s =- \xi^2$ to allow for an error estimate. The parameter $\xi$ is later varied in the interval $0.75 < \xi<2$.

\newpage
\noindent
The $C_{n0}$ have been calculated to second order for the condensates and, as given in {\cite{PP2}}, are 
\[ \begin{array}{rclrclrcl}
   p_{00}^{L+T} & = & 0\,,\qquad & p_{10}^{L+T} &=& 1\,,\qquad & p_{20}^{L+T}& = & \displaystyle \frac{7}{6} \,, \\ &&&&&&&& \nonumber\\
   r_{00}^{L+T} & = & 0\,,\qquad & r_{10}^{L+T} &=& 0\,,\qquad & r_{20}^{L+T}& = & \displaystyle -\frac{5}{3}+\frac{8}{3} \zeta(3) \,, \\&&&&&&&& \nonumber\\
   q_{00}^{L+T} & = & 1\,,\qquad & q_{10}^{L+T} &=& -1\,,\qquad  & q_{20}^{L+T}& = & \displaystyle -\frac{131}{24} \,, \\&&&&&&&& \nonumber\\
   t_{00}^{L+T} & = & 0\,,\qquad & t_{10}^{L+T} &=& 1\,,\qquad & t_{20}^{L+T}& = & \displaystyle \frac{17}{2} \,. \\
\end{array}  \]
We need scaling of these coefficients as an error estimate, but to the order we are calculating there are only terms 
linear in the logarithms. The nonzero ones are:
\[ \begin{array}{rclrclrcl}
r_{21}^{L+T} & = &\displaystyle -\frac{2}{3}\,,\qquad & q_{21}^{L+T} &=&\displaystyle \frac{9}{4} \,,\qquad & t_{21}^{L+T}& = &\displaystyle -\frac{9}{4} \,. \\
\end{array}  \]
On the other hand the mass coefficients have only been calculated up to first order, which is, including the scaling behavior:
 \[ \begin{array}{rclrcl}
  h_{0}^{L+T} & = & 1-\frac{1}{2}L\,,\qquad & h_{1}^{L+T} &=&\displaystyle \frac{25}{4} - 2\zeta(3)-\frac{25}{6} L + L^2\,, \\&&&&& \nonumber\\
  k_{0}^{L+T} & = & 0\,,\qquad & k_{1}^{L+T} &=&\displaystyle 1-\frac{2}{5}L \,, \\&&&&& \nonumber\\
  g_{0}^{L+T} & = & 1\,,\qquad & g_{1}^{L+T} &=&\displaystyle \frac{94}{9}-\frac{4}{3}\zeta(3)-4L \,, \\&&&&& \nonumber\\
  j_{0}^{L+T} & = & 0\,,\qquad & j_{1}^{L+T} &=&\displaystyle 0\,, \\ &&&&& \nonumber\\
  u_{0}^{L+T} & = & 0\,,\qquad & u_{1}^{L+T} &=&\displaystyle 0\,. \\ 
 \end{array}  \]
where $L = \ln \frac{-q^2}{\mu^2}$. The longitudinal coefficients are
\[ \begin{array}{rclrclrcl}
  h_{0}^{L} & = & 1-\frac{1}{2}L\,,\qquad & h_{1}^{L} &=&\displaystyle \frac{41}{6}-2\zeta(3) -\frac{14}{3}L + L^2\,, \\&&&&& \nonumber\\
  k_{0}^{L} & = & 1-\frac{1}{3}L\,,\qquad & k_{1}^{L} &=&\displaystyle 8-\frac{4}{3}\zeta(3) - \frac{40}{9}L+ \frac{2}{3}L^2\,, \\&&&&& \nonumber\\
  j_{0}^{L} & = & 0\,,\qquad & j_{1}^{L} &=& 0\,. \\ 
  \end{array}  \]

%% file: taudecay.tex
\chapter{Proof of the Formula for $\tau$-Decays}
\label{proof}

\hspace*{\fill}\begin{minipage}[t]{8.2cm}
\begin{footnotesize}
{\em Know how to solve every problem that has ever been solved.}\\
(Richard P. Feynman)\\
\end{footnotesize}
\end{minipage}

\noindent
The central formula of this work (this is by no means to suggest that it is new, but all that we are
doing relies on it) is (\ref{mainf}) that we have stated without proof. 
It was originally derived by Tsai in 1971 {\cite{tsai}} in the framework of Current Algebras.
In this chapter we present a somewhat more elaborate and different approach that basically relies on the optical theorem and 
the Cutkosky cutting rules that follow from it. 
In any case of doubt we will be using the 
notation of Peskin and Schroeder {\cite{PS}}.

First, remember that we have decomposed the correlator as follows:
\begin{equation}
\Pi^{\mu\nu}(q)  =  (q^{\mu}q^{\nu} - q^2 g^{\mu\nu})\Pi^{(T)}(q^2) + q^{\mu}q^{\nu} \Pi^{(L)}(q^2) \, .
\end{equation} 
We shall first be concerned with the transversal part, and we will see that the longitudinal component can then be 
easily inferred by the calculation.
The general formula for a decay rate $\Gamma$ is:
\begin{equation}
\Gamma = \frac{1}{2M_A} \int \left( \prod_{f} \, \frac{d^3 p}{(2 \pi)^3}\frac{1}{2E_f} \right) \Big|\mathcal{M} \Big(A 
\rightarrow \sum\limits_f f \Big) \Big|^2 (2\pi)^4
\delta(p_A-\sum\limits_f p_f) \, ,
\end{equation}
where $ |\mathcal{M}|^2$ is the squared decay matrix element, and we have let a particle $A$ decay into final states $f$.

\begin{figure}[bt]
\psfrag{2Im}{2 Im}
\psfrag{tau}{$\tau$}
\psfrag{nutau}{$\nu_{\tau}$}
\psfrag{a}{a)}
\psfrag{b}{b)}
\psfrag{q}{$q$}
\begin{center}
\epsfig{file=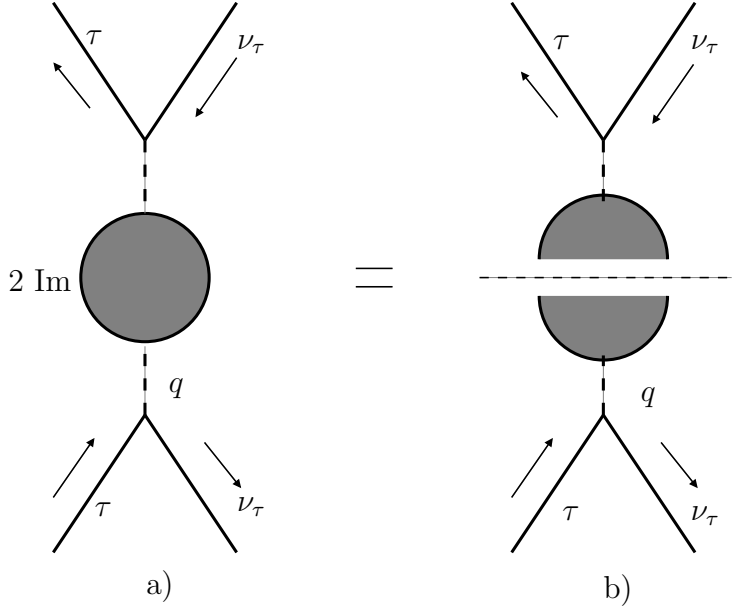,height=8cm,clip=}
\begin{minipage}[t]{12cm}
\caption{\label{taudec}The squared matrix element for $\tau$ decays into \mbox{hadrons b}) can be calculated as the imaginary part of the connected 
diagram a); a phase space integral as well as a sum over all possible hadronic states is understood. }
\end{minipage}
\end{center}
\end{figure}
\noindent
According to the optical theorem this squared matrix element as in Fig.~{\ref{taudec}} b) is equal
to twice the imaginary part of the doubled diagram of Fig.~{\ref{taudec}} a) -  these are the aforementioned Cutkosky Rules. 
In this process a we have to 
sum over all possible hadronic final states and integrate over their respective phase space factors. Then the delta function disappears
and only the phase space integration over the momentum of the neutrino is left. First, however, we need to calculate the 
matrix element for Fig.~{\ref{taudec}} a), which is:
\begin{eqnarray}
\label{Matrixel}
i \mathcal{M} & = & (-i)^2 \frac{g^2}{2} {\bar u}(p_{\tau}) \gamma_{\mu} \left( \frac{1-\gamma^5}{2} \right) u(p_{\nu}) \frac{-i}{M_W^2} 
\left(\frac{i}{4} \frac{g^2}{2} \Pi^{\mu\nu}(q)\right) \\ \nonumber
 &   & \quad \times \frac{-i}{M_W^2}{\bar u}(p_{\nu}) \gamma_{\nu} \left( \frac{1-\gamma^5}{2} \right) u(p_{\tau})\,.
\end{eqnarray}
Here we have used the subscript $\nu$ for a Lorentz index and to denote the momentum of the neutrino. It should be obvious from the context, however, 
which of the two is meant in all steps. 

Let us discuss the treatment of the spinors now and for that drop all the irrelevant constants. We need to sum over the final and 
average over the initial spin:
\begin{eqnarray}
 &   & \frac{1}{2} \sum\limits_{\mathrm{spins}}   {\bar u}(p_{\tau}) \gamma_{\mu} \left( \frac{1-\gamma^5}{2} \right) u(p_{\nu}){\bar u}(p_{\nu}) \gamma_{\nu} \left( \frac{1-\gamma^5}{2} \right) u(p_{\tau})\\
 & = & \frac{1}{2} \mathrm{Tr} \left[ \!\not\!p_{\tau}  \gamma_{\mu} \left( \frac{1-\gamma^5}{2} \right) \!\not\! p_{\nu}    \gamma_{\nu} \left( \frac{1-\gamma^5}{2} \right) \right]\\
 & = & \frac{1}{2}\mathrm{Tr} \left[ \!\not\!p_{\tau}  \gamma_{\mu}  \!\not\!p_{\nu}    \gamma_{\nu} \left( \frac{1-\gamma^5}{2} \right) \right]\\
 & = & \left(  p_{\tau}^{\mu} p_{\nu}^{\nu}  +  p_{\tau}^{\nu}  p_{\nu}^{\mu} - g^{\mu\nu}  p_{\nu} \cdot p_{\tau}
 -i\varepsilon^{\alpha\mu\beta\nu} p_{\tau\alpha} p_{\nu\beta} \right)\,.
\end{eqnarray}
This expression has to be dotted into the matrix element (\ref{Matrixel}). In doing so the $\varepsilon^{\alpha\mu\beta\nu}$ 
term will drop out due to symmetry and we obtain:
\begin{eqnarray}
\label{diff}
\frac{1}{2}\sum\limits_{\mathrm{spins}} |\mathcal{M}|^2 & = & \frac{g^4}{16 M_W^4} \left(  p_{\tau}^{\mu} p_{\nu}^{\nu}  +  p_{\tau}^{\nu}  p_{\nu}^{\mu} - 
g^{\mu\nu}  p_{\nu} \cdot p_{\tau} \right)   (q_{\mu}q_{\nu}- s \, g_{\mu\nu})\Pi^{(T)}(s)\qquad \\\nonumber
 & = & \frac{g^4}{16 M_W^4} \left[ s \, p_{+} \cdot p_{\tau} + 2(p_{+}q)(q p_{\tau}) \right]\Pi^{(T)}(s)\,.
\end{eqnarray} 
In the case of massless neutrinos\footnote{The following relations are most easy to see in the CMS}
\begin{equation}
2p_{\tau} p_{\nu} = 2 p_{\nu} q = M_{\tau}^2 - s, \qquad 2p_{\tau} q =M_{\tau}^2 + s \, ,
\end{equation}
so that the after taking the imaginary part and multiplying with the necessary numerical factors our Matrix element is:
\begin{equation}
\frac{g^4}{16} \frac{M_{\tau}^4}{M_{W}^4} \left(1-\frac{s}{M_{\tau}^2}\right) \left(1 + 2 \frac{s}{M_{\tau}^2} \right) \mathrm{Im} \Pi^{(T)}(s)\,.
\end{equation}
Finally we need to multiply with the phase space factor. Changing the integration variable from $p_{\nu}$ to $s$ we see from
\begin{equation}
\left| p_{\nu} \right| = \frac{1}{2} \frac{M_{\tau}^2-s}{M_{\tau}}
\end{equation}
that $dp_{\nu} = -ds/2\Mtau$. Additionally for the neutrino $p = E$ and our total phase space factor turns out to be 
\begin{equation}
 -\int  \frac{d\Omega}{4 \, 2} \frac{1}{\left(2 \pi \right)^3} ds \left( \frac{M_{\tau}^2-s}{M_{\tau}^2} \right)\,,
\end{equation}
and the minus sign vanishes if we choose the appropriate order of integration bounds, i.e. integrate from 0 to $\Mtau^2$.
Throwing all of this together and performing the angular integral the decay rate is 
\begin{equation}
\Gamma = \frac{1}{2\Mtau} \int\limits_{0}^{\Mtau^2} \frac{ds}{\Mtau^2} \frac{g^4}{2^8 \pi^2} \frac{\Mtau^6}{M_{W}^4} \left( 1 -\frac{s}{\Mtau^2} \right)^2 \left(1 + 2 \frac{s}{M_{\tau}^2} \right) \mathrm{Im}\Pi^{(T)}(s) \,.
\end{equation}
At this point we can go back to include the longitudinal part. The calculation is in principle similar, except for Eq. (\ref{diff}).
If we take only the $q^{\mu}q^{\nu}$ term the second line becomes
\begin{equation}
\frac{g^4}{16 M_{W}}\left[-s \, p_{+} \cdot p_{\tau} + 2(p_{+}q)(q p_{\tau}) \right]\Pi^{(L)}(s)\,.
\end{equation} 
Adding this to the result for the transverse correlator we arrive at the final result for the decay rate of the $\tau$ into 
hadrons:
\begin{eqnarray}
\Gamma (\tau \rightarrow \mathrm{hadrons}) & = & \frac{1}{2\Mtau} \int\limits_{0}^{\Mtau^2} \frac{ds}{\Mtau^2} \frac{g^4}{2^8 \pi^2} \frac{\Mtau^6}{M_{W}^4}
\left( 1 -\frac{s}{\Mtau^2} \right)^2\\ \nonumber 
 & & \qquad \times \left[ \left(1 + 2 \frac{s}{M_{\tau}^2} \right)
 \mathrm{Im}\Pi^{(T)}(s) +  \mathrm{Im}\Pi^{(L)}(s) \right] \,.
\end{eqnarray}
The decay rate of the $\tau$ into leptons is
\begin{equation}
\Gamma(\tau \rightarrow e \nu_{\tau} {\bar{\nu_e}}) =  \frac{G_F^2 \Mtau^5}{192 \pi^3}\,,
\end{equation}
where the Fermi constant $G_F$ is defined as $G_F/\sqrt{2} = g^2/(8 M_W^2)$, so that we find 
\begin{equation}
\Gamma(\tau \rightarrow e \nu_{\tau} {\bar{\nu_e}}) =  \frac{g^4 \Mtau^5}{32 \cdot 192 M_{W}^4 \pi^3}\,.
\end{equation}
Finally we can now give the expression for $R_{\tau}$
\begin{equation}
R_{\tau} = 12 \pi \int\limits_{0}^{\Mtau^2} \frac{ds}{\Mtau^2} \left( 1 -\frac{s}{\Mtau^2} \right) \left[ \left(1 + 2 \frac{s}{M_{\tau}^2} 
\right)
 \mathrm{Im}\Pi^{(T)}(s) +  \mathrm{Im}\Pi^{(L)}(s) \right] \,,
\end{equation}
which is the desired result.